\documentclass[traditabstract]{aa}  
\usepackage{graphicx,natbib, txfonts, threeparttable, colortbl}
\usepackage[normalem]{ulem}
\bibpunct{(}{)}{;}{a}{}{,}    % To get A&A format
\def\eq#1{\begin{equation} #1 \end{equation}}
\def\eqn#1{\begin{eqnarray} #1 \end{eqnarray}}
\def\kms {\hbox{km\,s$^{-1}$}}
\def\ergs {\hbox{erg\,s$^{-1}$}}
\newcommand{\e}[1]{\times 10^{#1}}
\newcommand{\iso}[1]{${}^{#1}$}

\newcommand{\wl}{$\lambda$}
\newcommand{\msun}{M$_\odot$}
\newcommand{\wll}{$\lambda \lambda$}

\def\Mo  {\hbox{$~\rm M_{\odot}$}}
\def\bestti {$1.5\e{-4}$\Mo}
\def\besttiwlim {$1.5_{-0.5}^{+0.5}\e{-4}\Mo$}
\setcitestyle{round,aysep={},yysep={,}}
\graphicspath{{/home/andersj/graphs/}{/home/andersj/graphs/MCspectra/}{/home/andersj/draw/}}

\begin{document}

\title{The \iso{44}Ti-powered spectrum of SN 1987A}

   \author{Anders Jerkstrand\inst{1,2} 
          \and         
          Claes Fransson\inst{1,2}
          \and
          Cecilia Kozma\inst{1,2}
         }

   \offprints{A. Jerkstrand, andersj@astro.su.se}

   \institute{Department of Astronomy,
              Stockholm University, Alba Nova University Centre, SE-106 91 Stockholm \\
              \email{andersj@astro.su.se}       	
   \and The Oskar Klein Centre, Stockholm University
   }
   \date{Accepted for publication in A\&A, March 15 2011.}

  \abstract{
   SN 1987A provides a unique opportunity to study the evolution of a supernova from explosion into very late phases. Owing to the rich chemical structure, the multitude of physical processes involved and extensive radiative transfer effects, detailed modeling is needed to interpret the emission from this and other supernovae.
   In this paper, we analyze the late-time (about eight years) Hubble Space Telescope spectrum of the SN 1987A ejecta, where $^{44}$Ti is the dominant power source. Based on an explosion model for a 19 \msun~progenitor, we compute a model spectrum by calculating the degradation of positrons and gamma-rays from the radioactive decays, solving the equations governing temperature, ionization balance and NLTE level populations, and treating the radiative transfer with a Monte Carlo technique. We obtain a UV/optical/NIR model spectrum that reproduces most of the lines in the observed spectrum with good accuracy. We find non-local radiative transfer in atomic lines to be an important process also at this late stage of the supernova, with $\sim$30\% of the emerging flux in the optical and NIR coming from scattering/fluorescence.  We investigate the question of where the positrons deposit their energy, and favor the scenario where they are locally trapped in the Fe/He clumps by a magnetic field. Energy deposition into these largely neutral Fe/He clumps makes Fe I lines prominent in the emerging spectrum. With the best available estimates for the dust extinction, we determine the amount of $^{44}$Ti produced in the explosion to be \besttiwlim.}


\maketitle

%===============================================

\section{Introduction}
The distance of only $\sim$50 kpc to SN 1987A makes a number of unique
observations possible. In particular, the supernova can 
after more than 20 years still be observed as resolved ejecta, glowing from
radioactive input. %Usually not an observable phase for supernovae, this is in many ways a unique astrophysical object.
About four years after exploding, the supernova entered the $^{44}$Ti-dominated phase. Since then, its emission has been that of a cool ($\sim$$10^2$ K) gas  powered by radioactive decays, complemented by freeze-out emission from the hydrogen envelope. Owing to the long life-time of $^{44}$Ti ($\sim$85 years), the character of the spectrum has changed little since this transition.%, although the absolute flux has decreased due to a decreasing contribution by freeze-out emission.

To understand and interpret the late-time emission from the ejecta, detailed spectral synthesis modeling is needed.
The spectral formation process in supernovae is complex, especially at late times when all important processes are non-thermal, but radiative transfer effects may still be important. High-energy gamma-rays and positrons from the radioactive decays deposit their energy into free and bound electrons, producing a population of fast electrons which heat, excite and ionize the gas. The ionizations produce secondary electrons that additionally contribute to these processes. The degradation process is quantified by solving the Boltzmann equation for each of the zones present in the supernova \citep[e.g.][KF92 hereafter]{Xu1991ED, Kozma1992}. Once the deposition in the various channels have been determined, the temperature, ionization structure and NLTE level populations can be computed by solving the equations for thermal and statistical equilibrium or their time-dependent variants \citep[e.g.][KF98 a, b hereafter]{deKool1998, Kozma1998I, Kozma1998II}. Because the ionization balance affects the solution for the radioactive deposition, the equation systems are coupled and a solution is found by iteration.

The equilibrium also depends on the radiation field, which should simultaneously be solved for. The strong velocity gradients in supernovae imply unique radiative transfer effects.
%The radiative transfer situation involves thousands of atomic lines, that together can provide high opacities because of the velocity gradient \textbf{in the ejecta}. 
As they travel through the ejecta, photons are continuously red-shifting with respect to the comoving frame, and are therefore exposed to absorption in lines over a broad wavelength range. Because especially iron-group elements have a large number of lines at UV/blue wavelengths, radiation here experiences a complex transfer and a significant fraction of it emerges by fluorescence as a quasi-continuum at longer wavelengths \citep{Li1996}. The efficient blocking by atomic lines is equivalent to an effective continuum opacity known as line blocking. The emerging UV emission may often be due to 'holes' in this line blocking where radiation can escape \citep{Mazzali1993}. Although the optical depths decrease over time, the multi-line transfer is important for modeling supernova spectra for many years or even decades after explosion. %In addition, the effect can actually increase its influence on the optical/NIR spectrum with time as all cooling lines eventually move into the far-IR \citep{Li1996}. 

The observational determination of the masses of the three main
radioactive isotopes \iso{56}Ni, \iso{57}Ni, and \iso{44}Ti constitutes
one of the main constraints on explosion models of core collapse
supernovae. The production of these isotopes is sensitive to both density and temperature, and thus
to the explosion dynamics \citep[e.g.][]{Woosley1991}. During the first $\sim$500 days the decay of \iso{56}Co (which is the product of rapid \iso{56}Ni decay) dominated the
bolometric light curve of SN 1987A. From the light curves, the \iso{56}Ni mass could
be determined to be $0.069\ \Mo$
\citep{Bouchet1991a}. The determination of the 
\iso{57}Ni mass was complicated by the fact that when this isotope became important, the bolometric light curve
was already affected by the delayed release of ionization energy
(freeze-out). A time-dependent modeling that took this into account resulted in an estimated \iso{57}Ni mass of $\sim$$3.3\e{-3}
\Mo$ \citep[][FK93 hereafter]{Fransson1993}, which agrees well with the mass derived from
observations of the infrared Co II/Fe II lines and the gamma-rays emitted in the decay \citep{Varani1990,Kurfess1992}.

The determination of the \iso{44}Ti-mass is
complicated by a number of factors, the first being the freeze-out contribution just as for \iso{57}Ni.  Second, at late times the low temperature of the gas and the presence of dust lead to most of the deposited energy being re-radiated in the largely unobservable far-infrared, 
which prohibits a determination of the bolometric luminosity. The \iso{44}Ti-mass must therefore be determined
from detailed modeling of the fraction of the energy that emerges in the (observable) UV/optical/NIR bands, and must consider the time-dependent freeze-out. From this type of modeling, \citet[][FK02 hereafter]{Fransson2002} determined the \iso{44}Ti-mass to $(0.5-2)\e{-4}$ \msun, which agrees with the $(1-2)\e{-4}$ \msun~estimated from a nebular analysis of the eight-year spectrum by \citet[][C97 hereafter]{Chugai1997}. \citet{Lundqvist2001} derived an upper limit of $1.1\e{-4}$ \msun~, based on the non-detection of the [Fe II] 26 $\mu$m line by ISO, which rested on the assumption of local positron deposition and the absence of any dust cooling.

Another possibility for determining the \iso{44}Ti-mass is by direct detection of the gamma-rays produced in the decay. Only in one case, Cas A, has such a detection been made \citep{Iyudin1994,Vink2001}, with the mass estimated to
$1.6^{+0.6}_{-0.3} \times 10^{-4} \Mo$ \citep{Renaud2006}. For Cas A we do not
know the masses of the \iso{56}Ni and \iso{57}Ni isotopes though.  
For SN 1987A, the detection limits by INTEGRAL
resulted in only a weak upper limit of $1\e{-3}\ \Mo$ for the \iso{44}Ti-mass
\citep{Shtykovskiy2005}. 

In this paper we present a detailed model of the UV/optical/NIR spectrum of SN 1987A at an age of eight years, using a 19 \msun~explosion model as input. Using a new code with a detailed radiative transfer treatment, our objectives are
to see if the spectral model can reproduce the main features in the observed spectrum, to understand the contributions by various zones and elements to the spectral formation process, to quantify the importance of the (non-local) line transfer, and to determine the \iso{44}Ti-mass as accurately as possible. Previous discussions of the late-time ejecta spectra can be found, e.g., in \citet{Wang1996} and C97. Our paper also serves to describe the code we developed for reference in future modeling.

%The paper is arranged as follows; In \textsection \ref{sec:modeling} we describe the modeling, both in terms of the physical processes treated and the explosion model used. The observation used are described in \textsection \ref{sec:TOS}. We describe the results of the modeling in \textsection \ref{sec:results}, and \textsection \ref{sec:discussion} contains a discussion of them. Finally, we list our conclusions in \textsection \ref{sec:conclusions}.

\section{Modeling}
\label{sec:modeling}

In KF98 a, b, a self-consistent model for the nebular phase spectrum of SN 1987A was presented. This included
a detailed calculation of the gamma-ray/positron thermalization and a
determination of the time-dependent ionization, excitation, and temperature
structure of the ejecta. Comparing the resulting spectra from two different explosion models, the evolution of all major emission lines
were analyzed. \citet{Fransson2002} contains an update of this work, where
the modeling of some of the most important lines and
the broad band photometry were discussed. 

A shortcoming in these models was that (non-local) radiative transfer in atomic lines was not included, which introduced an
uncertainty for the internal radiation field and thereby for the ionization balance, and also for the emerging spectrum. 
Here, we calculate in detail the line scattering and fluorescence throughout the spectrum (Sect. \ref{sec:RT}), which significantly improves the accuracy of the model. In addition, we updated the atomic
physics by adding new atomic data and including more and larger model atoms.

While we do a more accurate radiative transfer calculation, this
comes at the expense of time-dependent effects as our model assumes steady-state. In
FK93 it was shown that in the envelope the recombination time scale became comparable to the radioactive
decay time scale at $\sim$800 days. From that time on, much of the
ionization energy was not radiated instantaneously but on a longer time scale, and a time-dependent calculation was
therefore necessary. In the
\iso{44}Ti-dominated phase the radioactive time scale becomes much
longer, $\sim$85 years \citep{Ahmad2006}, and is less
relevant. Instead, it is the adiabatic expansion time scale, i.e. the age of
the supernova, that is the most relevant, and should be compared
to the recombination and cooling time scales. In zones where a time-dependent treatment is necessary, we use the solutions obtained by an updated version of the time-dependent code used in KF98 a, b for
the ionization balance. As we show below (Sect. \ref{Res:Tandion}), this is only necessary in the hydrogen envelope.

\subsection{Ejecta model}
\label{sec:em}
We base our ejecta model on the 19 \msun~explosion model computed by \citet[][WH07 hereafter]{Woosley2007}. From the distribution of elements in this model, we construct seven types of zones; Fe/He, Si/S, O/Si/S, O/Ne/Mg, O/C, He, and H, where the zones are named after their dominant elements. The mass cuts for the zones are taken to be where the most common or second most common element changes. %, obtained from A. Heger (priv. comm.). 
The mass of the Fe/He zone is adjusted to give a total \iso{56}Ni mass of 0.072 \msun~in the ejecta (the value of 0.069 \msun~from \cite{Bouchet1991a} corrected for the slighly larger distance we use here), and the mass fractions of \iso{57}Co and \iso{44}Ti in this zone are then adjusted to give a total \iso{57}Co mass of of $3.3\e{-3}$ (FK93) and a total \iso{44}Ti mass of \bestti, which we later show to be our favored value for this isotope. The H zone mass is adjusted to match the observed estimate of $\sim$8 \msun~(KF98 b). Finally, we replace the He/C zone in the explosion model with a He/N zone, that has not experienced any outward mixing of carbon. This is also based on modeling in KF98 b, where it was found that a significant He/C zone would result in [C~I] lines much stronger than the observed ones.
 
The masses and compositions of the zones in the model are given in Table \ref{table:chem}. Abundances are taken as the value in the middle of each zone. While this does not exactly conserve the total element masses, the differences are small compared to using zone-averaged values, and the middle values should give a more consistent description for the typical composition.
Because the WH07 models are based on solar metallicity progenitors, we replace the abundances of all elements with mass fractions differing by less than a factor 1.25 from the solar values with the LMC abundances taken from \citet{Russell1992}. A comparison of models with different metallicities in \citet[][WW95 hereafter]{WW95} shows that most nucleosynthetic yields only weakly depend on metallicity, so the abundances for the processed elements should be very similar for solar and LMC metallicity progenitors. Observations of the circumstellar material indicate that CNO-burning products have been mixed out from the helium core into the hydrogen envelope \citep{Fransson1989CNO}. We use the abundances of He, N and O for the envelope derived by \citet{Mattila2010} and of C by \citet{Lundqvist1996} (see Table \ref{table:chem}).%elements in the hydrogen envelope to the values derived from the inner circumstellar ring (see Table \ref{table:chem}).
 
We model the supernova as a spherically symmetric, homologously expanding nebula, divided into a core
and an envelope. The core contains the heavy element zones (Fe/He, Si/S, O/Si/S, O/Ne/Mg, O/C), and  
fractions of the He and H zones, which are mixed inwards during the explosion. These fractions were chosen to be 25\% (see Sect. \ref{sec:thecore}). The envelope contains the remaining 75\% of the He and H zones. 

The extent of the core can be estimated from the widths of 
the lines that are emitted mainly from the heavy element zones, for example the iron lines and [O~I] \wll 6300, 6364. The oxygen lines have maximum expansion velocities of $\sim$1700 \kms, whereas the iron, cobalt, and nickel lines extend to $\sim$2500 \kms, with some contribution even out to $\sim$ 3500 \kms \citep[see][for a review]{McCray1993}. 
%At the same time, the bulk of the material is likely to have velocities around $\sim$ 1000-1500 \kms \citep[see e.g. the simulations by][]{Kifonidis2006}. 
Our single-core model forces a compromise value, %$, %that is a compromise between getting the bulk of the matter at the the right velocity ($V_{\rm core}\sim 1500-2500\kms$) and getting wide enough tails on the line profiles ($V_{\rm core}\sim 3500 \kms$). We consider the first criteria more important 
which we choose as $V_{\rm core}=2000~$\kms, the same as in KF98 a, b. A higher core velocity means less gamma-ray deposition, but because the positrons dominate the energy input at late times, this choice is not critical for the energy budget here. 

\subsubsection{The core}
\label{sec:thecore}

\begin{table*}
\caption{Chemical compositions (mass fractions) of the zones used in the model.}
\label{table:chem}
\centering

\begin{tabular}{r l l l l l l l l}
  \hline\hline
Zone  & Fe/He & Si/S & O/Si/S & O/Ne/Mg & O/C & He & H\\
 \hline
Mass (\msun): & 0.092 & 0.14 & 0.16 & 1.9 & 0.58 & 1.3 & 8.0\\
\hline
Mass fractions:\\ 
\hline
$^{56}$Ni & 0.75 & 0.022 & 0 & 0 & 0 & 0 & 0\\
$^{57}$Co & 0.034 & $9.0\e{-4}$ & $1.0\e{-5}$ & 0 & 0 & 0 & 0 \\
$^{44}$Ti & $1.4\e{-3}$ & $5.2\e{-5}$ & $1.6\e{-5}$ & 0 & 0 & 0 & 0\\
\hline 
H             & $2.3\e{-6}$ & $4.2\e{-8}$ & $8.7\e{-9}$ & $4.6\e{-10}$ & $2.1\e{-10}$ & $2.5\e{-7}$ & 0.60\\
He            & 0.14        & $8.3\e{-6}$ & $4.3\e{-6}$ & $2.0\e{-6}$  & 0.078        & 0.99         & 0.40  \\
C             & $3.6\e{-7}$ & $1.4\e{-6}$ & $8.5\e{-5}$ & 0.031        & 0.28         & $3.0\e{-4}$    & $2.4\e{-4}$** \\
N             & $1.4\e{-6}$ & $1.1\e{-9}$ & $1.8\e{-5}$ & $3.2\e{-5}$  & $4.7\e{-6}$  & $9.0\e{-3}$  & $2.2\e{-3}$** \\
O             & $9.7\e{-6}$ & $1.6\e{-5}$ & 0.46        & 0.70         & 0.62         & $1.6\e{-4}$  & $1.8\e{-3}$**\\
Ne            & $1.0\e{-5}$ & $1.6\e{-6}$ & $1.4\e{-4}$ & 0.22         & 0.018        & $1.1\e{-3}$  & $6.1\e{-4}$*\\
Na            & $6.0\e{-7}$ & $1.0\e{-6}$ & $1.2\e{-6}$ & $7.0\e{-3}$  & $1.9\e{-4}$  & $1.8\e{-4}$  & $1.6\e{-5}$*\\
Mg            & $1.2\e{-5}$ & $1.6\e{-4}$ & $7.4\e{-4}$ & 0.042        & $4.9\e{-3}$  & $5.4\e{-4}$* & $5.4\e{-4}$*\\
Al            & $2.1\e{-5}$ & $2.6\e{-4}$ & $1.9\e{-4}$ & $4.0\e{-3}$  & $3.0\e{-5}$* & $3.0\e{-5}$* & $3.0\e{-5}$*\\
Si            & $1.6\e{-4}$ & 0.44        & 0.27        & $3.1\e{-3}$  & $2.8\e{-4}$* & $2.8\e{-4}$* & $2.8\e{-4}$*\\
S             & $1.2\e{-4}$ & 0.39        & 0.21        & $2.6\e{-4}$  & $1.2\e{-4}$* & $1.2\e{-4}$* & $1.2\e{-4}$*\\
Ar            & $1.2\e{-4}$ & 0.051       & 0.043       & $8.4\e{-5}$  & $5.4\e{-5}$* & $5.4\e{-5}$* & $5.4\e{-5}$*\\
Ca            & $1.3\e{-3}$ & 0.031       & 0.012       & $3.1\e{-5}$  & $2.3\e{-5}$* & $2.3\e{-5}$* & $2.3\e{-5}$*\\
Sc            & $2.3\e{-7}$ & $2.3\e{-7}$ & $1.2\e{-6}$ & $1.6\e{-6}$  & $5.6\e{-7}$  & $1.4\e{-8}$* & $1.4\e{-8}$*\\
Ti (stable)   & $8.6\e{-4}$ & $2.9\e{-4}$ & $8.4\e{-5}$ & $6.2\e{-6}$  & $5.3\e{-6}$  & $2.3\e{-6}$* & $2.3\e{-6}$* \\
V             & $2.1\e{-5}$ & $1.3\e{-4}$ & $3.0\e{-6}$ & $6.4\e{-7}$  & $4.5\e{-7}$* & $4.5\e{-7}$* & $4.5\e{-7}$*\\
Cr            & $1.3\e{-3}$ & $3.2\e{-3}$ & $4.1\e{-5}$ & $1.3\e{-5}$  & $1.2\e{-5}$* & $1.2\e{-5}$* & $1.2\e{-5}$*\\
Mn            & $1.0\e{-5}$ & $6.6\e{-4}$ & $1.9\e{-6}$ & $3.8\e{-6}$  & $6.6\e{-6}$* & $6.6\e{-6}$* & $6.6\e{-6}$*\\
Fe (stable)   & $2.3\e{-3}$ & 0.050       & $4.3\e{-4}$ & $6.9\e{-4}$  & $7.2\e{-4}$* & $7.2\e{-4}$* & $7.2\e{-4}$*\\
Co (stable)   & $3.2\e{-8}$ & $5.5\e{-9}$ & $1.4\e{-6}$ & $1.9\e{-4}$  & $1.3\e{-4}$  & $2.4\e{-6}$* & $2.4\e{-6}$*\\
Ni (stable)   & 0.037       & $2.4\e{-3}$ & $1.2\e{-3}$ & $5.6\e{-4}$  & $2.3\e{-4}$  & $4.8\e{-5}$* & $4.8\e{-5}$*\\
\hline
Mean atomic weight & 19.4 & 31.6 & 21.5 & 16.9 & 12.1 & 4.0 & 1.4 \\
\hline
 \end{tabular}
\tablefoot{
Primordial LMC abundances are marked with an asterisk, and CNO abundances derived from observations of the inner circumstellar ring are marked with two asterisks.  See also Sect. \ref{sec:em}. \\
       %\item[b] Grows as $^{56}$Ni and $^{57}$Co decay.  
       %\tablefoottext{*}{Primordial LMC abundance \citep{Russell1992}.}
}
\label{table:zoneabund}
\end{table*}

\begin{figure}[htb]
\centering
%\resizebox{\hsize}{!}{\includegraphics{./ejectamodel_new.eps}}
\includegraphics[width=0.8\linewidth,clip=true]{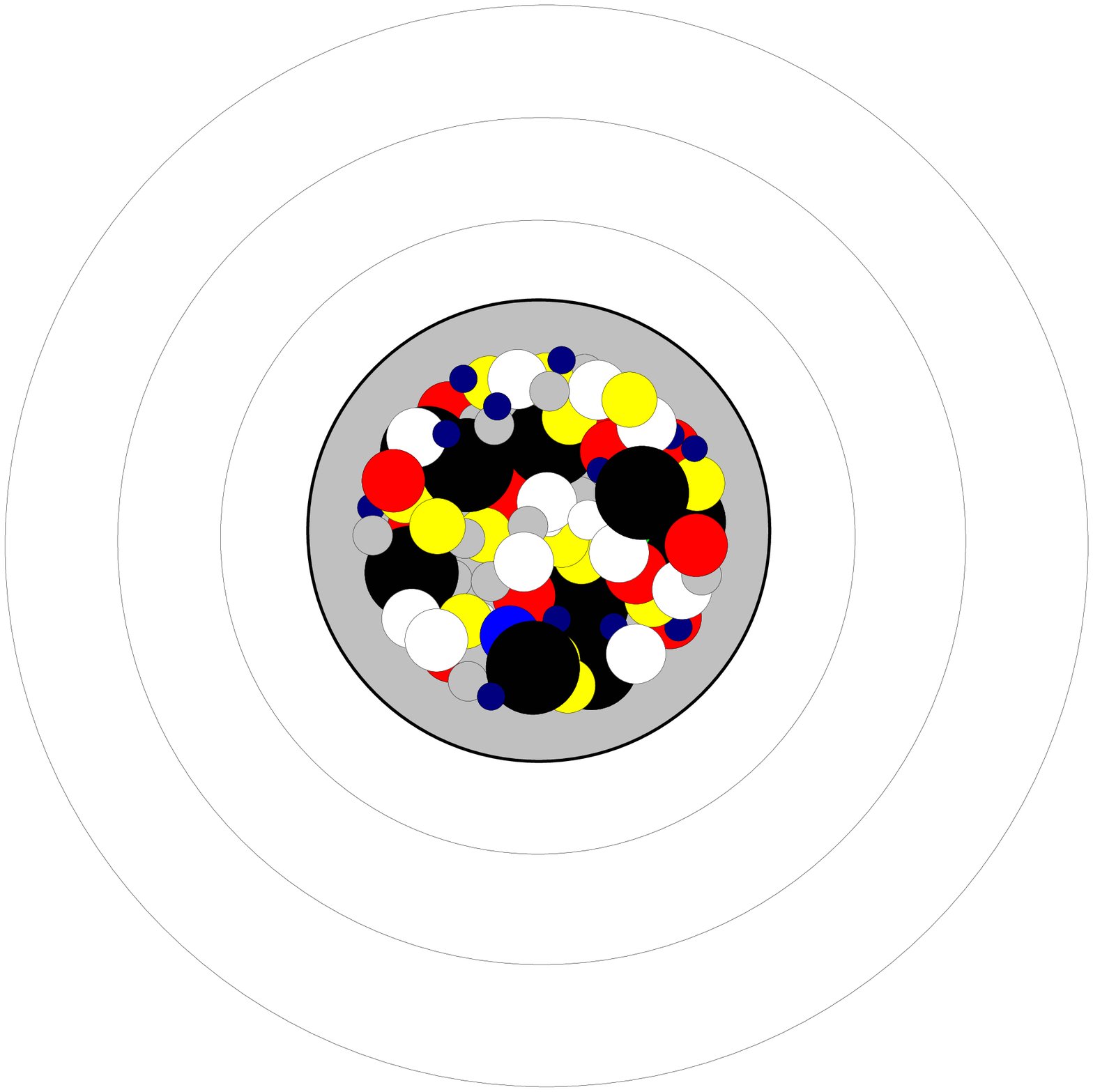}
\caption{Schematic structure of the used ejecta model. The core consists of several chemically distinct zones distributed
as $N_{\rm cl}$ clumps each. The envelope consists of an initial He zone (gray) followed by shells of H-rich gas (white). The densities of the H shells are decreasing outward.}
\label{fig:ejectamodel}
\end{figure}

We assume that the zones in the core experience complete macroscopic mixing, where they fragment into $N_{\rm cl}$ clumps that become completely mixed in velocity space (see Fig. \ref{fig:ejectamodel}). Support for this macroscopic mixing comes from theoretical considerations \citep[e.g.][]{Mueller1991}, multi-dimensional simulations \citep[e.g.][]{Herant1991,Mueller1991,Kifonidis2006,Hammer2010}, the early emergence of X-rays and gamma-rays \citep[][]{Dotani1987, Pinto1988}, and from the similar line profiles of many of the elements \citep[e.g.][]{Spyromilio1990}. 
We use a value of $N_{\rm cl}=100$ based on the analysis of the iron clumps in \citet{Li1993iron}. In Sect. \ref{sec:fragmentation} we find that the model spectrum is insensitive to this choice. We further assume that no microscopic mixing occurs \citep[e.g.][]{Fransson1989, Fryxell1991}, so that the zones retain their compositional integrity. 

The clumpy structure of the core calls for a special handling of the computational grid. We use a gridless domain for the core region; when a clump emits photons, the position of the clump within the core is selected by a random draw. When we follow photons through the core (Sect. \ref{sec:RT}), exiting one clump leads to a random draw that determines which type of clump to encounter next. We set the probability of encountering any given type of zone to be proportional to the total surface area of that zone type. A random number also selects the impact parameter on the new clump. 
%The probability of impacting a sphere of radius $R$ at angle $\theta$ is given by
%\begin{equation}
%P(\theta)d\theta = \frac{2\pi h dh}{\pi R^2}
%\end{equation}
%Since $sin(\theta)=h/R$ we get $dh = R cos(\theta)$ and
%\begin{equation}
%P(\theta)d\theta = 2 sin(\theta) cos(\theta)
%\end{equation} 
%The path length through the clump at impact angle $\theta$ is $2R\cos(\theta)$. The mean path length is then
%\begin{equation}
%<L> = \int_0^{\pi/2} 2 \sin(\theta) \cos(\theta) 2 R\cos(\theta) d\theta = 4/3R
%\end{equation}
The path traveled through zone $i$ is then proportional to $A_i\cdot R_i \propto V_i$, which recovers the desired property that a photon should spend its time in a given zone in proportion to the filling factor of the zone, $\epsilon_i$.  

The core structure is a 'virtual' one in the sense
that there is no fixed density grid; we have a set of spherical clumps
which together make up the total volume of the core, but which obviously cannot be
exactly fit together since they are all spheres. There are several advantages to this
virtual grid. The extension of the Monte Carlo philosophy to encompass also
a randomization of the grid is a powerful and appealing way to make computations in clumpy media such as this. 
While the radiative transfer can still be done realistically by the use of surface and impact probabilities as described above,
a virtual grid naturally offers a way to capture the situation of macroscopic mixing and to
parametrize the degree of fragmentation of the core. Finally, the line profiles automatically become smooth rather
than jagged, as when using shells. %The main disadvantage is that the structure loses realism as $N_{\rm cl}\rightarrow 1$; we then get single clumps which nevertheless are 'everywhere'. We avoid running models with $N_{\rm cl}$ smaller than $\sim$ 10. 

The filling factors for the oxygen zones are set from the oxygen number density derived from the evolution of the [O~I]\ 6300 \AA/6364 \AA~line ratio, $n_{\rm O}=6.2\e{10}~\mbox{cm}^{-3}$~at 100 days \citep{Spyromilio1991, Li1992}. This allocates $\sim$10\% of the core volume to oxygen clumps. 

The iron clumps are believed to expand because of their content of radioactive materials \citep{Woosley1988, Herant1991,Basko1994}, which was verified for SN 1987A by modeling of the cooling lines from the iron clumps \citep{Li1993iron}. The iron zone was found to have a filling factor of $\epsilon_{\rm Fe/He}\gtrsim 0.3$, assuming $V_{\rm core}=2500$ \kms. The same density limit using  $V_{\rm core}=2000$ \kms corresponds to $\epsilon_{\rm Fe/He}\gtrsim 0.6$. We use a value of 0.6 in our modeling here. 

In the explosion model, the Si/S zone contains $\sim$5\% of the $^{56}$Ni, and can therefore be expected to expand as well.  For this zone, we therefore use a density in between the oxygen zones and the iron zone, chosen to be ten times lower than the oxygen zone density. This gives a filling factor of $\sim$3\%. In Sect. \ref{sec:lineids}, we find support from our modeling for a low density in this zone.  

Hydrodynamical simulations by \citet{Fryxell1991} find $\sim$1 \msun~of hydrogen mixed inside 2000 \kms, while \citet{Herant1992} find $\sim$2 \msun. From modeling of the H$\alpha$ line in SN 1987A, KF98 b estimate the presence of 2.2 \msun~of H-zone gas inside 2000 \kms. Here, we take 2 \msun~of H-zone material to be mixed inside the core, which corresponds to $\sim$25\% of the total H zone mass. We assume the same fraction (25\%) of the He zone to be mixed into the core as well, giving a He-zone core component of 0.3~\msun~and a He-envelope component of 1.0~\msun. 
%With total masses of  and $M_{\rm H}=8~M_\odot$ (see below) we then get  and $M_{\rm H}^{\rm core}=2~M_\odot$. The masses a%nd% filling factors of all the core zones can be seen in Table \ref{table:zonemasses}. 
The total mass of the core, including the mixed-in He and H, is then 5.2 \msun.

The filling factors for the H and He zones in the core are set by allocating the remaining volume so that these zones obtain equal number densities. This gives the H component a filling factor of $\sim$26\% and the He component a filling factor of $\sim$2\%. See also KF98 a for further discussions regarding filling factors.
  
\subsubsection{The envelope}
Outside the core we attach an envelope consisting of an inner He zone
followed by logarithmically spaced (constant $dR/R$) H-zone shells. As mentioned above, the total He-zone mass is 1.3 \msun, of which we put $1.0$ \msun~in the envelope and $0.3$ \msun~in the core. The total H-zone mass is taken as 8 \msun~from H$\alpha$ modeling (KF98 b), of which we put $6$ \msun~in the envelope and $2$ \msun~in the core (see Sect. \ref{sec:thecore}).% Since we assume that 25\% of the envelope gets mixed into the core, the total masses for the envelope is 1.0 $M_\odot$ for the He zone and 6.0 $M_\odot$ for the H zone.

We use a density distribution of the H-envelope based on hydrodynamical modeling by \citet{Shigeyama1990} (model 14E1). The validity of this model is confirmed by modeling of the Mg~II \wl2800 and Mg~I \wl2852 features in C97. The density profile can be described as gradually steepening from an initial -2 power law to an asymptotic $\sim$-8 for $V>5000$~\kms. We terminate the envelope at a velocity of 10,000 \kms.

\begin{table*}
\caption{Properties of the zones in the model.}
\centering
 \begin{tabular}{l | l l l l l l}
  \hline\hline
  Zone & Mass & $V_{\rm in}$ & $V_{\rm out}$ & Filling factor ($\epsilon$) & Density & Rec. time\\
                &  (\msun)  &  (\kms) & (\kms) &           & (cm$^{-3}$) & (years)\\
\hline
  Fe/He & 0.092 & 0  & 2000 & 0.6         & $1.8\e{4}$ & 2.5\\
  Si/S & 0.14   & 0  & 2000 & $2.9\e{-2}$ & $3.5\e{5}$ & 0.8 \\
  O/Si/S & 0.16 & 0  & 2000 & $4.1\e{-3}$ & $4.2\e{6}$ & 0.4\\
  O/Ne/Mg & 1.9 & 0  & 2000 & $7.3\e{-2}$ & $3.5\e{6}$ & 0.9 \\
  O/C & 0.58 & 0     & 2000 & $2.0\e{-2}$ & $5.6\e{6}$ & 0.9\\
  He-core & 0.32 & 0 & 2000 & $1.5\e{-2}$ & $1.3\e{7}$ & 0.5\\
  H-core & 2 &  0    & 2000 & 0.26 &  $1.3\e{7}$ & 2.1\\ 
 \hline
  He-env & 1.0       & 2000 & 2200 & 1 & $1.7\e{6}$ & 1.3\\
  H-env-1  & 1.9     & 2200 & 2730 & 1 & $2.5\e{6}$ & 6.9\\
  H-env-2 & 2.0      & 2730 & 3390 & 1 & $1.4\e{6}$ & 10 \\
  H-env-3 & 1.2      & 3390 & 4210 & 1 & $4.3\e{5}$ & 17 \\
  H-env-4 & 0.61     & 4210 & 5230 & 1 & $1.2\e{5}$ & 29 \\
  H-env-5 & 0.24     & 5230 & 6490 & 1 & $2.4\e{4}$ & 49\\ 
  H-env-6 & 0.083     & 6490 & 8050 & 1 & $4.3\e{3}$ & 105\\
  H-env-7 & 0.028     & 8050 & 10000 & 1 & $7.6\e{2}$ & 332\\ 
\hline
 \end{tabular}
\tablefoot{
The last column shows the recombination time obtained from a steady state solution (see Sect. \ref{Res:Tandion}).}
\label{table:zonemasses}
\end{table*}

\subsection{Energy deposition and degradation}
\label{sec:edep}
We calculate the deposition of gamma-rays emitted by the clumps containing \iso{56}Co, \iso{57}Co and \iso{44}Ti. Based on the simulations by \citet{Colgate1980, Woosley1989} and \citet{Fransson1989} we assume absorption with effective opacities of $\kappa(^{56}\mbox{Co})=0.030\left(\frac{\bar{Z}}{\bar{A}}/0.5\right)$,~$\kappa(^{57}\mbox{Co})=0.072\left(\frac{\bar{Z}}{\bar{A}}/0.5\right)$ and $\kappa(^{44}\mbox{Ti}) = 0.040\left(\frac{\bar{Z}}{\bar{A}}/0.5\right)$~cm$^{2}$g$^{-1}$, where $\bar{Z}$ and $\bar{A}$ are the average nuclear charges and atomic weights in each zone. We do not include other radioactive isotopes such as \iso{60}Co and \iso{22}Na, whose abundances are highly uncertain. We discuss the possible effects of these in Sect. \ref{sec:discussion}. 

From each zone containing radioactive material, we emit and track \iso{56}Co, \iso{57}Co, and \iso{44}Ti gamma-ray packets
%\begin{equation}
%E_{\rm 56Co} = \frac{1}{N_{\rm \gamma}}8.69\e{41}e^{-(\rm t-8.8~d)/\tau_{56Co}}~\mbox{ergs}~,
%\end{equation}
%\begin{equation}
%E_{\rm 57Co} = \frac{1}{N_{\rm \gamma}}4.52\e{38}e^{-\rm t/\tau_{\rm 57Co}}~~\mbox{ergs},
%\end{equation}
%and
%\begin{equation}
%E_{\rm 44Ti} = \frac{1}{N_{\rm \gamma}}4.1\e{36}e^{-\rm t/\tau_{44Ti}}~~\mbox{ergs},
%\end{equation}
%\textbf{where $N_{\rm \gamma}$ is the number of packets to split the emission into, $\tau_{\rm 56Co}=111.3$ d, $\tau_{\rm 57Co}=391.2$ d and $\tau_{\rm 44Ti}=85$ y and $t=2875$ d. We emit and track these packets 
as described for the UV/optical/NIR radiation in Sect. \ref{sec:RT}, with the difference that these packets travel in straight paths until they reach the edge of the nebula, depositing a fraction $\exp{\left(-\kappa \rho_{\rm i} l\right)}$ of their energy for each traveled distance $l$, where $\rho_{\rm i}$ is the density of zone $i$. We use radioactive luminosities taken from KF98, with the \iso{44}Ti
values updated for a life-time of 85 years instead of 78 years.%Each packet obtains a random starting position and flight angle (Eqs. (\ref{eq:rstart}), (\ref{eq:dircos}) and (\ref{eq:dircostrans}) in Sect. \ref{sec:RT}), and is tracked along a straigh path until it escapes the ejecta, depositing energy in each zone it passes according to the opacities above.} 

After $\sim$2000 days,  $^{44}\mbox{Ti}$ completely dominates the deposition. The degradation of the deposited gamma-ray and positron energy into heating, ionizations and excitations is obtained by solving the Spencer-Fano equation using the routine
developed by KF92. For this routine, we updated the collisional cross sections for Fe~II \citep{Ramsbottom2005,Ramsbottom2007} and Ca I \citep{Samson2001}. Ionizations are generally assumed to leave the ions in their ground states, except for O II, for which we compute specific rates to the first few excited states. 

\begin{figure}[htb]
\centering
\includegraphics[width=0.8\linewidth]{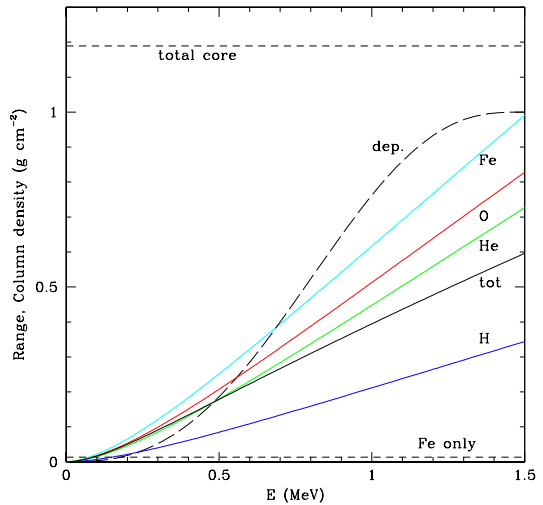}
\caption{Positron stopping range for different elements as a function of kinetic energy. The solid black curve denoted 'tot' shows the range weighted by the different zones in the core. The lower horizontal dashed line marks the column density of the Fe/He zone (assuming it to exist as a single clump), and the upper that of the full core for $V_{\rm core} = 1800 ~ \kms$. The long-dashed curve shows the cumulative positron distribution.}
\label{fig:stopping_power}
\end{figure}

In our standard model, the positrons are assumed to be absorbed
on the spot. As noted before, e.g. in C97, this requires a magnetic field.  The average positron
energy in the \iso{44}Sc decay is 0.6 MeV and the maximum energy of
the positrons is 1.47 MeV \citep{Browne1986}. The energy loss per unit
length, $dE/dx$, is dominated by excitations and
ionizations and is given by the Bethe formula \eq{ {dE\over dx} = -{4
    \pi r_0^2 m_{\rm e} c^2 \rho \over m_{\rm u}\beta^2} {Z\over A} \left[
    \ln({\beta^2 m_{\rm e} c^2\over I}) - f_z \right]~,
%where \eq{f_z = {\ln
%    2 \over 2} - {\beta^2 \over 12} \left[ {23 \over 2} + {7 \over
%      (\gamma + 1)} + {5 \over(\gamma + 1)^2} + {2 \over (\gamma +
%      1)^3} \right] 
\label{eq:bethe}}
%% \eq{
%% d_s = {3.36 \over \rho} {E \over m_{\rm e} c^2} {A \over Z} \left[\ln(E/I)\right]^{-1} \\ \rm cm
%% \label{eq:bethe}
%% }
%% \citep[e.g.][]{Ruiz-Lapuente1998}. 
%\citep{Rohrlich1954}. 
where $r_0$ is the electron radius, $\beta = V/c$, $I$ is the effective ionization potential of the atom, $f_z$ is a relativistic correction factor, and all other symbols have their usual meaning. 
%The effective excitation/ionization potential is
%given by
%\eq{ I = 11.2 + 11.7 Z \ \rm eV} for $Z \le 13$ and
%\eq{ I = 52.8 + 8.71 Z \ \rm eV} for $Z > 13$.
%\eq{I = 9.1 Z (1. + {1.9 \over Z^{2/3}}) \ \rm eV. }  
%\citep{Dalton1968}. 
In Fig. \ref{fig:stopping_power} we show the stopping range, $\rho E/
\frac{dE}{dx}$ (g cm$^{-2}$), for the most important elements in the core
together with the total range, averaged over the different 
zones in the core. 

These ranges should be compared to the mass column density $D$ of the various zones in the core and envelope. For the total core, this is given by
\eqn{D =
%&=& {3 \left(M_{\rm core} \over 4 \pi (V_{\rm core} t)^3}\cr = 7.6\e{-18}
  0.93\left(\frac{M_{\rm core}}{5 \Mo} \right) \left({V_{\rm core} \over 2000
    \ \kms }\right)^{-2} \left({t \over \ 8 \ \rm yrs }\right)^{-2}
  \ \rm g \ cm^{-2}~. } 
The column density for a core with $V_{\rm core}=1800$ \kms at 7.87 years is shown as
the upper dashed line in Fig. \ref{fig:stopping_power}. For comparison
we also show the column density of the Fe/He zone as the lower
horizontal line. 

That the column density of the Fe/He zone is nearly two orders of magnitude lower than the stopping range for Fe and He means that in the absence of a sufficiently strong magnetic field the positrons will not be trapped in this zone, even if it is not strongly fragmented into many small pieces. The figure also shows, however, that the total core has a sufficiently high column density to trap the positrons even without a magnetic field. There should therefore be no positrons entering the envelope at this or earlier epochs.

%% With $A/Z \approx 2$, $E \sim 0.5$ MeV, $I \sim 10$ eV {\bf Not
%% correct see RLS} we get for the positron range
%% \eqn{
%% d_s = 8.0\e{16}
%% \left({M_{core} \over 4 \ \Mo}\right)^{-1} \left({V_{core} \over 2000 \ \kms }\right)^{3}
%% \left({t \over \ 10 \ \rm yrs }\right)^{3} \ \rm \ cm
%% } This is comparable to the radius of core, $\sim 6.3\e{16} (V_{core}/2000
%% \ \kms) (t / 10 \ \rm yrs)$ cm, at the age discussed here. With the adopted parameters  the positrons 
%% are trapped in the core up to $\sim$ 10 years after explosion.
%Clumping will not affect this conclusion much. The mass column density scales as $\rho^{2/3}$ 
%and the size of the clumps scale inversely with the density. 

%Clumping will not affect this conclusion much, as can be seen from the
%following argument. Consider a large number, $N$, of clumps, each with a
%radius $R_{\rm c}$ and density $\rho_{\rm c}$. The mean free path between
%different clumps is $\lambda = 1/n_{\rm c} \sigma = \mathcal{V} / \pi N R_{\rm c}^2$, where
%$\mathcal{V}$ is the total volume of the core, $\sigma $ is the cross section of each clump and $n_c$ the number density of% clumps. The average number of clumps
%intersected by a ray from the core is then $R/\lambda$, so that the
%total column density of all clumps from the core is $R/\lambda \times 4 R_{\rm c}
%\rho_{\rm c} /3 = 4 \pi / 3  N R_{\rm c}^3 \rho_{\rm c} /  \mathcal{V} = M R /  \mathcal{V}$, independent of
%the number of clumps and their size. This obviously breaks down if the
%number of clumps is small. {\bf Do you buy this? }

The presence of magnetic fields in supernova ejecta has been suggested many times, see e.g. \citet{Axelrod1980}.
\cite{Colgate1980} argue that any magnetic field present at the time of the explosion becomes radially 'combed' and will therefore not be important for the trapping \citep[see also][]{Ruiz-Lapuente1998}. 
While the positrons should be confined to the core at eight years, it is therefore not clear whether they are absorbed by the zones producing them (which is mainly the Fe/He zone, but also the Si/S zone), or if they can propagate into the other core zones before they are absorbed. We will later investigate this question by comparing the spectra produced in the different scenarios (Sect. \ref{sec:pos}).

\subsection{Statistical and thermal equilibrium}  
\label{sec:sate}
For each zone we solve the equations for statistical equilibrium with
respect to the ionization and excitation structure, as well as the
equation of thermal equilibrium. These equations are essentially the ones described in KF98 a, with some
modifications and improved rate calculations described below. For each iteration, the algorithm deployed is to first compute the ionization balance, then iterate solutions to excitation structure and thermal equilibrium until the temperature stabilizes, and finally to calculate the radiative transfer (Sect. \ref{sec:RT}). From the radiative transfer, new photoionization rates, photoexcitation rates (although these are set to zero in this work, see Sect. \ref{sec:excstr}), and radiative heating rates are calculated, and used for the new solutions for ionization, excitation and temperature in the next iteration. Iteration continues until the maximum change in any zone temperature, electron fraction or ion abundance is less than 1\%. We then also verify that the change in the emerging radiation field is neglegible. 

\subsubsection{Ionization balance}

Steady-state for the ionization balance is appropriate as long as the recombination time-scale $(1/n_{\rm e}\alpha)$ is short compared to the evolutionary and radioactive time scales. For the densities at the epoch studied here, this is satisfied (Sect. \ref{Res:Tandion}) for all core zones as well as for the envelope helium zone. The hydrogen envelope, however, experiences freeze-out after $\sim$800 days (FK93). For the envelope, we therefore use the solutions for the hydrogen ionization fraction from a time-dependent calculation, based on the model in KF98 a, given in Table \ref{table:freeze}. We do, however, use steady-state solutions for the other elements in the envelope, because the time-dependent model lacks photoionizations by line radiation, and may therefore significantly underestimate the ionization degree for some of them. We also take the hydrogen envelope temperatures from the time-dependent solution, because temperature and ionization balance are coupled. 

\begin{table}[h!]

\caption{The degree of hydrogen ionization, and the temperature in the hydrogen envelope at day 2875, taken from a time-dependent calculation based on the model in KF98 a, b.}
\centering
\begin{tabular}{r l l l}
\hline\hline
Velocity &  Number density & n(H II)/n(H I) & Temperature\\
 (\kms) & (cm$^{-3}$) &- & (K)\\
\hline
 core & $1.7\e{7}$ & $6.5\e{-5}$ & 119\\ 
 2000 & $3.2\e{6}$ & $2.4\e{-4}$ & 114\\
 2400 & $2.5\e{6}$ & $2.5\e{-4}$ & 101\\
 2800 & $1.5\e{6}$ & $3.3\e{-4}$ & 93\\
 3300 & $5.4\e{5}$ & $7.3\e{-4}$ & 89\\
 4100 & $2.4\e{5}$ & $1.3\e{-3}$ & 85\\
 4440 & $1.2\e{5}$ & $2.2\e{-3}$ & 84\\
 5000 & $4.6\e{4}$ & $4.4\e{-3}$ & 84\\
 5500 & $2.2\e{4}$ & $7.5\e{-3}$ & 83\\
 6000 & $1.1\e{4}$ & $1.2\e{-2}$ & 81\\
 6500 & $5.4\e{3}$ & $1.6\e{-2}$ & 76\\
 7000 & $3.0\e{3}$ & $2.0\e{-2}$ & 73\\
 7500 & $1.6\e{3}$ & $2.4\e{-2}$ & 70\\
 8000 & $9.5\e{2}$ & $2.8\e{-2}$ & 67\\
 8500 & $5.7\e{2}$ & $3.2\e{-2}$ & 64\\
 9000 & $3.5\e{2}$ & $3.5\e{-2}$ & 64\\
 9500 & $2.2\e{2}$ & $3.8\e{-2}$ & 62\\
 10,000 & $1.4\e{2}$ & $4.2\e{-2}$ & 60\\
\hline
\end{tabular}
\tablefoot{
Note that the zones here are somewhat differently structured than those we use for the calculations in this paper (Table \ref{table:zonemasses}).
}
\label{table:freeze}
\end{table}

For all other zones, the ionization balance is computed from the steady-state variant of Eq. (10) (and its generalized version) in KF98 a. We include the first three ionization stages of all elements listed in Sect. \ref{sec:excstr}. The main improvement in our treatment here is that we now compute detailed photoionization rates from the Monte Carlo simulation (Sect. \ref{sec:RT}). For zone $i$ and atom $k$, the photoionization rate is
\begin{equation}
P_{\rm i,k} = \frac{\sum_j N_{\rm i,k, j}}{V_{\rm i} \sum_j n_{\rm i,k,j}^*}~,
\end{equation}
where $j$ is the level, $N_{\rm i,k,j}$ is the total number of photoionizations for zone $i$, atom $k$, level $j$, in the Monte Carlo simulation, $V_{\rm i}$ is the zone volume and  $n_{\rm i,k,j}^*$ is the previous solution to the number density of level $j$.

\subsubsection{Excitation structure}
\label{sec:excstr}

Given the ion abundances, steady-state is appropriate for the atomic level populations since the collisional and radiative time-scales are short. We compute NLTE solutions for the neutral and singly ionized
elements of H, He, C, N, O, Ne, Na, Mg, Al, Si, S, Ar, Ca, Sc, Ti, V, Cr, Mn, Fe, Co, and Ni, as well as for doubly ionized Fe, using model atoms that include fine-structure levels.  %For the other doubly ionized elements, only the ground states are included (in the ionization equilibrium).
With their large number of
resonance lines in the UV and optical, iron-group elements are especially
important for the radiative transfer, which motivated us to include also those with small abundances. Appendix \ref{app:atomicdata} gives references for the atomic data that we use.

For the line optical depths, we use the Sobolev approximation \citep{Sobolev1957}
\begin{equation}
\tau_{\rm ij} = \frac{A_{\rm ji}\lambda_{\rm ji}^3}{8\pi}\frac{g_{\rm j}}{g_{\rm i}}t\left(n_{\rm i}-\frac{g_{\rm i}}{g_{\rm j}}n_{\rm j}\right)~,
%\tau_{S} = \frac{A_{\rm ji}\lambda_{\rm ji}^3}{8\pi}\frac{g_{\rm j}}{g_{\rm i}}t\left(n_{\rm i}-\frac{g_{\rm i}}{g_{\rm j}}n_{\rm j}\right)~,
\end{equation}
with the corresponding escape probabilities $\beta_{\rm ij}^{\rm S}$
\begin{equation}
\beta_{\rm ij}^{\rm S} = \frac{1-e^{-\tau_{\rm ij}}}{\tau_{\rm ij}}~,
%\beta_{S} = \frac{1-e^{-\tau_{\rm S}}}{\tau_{\rm S}}~,
\end{equation}
%\end{equation}
where $i$ and $j$ are the lower and upper levels, respectively. The Sobolev approximation is valid as long as the region of line interaction is much smaller than the region over which physical conditions change, which is generally fulfilled in supernovae. The errors caused by line-overlap should be small in the nebular phase \citep{Li1996}. For lines with small $\beta_{\rm ij}^{\rm S}$, we also take into account continuum destruction probabilities \citep{Hummer1985,Chugai1987}, so that the effective escape probability is
\begin{equation}
\beta_{\rm ij}^{\rm eff} = \beta_{\rm ij}^{\rm S} + \beta_{\rm ij}^{\rm C}. %\frac{\beta_{\rm S} + \beta_{\rm C}}{1 + \beta_{\rm C}}
%\beta_{\rm eff} = \beta_{\rm i,j}^{\rm S} + \beta_{\rm i,j}^{\rm C}. %\frac{\beta_{\rm S} + \beta_{\rm C}}{1 + \beta_{\rm C}}
\end{equation}
where the $\beta_{\rm ij}^{\rm C}$ probabilities are calculated as in KF98 a. The effective radiative deexcitation rates used in the NLTE equations are then $A_{\rm ji}\beta_{\rm ij}^{\rm eff}$.

The statistical equilibrium is obtained by solving Eq. (15) in KF98 a. Deexcitation of the He I(2$^3$S) state also occurs through Penning ionizations of H I. Our treatment here differs mainly in that we now compute detailed photoionization rates $P_{\rm i,k,j}$ from the Monte Carlo calculation (Sect. \ref{sec:RT})
% for level $i$ in ion $k$ is
%\begin{equation}
%n_{k,i} R^{\rm out}_{k,i} = R^{\rm in}_{k,i}
%\end{equation}
%where $n_{k,i}$ is the number density of the level and
%\begin{eqnarray}
%R^{\rm out}_{k,i} &=& P_{k,i} + \Gamma_{k,i} + \alpha_{i,k-1}n_e  + \sum_{j<i} \left(A_{ij}\beta_{ij} + C_{ij}n_e + P_{ij}\right)\nonumber  \\
%       &+& \sum_{j>i} \left(C_{ij}n_e + \Gamma_{ij} + P_{ij}\right) + \sum_l n_l \xi^{CT}_{k,l}
%\end{eqnarray}
%and
%\begin{eqnarray}
%R^{in}_{k,i} &=& \sum_{j<i} n_{k,j}\left(C_{ji}n_e + \Gamma_{ji}\right) + \sum_{j>i} n_{k,j}\left(A_{ji}\beta_{ji} + C_{ji}n_e + \Gamma_{ji}\right) \nonumber \\
%   &+& n_{k+1}\left(\alpha_{k,i}n_e + \sum_l n_l \xi^{CT}_{k+1,l}\right) \nonumber \\
%   &+& n_{k-1}\left(P_{i,k-1} + \Gamma_{i,k-1} + \sum_{k'} n_{k'} \xi^{CT}_{k-1,l}\right)
%\end{eqnarray}
%where $P_{k,i}$ is the photoionization rate, $\Gamma_{k,i}$ is the non-thermal ionization rate, $\alpha_{i,k-1}$ is the recombination coefficient, $C_{ij}$ is the collision rate, $P_{ij}$ is the photoexcitation/deexcitation rate, $\Gamma_{ij}$ is the non-thermal excitation rate, $\xi^{CT}_{k,l}$ is the charge transfer rate, and all other symbols have their usual meaning. The photoionization rate is computed as
\begin{equation}
P_{\rm i,k,j} = \frac{N_{\rm i,k,j}}{V_{\rm i} n_{\rm i,k,j}^*}~.
\end{equation}
%where $N_{\rm i,k,j}$ is the number of photons that were photoabsorbed per second in zone $i$ by level $j$ in atom $k$ in the Monte Carlo simulation (Sect. \ref{sec:RT}), $V_{\rm i}$ is the volume of the zone, and 
%where $n_{\rm i,k,j}^*$ is the previous solution to the level population.}

As in KF98, we include non-thermal ionizations and photoionizations only to the first level of the ionized element. The only exception is for O II, where non-thermal ionizations to the excited states $^2D_{\rm o}$, $^2P_{\rm o}$ and $^4P$ are also included. Non-thermal ionizations are included only from the ground multiplets in the atoms, whereas photoionizations can occur from excited states where cross sections are available.%, and are then computed as
%\begin{equation}
%\Gamma_{\rm k,i} = \frac{\dot{E}\eta_{\rm k}}{\chi_{\rm k}n_{\rm k,GM}}
%\end{equation}
%where $\dot{E}$ is the total deposition rate, $\eta_k$ is the fraction going to ionizations of ion $k$, $\chi_k$ is the (ground state) ionization potential of ion $k$, and $n_{k,GM}$ is the number density of the ground statep populations.

%Regarding photoexcitations, we compute
%\begin{equation}
%P_{ij} = \frac{\left(1-e^{\tau_{ij}}\right)N_{ij}}{V n_{k,i}^*}
%\end{equation}
%where $N_{ij}$ is the total number of photons that came into resonance with the transition in the MC simulation (REF).
%\begin{equation}
%N_{ij} = \left(1-e^{\tau_{ij}}\right)\sum N_{ij}
%\end{equation}
In this work, we set all photoexcitation/deexcitation rates to zero. Although these can in principle be computed from the radiative transfer calculation and be included in the population equations, this would force a slow single-stepping Monte Carlo transfer. We also aim to study the scattering/fluorescence process in detail (Sect. \ref{sec:sprop}), which is easier if we %do not include the radiative excitations already when computing emissivities from the NLTE solutions. ]
can track the Monte Carlo packets through all scattering events until they escape. Finally, this approximation also allows us to compute faster (approximate) NLTE solutions including a fewer number of states. At these late times, photoexcitations do not strongly influence the solutions beyond the direct scattering/fluorescence.

For the treatment of charge transfer, see %the charge transfer rates, we assume all reactions to occur from and to the ground states. The only exception is for the O II + C I $\rightarrow$ O I + C II reaction, see 
Sect. \ref{sec:ct}.%which we assume to go to the resonant 2p($^1$D) state in O I. This is important for the emergent [O I]~\wll 6300, 6364 emission (Sect. \ref{sec:onemg}).}
%An important point is that we allow photoexcitations to decrease the populations of the absorbing levels, but we \emph{dont} allow them to increase the population of the upper level. The reason for this is that the scattering/fluorescence following photon absorptions is computed in the MC simulation, and would be counted twice if also included in the NLTE solutions. We prefer this method since it allows to keep the NLTE solutions smaller and more stable, and also allows the scattering/fluorescence process to be studied in detail. The approximate NLTE solutions corresponds to the limit of deexcitations happening instantanously, making a photoexcitation move the atom from the absorbing state to the ground state. See also sec XX.

%The nonthermal excitation rate is computed as
%\begin{equation}
%\Gamma_{i,j} = \frac{\dot{E}\eta^{exc}_{ij}}{\chi_{ij}n_{k,i}^{**}}
%\end{equation} 
%where $\eta^{exc}_{i,j}$ is the fraction going into excitations of the level, $\chi_{ij}$ is the transition energy. I use nk instead of ni here..check..would underestimate the nonthermal excitations.

%The charge transfer rates are taken from a wide range of sources, see REF

%\textbf{For the $2^3S$ and $2s^1S$ states in He I, we also include deexcitations due to Penning ionizations:
%\begin{equation}
%\mbox{He I (2}^3\mbox{S or 2s}^1\mbox{S) + H I} \rightarrow \mbox{He I (GS) + H II + e}^- + 6.2 \mbox{eV}
%\end{equation}
%with a rate of $7.5\e{-9}\left(T/300\right)^{0.5}$ (REF).}

\subsubsection{Thermal balance}

For the core zones, the temperature is computed by balancing heating with cooling. Steady-state is appropriate as long as adiabatic cooling is neglegible. This may not be fulfilled at late times for the oxygen, helium, and hydrogen zones (KF98 a). However, we compared the steady-state temperatures in these zones with those obtained from the time-dependent model (KF98 a, b), and they are found to be very similar. We therefore use the steady-state approximation for all core zones. For the envelope zones, we use the temperatures calculated in the time-dependent model. These are given in Table \ref{table:freeze}. 

Heating has contributions from non-thermal deposition, photoionizations, collisional deexcitations (occuring in the Monte Carlo calculation), charge transfer reactions, free-free absorptions, and Penning ionizations
\begin{equation}
H = H_{\rm nt} + H_{\rm pi} + H_{\rm cd} + H_{\rm ct} + H_{\rm ff} + H_{\rm pe}~.
\end{equation}
%The non-thermal heating $H_{nt}$ is
%\begin{equation}
%H_{nt} = \dot{E}_{nt}\eta_{heat}
%\end{equation}
The terms $H_{\rm pi}$, $H_{\rm cd}$ and $H_{\rm ff}$ are computed from the Monte Carlo simulation (Sect. \ref{sec:RT}) by summing all contributions from Eqs. (\ref{eq:piheating}), (\ref{eq:ffheating}) and (\ref{eq:collheat}).
%\begin{equation}
%H_{\rm pi} = \frac{1}{V}\sum_{\rm k,i,j} E_{\rm j} \left(1 - \frac{\chi_{\rm k,i}}{hc/\lambda_{\rm j}}\right),
%\end{equation}
%where summation is done over all packets (energy $E_{\rm j}$, wavelength $\lambda_{\rm j}$) absorbed by level ($k,i$) with ionization potential $\chi_{\rm k,j}$. The collisional deexcitation heating is computed by summing all events according to Eq. (\ref{eq:collheat}). The free-free heating $H_{\rm ff}$ is obtained by summing the total energy absorbed by free-free opacity.
%\begin{equation}
%H_{\rm ff} = \frac{1}{V}\sum E_{\rm j}^{\rm ff}
%\end{equation}
The charge transfer heating is \citep{Kingdon1996}
\begin{equation}
H_{\rm ct} = \sum_{\rm k,k'}n_{\rm k} n_{\rm k'}\xi^{\rm CT}_{\rm k,k'}\Delta E_{\rm k,k'}~,
\end{equation}
where $\xi^{\rm CT}_{\rm k,k'}$ is the reaction rate between atoms $k$ and $k'$, and $\Delta E_{\rm k,k'}$ is the energy defect of the reaction, which is negative for endothermic reactions, which act as coolers. 

The heating from Penning ionizations is, based on the rate in \citet{Bell1970}
\begin{equation}
%H_{\rm pe} = 7.5\e{-21}\left(\frac{T}{300~K}\right)^{0.5}n_{\rm He I (2^3 S)}~n_{\rm HI}~\mbox{ergs s}^{-1}~.
H_{\rm pe} = 7.5\e{-21}n_{\rm He I (2^3 S)}~n_{\rm HI}~.%~\mbox{ergs s}^{-1}~.
\end{equation}

Cooling is computed from Eqs. (3)-(6) in KF98 a. It has contributions from line cooling, recombination cooling, and free-free emission.
%\begin{equation}
%C = C_{\rm lines} + C_{\rm rec} + C_{\rm ff}
%\end{equation}
%where 
%\begin{equation}
%C_{lines} = n_e\sum_{i,j>i}\left(n_iC_{ij}(T) - n_j C_{ji}(T)\right)E_{ij} 
%\end{equation}
%The recombination cooling is approximated as
%\begin{equation}
%C_{rec} = \sum_k n_{k+1}n_e\alpha_k\cdot 0.8kT 
%\end{equation}
%and 
%\begin{equation}
%C_{ff} = \sum_k 1.426\e{-27}T^{1/2}Z_k^2 n_e n_k g_{ff} 
%\end{equation}
%The adiabatic cooling is $2T/t$, why is it not possible to include this is a steady state model? Its the cooling per second due to the expansion?? From eq 2 in KF98, setting dT/dt=0 gives
%\begin{equation}
%H = C + 3k(1+x_e)n\frac{T}{t}
%\end{equation}
%what is k here? If adiabatic cooling dominates, T goes as $t^{-2}$.

\subsection{Charge transfer}
\label{sec:ct}

We use charge transfer rates taken from \citet{Rutherford1971}, \citet{Rutherford1972}, \citet{Arnaud1985}, \citet{Pequignot1986}, \citet{Kimura1993}, \citet{Swartz1994}, \citet{Kingdon1996}, \citet{Stancil1998} and \citet{Zhao2004}. %For reactions involving iron-group elements, the rich level structure makes resonances and high reaction rates probable. 
For reactions where no measured or calculated rates exist, we use the recipe for estimating rates given by \citet{Pequignot1986}. 

The many unknown or uncertain rates introduce an uncertainty for the level of ionization for some of the elements. For some zones, this prohibits a definite determination of which elements that re-emit the non-thermal and radiative ionization energy.  We discuss this and the potential effects on the spectrum in Sect. \ref{Res:Tandion} and \ref{sec:eoct}.

%When a reaction occurs, we assume that the net energy released or absorbed is exchanged with the thermal pool, causing heating or cooling of the gas. 
%Since the end states of the products are often unknown, this introduces an uncertainty for the temperature of the gas. The energy flows are by no means unimportant; a single or a set of charge transfer reactions that move the ionizations from element $i$ with ionization potential $E_i$ to element $j$ with ionization potential $E_j$ will have converted a fraction $E_j/E_i$ of the ionization energy to heat. Clearly $E_j/E_i$ can be of order 1/2 or larger (consider for instance if H ionizations would be moved to Ca or Na ionizations).
When the final states are unknown, we assume these to be the ground states of the elements. An exception to this is the important O II + C I $\rightarrow$ O I + C II reaction, where we assume that the oxygen atom is left in the excited 2p($^1$D) state, which is the one closest to resonance with an energy defect of 0.4 eV (Sect. \ref{sec:onemg}). We also assume that the reaction Ca I + O II $\rightarrow$ Ca II + O I occurs to the excited 5p($^2$P) state in Ca II \citep{Rutherford1972}.

\subsection{Radiative transfer}
\label{sec:RT}

Radiative transfer is treated with a Monte Carlo technique. The application of this method for dealing with differentially expanding astrophysical envelopes was described by \citet{Abbott1985}, where it was applied to the case of stellar winds. \citet{Lucy1987} and \citet{Mazzali1993} describe the method for computing photospheric-phase supernova spectra. The main difficulty working with the equation of radiative transfer is the complexity introduced by the line overlaps that are caused by the differential velocity field. This gives a non-local coupling for which a Monte Carlo treatment is better suited. In addition, Monte Carlo radiative transfer is easily generalized to 2D and 3D, and is well suited for parallelization.

The early Monte Carlo codes mentioned above allowed for electron and resonance line scattering, where deexcitations occured in the same transition as the absorption. \citet{Lucy1999} and \citet{Mazzali2000} improved on this treatment by including fluorescence in all transitions directly connected to the (initial) upper level. \citet{Kasen2006} used the same treatment for the fluorescence in computing photospheric-phase, time-dependent spectra for Type Ia supernovae. 
To allow for a more exact treatment of the fluorescence, but still retain the properties of indivisibility and indestructibility for the photon packets, a Monte Carlo formalism based on the concept of macro-atoms was developed in a series of papers by \citet{Lucy2002,Lucy2003,Lucy2005}. Applications and further developments of this method are described in e.g. \citet{Sim2007a}, \citet{Sim2007b} and \citet{Kromer2009}.

Here, we implemented a Monte Carlo algorithm with a similarly high level of physical realism by computing the complete deexcitation cascade following each photon packet absorption. A new photon packet is created for each radiative deexcitation in the cascade, with collisional deexcitations being allowed for as well. The (potential) creation of several packets following the absorption of one packet is handled by a recursive technique. We find that this works well and have therefore not enforced the property of indivisibility otherwise commonly used to simplify the bookkeeping \citep[e.g.][]{Lucy2002}. Also, the property of indestructibility devised in those papers has its main advantage as a Lambda accelerator in the photospheric phase, and is not enforced here. %It enforces radiative equilibrium (which is a perturbed operator prior to convergence) which is appropriate in a Schuster-Schwarzschild treatment where no energy is deposited in the atmosphere.}

\subsubsection{Packet emission}
Below, quantities are denoted by primed symbols in the comoving frame, and by unprimed symbols in the rest frame (observer frame). We ignore terms of order $(V/c)$ or higher in the transfer, except for the critical Doppler shifts. The letter $z$ denotes a random number between 0 and 1, newly sampled for each computation.

Total emissivities $j_{\lambda'}'(i,j)$ from lines and continua (two-photon, free-bound and free-free) are obtained for each zone $i$ and wavelength bin$j$ from solutions to the ionization balance, temperature and NLTE level populations (Sect. \ref{sec:sate}). Photon packets are then created with energy
\begin{equation}
E_{\rm i,j}' = \frac{4\pi j'_{\rm \lambda'}(i,j) V \Delta \lambda_{\rm j}'}{N_{\rm i,j}}~,
\end{equation}
where $V$ is the volume of the zone, $\Delta \lambda'_{\rm j}$ is the width of the wavelength bin, and $N_{\rm i,j}$ is the number of packets to split the emission into. The wavelength $\lambda'$ is chosen as the maximum wavelength of bin $j$. This choice ensures that no line self-absorptions, which are already included in the Sobolev escape formalism, occur in the transfer. We currently use a uniform weighting for the number of packets emitted per zone and wavelength bin, $N_{\rm i,j} = N_0$. From inspection of the emerging spectrum (which we also smooth, see Sect. \ref{sec:emspec}), we find $N_0\sim$100  to be sufficient for obtaining neglegible Monte Carlo noise, for binning width $\Delta \lambda'/\lambda'=10^{-3}$. %The energy for a given packet emitted from zone $i$ at wavelength $\lambda$ in wavelength bin $j$ is
%For the iterations to achive convergence in state parameters, we typically use a smaller value $N_0\sim$10.} %With $\lambda_{\rm min}=500~\AA$ and $\lambda_{\rm max}=21,000~\AA$ 
%The total number of packets emitted per iteration is then $\sim 5\e{5}$.}

The starting radius of the packet is sampled from
\begin{equation}
r = \left[r_{\rm i,in}^3 + z\left(r_{\rm i,out}^3-r_{\rm i,in}^3\right)\right]^{1/3}~,
\label{eq:rstart}
\end{equation}
which corresponds to uniform sampling in a shell bounded by $r_{\rm i,in}=V_{\rm i,in}t$ and $r_{\rm i,out}=V_{\rm i,out}t$. The shell velocities are listed in Table \ref{table:zonemasses}.

Isotropic emission corresponds to sampling a direction cosine $\mu$ from
\begin{equation}
\mu = -1 + 2z~.
\label{eq:dircos}
\end{equation}
%and ignoring aberration we take $\mu=\mu'$.}
%which is transformed to the rest frame direction $\mu$ by the aberration formula
%\begin{equation}
%\mu = \frac{\mu' + V/c}{1 + \mu' V/c}~,
%\label{eq:dircostrans}
%\end{equation}
%where $V$ is the expansion velocity of point $r$, $V=rt$.}
If the zone is a core zone, the internal radius in the (spherical) clump is also determined from% Eq. (\ref{eq:rstart}) using an inner radius of zero and an and outer radius of $R_{\rm i,clump}$
\begin{equation}
r_{\rm int} = z^{1/3} R_{\rm i,clump}~,
\label{eq:rclump}
\end{equation}
where $R_{\rm i,clump}$ is the radius of the clumps of zone type $i$. An internal direction cosine $\mu_{\rm int}$ is also randomly drawn according to Eq. (\ref{eq:dircos}).% and (\ref{eq:dircostrans}).}

\subsubsection{Packet propagation}
For each packet to be emitted, a random number is drawn to determine at which optical depth the packet should be absorbed (if at all)
\begin{equation}
\tau_{\rm abs} = -\ln(1-z)~.
\label{eq:lifetau}
\end{equation}
%The packet is then propagated with the accumulated optical depth continuously being increased due to continuum opacities, and discretely increased when the packets redshifts into a line. When the accumulated optical
%depth transcends the critical value $\tau_{\rm abs}$, an interaction with the process that caused this occurs. If the photon reaches a zone boundary before
%this happens, line and continuum opacities are retrieved for the new zone and propagation
%then continues. See Sect. \ref{sec:thecore} for how we handle the transfer in the grid-less core.
Next, the distance to the next critical point is computed.
The comoving wavelength of the photon packet is continuously red-shifting as the packet
travels through the differentially expanding ejecta. From the NLTE solutions, we have a wavelength ordered list of all lines with optical depth $\tau_{\rm ij} > 10^{-3}$. The velocity distance to the next line is computed as %(ignoring terms of order $(V/c)^2$)
\begin{equation}
\Delta V_{\rm line} = c\left(\lambda_{\rm line}^0/\lambda'-1\right)~,%}{\lambda'},
\label{eq:linedist}
\end{equation}
where $\lambda_{\rm line}^0$ is the (atom rest frame) wavelength of the the line closest to $\lambda'$ in that list.

%and we approximate $\Delta V_{\rm line} = \Delta V_{\rm line}'$.
The distance to the next ionization edge is computed as
\begin{equation}
\Delta V_{\rm ion.edge} = c\left(\lambda_{\rm ion.edge}^0/\lambda'-1\right)~,%}{\lambda'},
\end{equation}
where $\lambda_{\rm ion.edge}^0$ is the (atom rest frame) ionization edge closest in wavelength to $\lambda'$. %Again we approximate the rest frame distance with the comoving distance, $\Delta V_{\rm ion.edge} = \Delta V_{\rm ion.edge}'$.
Also, the distance to the next shell $\Delta V_{\rm shell}$ is computed, %from the law of cosines,
 as is the distance to the clump edge $\Delta V_{\rm clump}$, if the zone is a core zone.
%\begin{equation}
%\Delta V_{\rm shell} = -V \mu \pm \left[V^2\mu^2 -\left(V^2- V_{\rm shell}^2\right)\right]^{1/2}
%\end{equation}
%where the sign is chosen depending on whether impact occurs on the inner or outer shell edge, and $V_{\rm shell}$ is the corresponding velocity.
%If the zone is a core zone, the distance to the clump edge is similarly computed as
%\begin{equation}
%\Delta V_{\rm clump} = -V_{\rm int} \mu_{\rm int} +\left[V_{\rm int}^2\mu_{\rm int}^2 -\left(V_{\rm int}^2- V_{\rm clump}^2\right)\right]^{1/2}
%\label{eq:blobdist}
%\end{equation}
%where $V_{\rm int} = r_{\rm int}t$ and $V_{\rm clump} = R_{\rm i, clump}t$.}
The velocity distance to the next critical point is then chosen as the smallest of these four distances
\begin{equation}
\Delta V = \mbox{min} \left\{\Delta V_{\rm line},~\Delta V_{\rm ion.edge},~\Delta V_{\rm shell},~\Delta V_{\rm clump}\right\}~.
\end{equation}
The continuum optical depth for this distance is computed as
\begin{equation}
%\tau_{\rm cont} = \Delta V t \left(\alpha_{\rm i,es} + \alpha_{\rm i,bs} + \alpha_{\rm i,dust} + \sum_k \sum_j \alpha_{\rm pi}^{\rm k,j}\right)
\tau_{\rm cont} = \Delta V t \left(\alpha_{\rm i,dust} + \sum_{\rm k,j} \alpha_{\rm pi}^{\rm i,k,j}(\lambda') + \alpha_{\rm i,ff}(\lambda')\right)~,
\end{equation} 
where $\alpha_{\rm i,dust}$ is the absorption coefficient for the dust
\begin{eqnarray}
\alpha_{\rm i,dust} &=& \frac{\tau_{\rm d}}{V_{\rm core}t}~~~~~~~V_{\rm out}^i \le V_{\rm core} \nonumber \\
                  &=& 0~~~~~~~~~~~~~~~\mbox{otherwise},
\end{eqnarray}
 $\alpha_{\rm pi}^{\rm i,k,j}(\lambda')$ is the photoionization absorption coefficient,
%\begin{equation}
%\alpha_{\rm pi}^{\rm i,k,j}(\lambda') = n_{\rm i,k,j}\sigma_{\rm pi}^{\rm k,j}~,
%\end{equation}
and $\alpha_{\rm i,ff}(\lambda')$ is the free-free absorption coefficient.
%\begin{equation}
%\alpha_{\rm i,ff}(\lambda') = 3.7\e{8}T^{-1/2} n_{\rm e} \nu'^{-3}\left(1-e^{-\rm h\nu'/kT}\right)\bar{g}_{\rm ff}\sum Z_{\rm k}^2 n_{\rm k}~.
%\end{equation}
 %We ignore to differentiate between comoving and rest frame opacities. 
Electron scattering is neglegible at the epoch studied here, but can easily be included as well.

The accumulated optical depth $\tau_{\rm tot}$ is increased by $\tau_{\rm cont}$. If this causes $\tau_{\rm tot}$ to transcend $\tau_{\rm abs}$, the packet is continuum-absorbed, and propagation terminates. The total number of photoionizations for zone $i$, atom $k$, level $j$, is then increased by
\begin{equation}
%\Delta N_{\rm i,k,j} = \frac{\Delta V t\alpha_{\rm pi}^{\rm i,k,j}(\lambda')}{\tau_{\rm cont}} \frac{N_{\rm packet}}{\Delta t'}~,
\Delta N_{\rm i,k,j} = \frac{\Delta V t\alpha_{\rm pi}^{\rm i,k,j}(\lambda')}{\tau_{\rm cont}} N_{\rm packet}~,
\label{eq:deltaNikj}
\end{equation}
where $N_{\rm packet}$ is the number of photons in the packet
\begin{equation}
N_{\rm packet} = \frac{E'}{hc/\lambda'}~.
\end{equation}
 The photoelectric heating rate contribution is
\begin{equation}
%\Delta H_{\rm pi}^i = \frac{1}{V_{\rm i}\Delta t'}\sum_{\rm k,j} \Delta N_{\rm i,k,j}\left(\frac{hc}{\lambda'}-\chi_{\rm k,j}\right)~, 
\Delta H_{\rm pi}^i = \frac{1}{V_{\rm i}}\sum_{\rm k,j} \Delta N_{\rm i,k,j}\left(\frac{hc}{\lambda'}-\chi_{\rm k,j}\right)~, 
\label{eq:piheating}
\end{equation}
where $\chi_{\rm k,j}$ is the ionization potential for atom $k,$ level $j$. The free-free heating rate contribution is
\begin{equation}
%\Delta H_{\rm ff}^i = \frac{1}{V_{\rm i}\Delta t'} \frac{\Delta V t\alpha_{\rm i, ff}(\lambda')}{\tau_{\rm cont}}E'
\Delta H_{\rm ff}^i = \frac{1}{V_{\rm i}} \frac{\Delta V t\alpha_{\rm i, ff}(\lambda')}{\tau_{\rm cont}}E'~.
\label{eq:ffheating}
\end{equation}

If continuum absorption does not happen, the position, flight angle, wavelength and energy of the packet are now updated. The new position is
\begin{equation}
%r_{\rm new} = r + \Delta V t \mu
r_{\rm new} = \left[r^2 + (\Delta Vt)^2 +2 r \Delta V t \mu\right]^{1/2}~.
\end{equation}
The new flight angle is
\begin{equation}
\sin{\theta}_{\rm new} = \sin{\theta}\frac{r}{r_{\rm new}}~.
\end{equation}
The new comoving wavelength is 
\begin{equation}
%\lambda'_{\rm new} = \lambda' \gamma \left(1 + \frac{\Delta V}{c}\right).
\lambda'_{\rm new} = \lambda' \left(1 + \frac{\Delta V}{c}\right)~, 
\label{eq:redshift}
\end{equation}
%where $\lambda'$ is the comoving wavelength at a point separated by velocity $\Delta v$ from a frame where the comoving wavelength is $\lambda$. When moving from one point to another, we update the comoving photon wavelength according to Eq. (\ref{eq:redshift}), and 
and the new comoving energy is%, we take both the Doppler frequency shift and the differences between received and emitted time intervals into consideration (making it effectively a comoving power), and compute}
\begin{equation}
%E'_{\rm new} = E'\left[\gamma \left(1+\frac{\Delta V}{c}\right)\right]^{-2}.
E'_{\rm new} = E'\left(1+\frac{\Delta V}{c}\right)^{-1}~.
%E' = E\frac{\lambda}{\lambda'}
\end{equation}
%The time interval over which the photons in the packet arrive also transforms with one Doppler factor, so
%\begin{equation}
%%E'_{\rm new} = E'\left[\gamma \left(1+\frac{\Delta V}{c}\right)\right]^{-2}.
%\Delta t'_{\rm new} = \Delta t'\left(1+\frac{\Delta V}{c}\right)~.
%%E' = E\frac{\lambda}{\lambda'}
%\end{equation}
%the second Doppler factor correct also for the difference between emitted and received time intervals
%\begin{equation}
%\Delta t'_{\rm rec} = \gamma\left(1-\frac{v}{c}\right)\Delta t_{\rm emit}
%\end{equation}
%effectively letting the energy content of a packet represent to comoving power

If the new point $r_{\rm new}$ corresponds to the passage of an ionization edge or a shell edge, new velocity distances are computed, starting from Eq. (\ref{eq:linedist}), and the procedure is repeated. If $r_{\rm new}$ corresponds to entering a new core clump, its type is chosen from surface impact probabilities given by
\begin{equation}
p_{\rm i} = \frac{R_{\rm i,clump}^2}{\sum_j  R_{\rm j,clump}^2}~,
\end{equation}
which is valid as long as all clumps have the same number of fragments ($N_{\rm cl}$). An impact parameter is also drawn as
\begin{equation}
\mu_{\rm imp} = z~.
\end{equation}
Propagation then continues into the new clump, starting from Eq. (\ref{eq:linedist}).

\subsubsection{Line interaction}
If $r_{\rm new}$ corresponds to reaching a line with $\tau_{\rm ij} \ge 10^{-3}$, $\tau_{\rm tot}$ is increased by $\tau_{\rm line}$. If $\tau_{\rm tot}$ then transcends $\tau_{\rm abs}$, the packet is absorbed in the line.

All downward radiative and collisional transition rates are then computed. If $u$ is the upper level of the line, the radiative deexcitation rates are ($i<u$)
\begin{equation}
R_{\rm u,i} = A_{\rm u,i}\beta_{\rm u,i}^{\rm eff}~.
\label{eq:raddeexc}
\end{equation}
%where $\beta_{u,i}^{\rm eff}$ is the effective loss probability. 
The collisional deexcitation rates are \citep{OB}
\begin{equation}
C_{\rm u,i} = \frac{8.629\e{-6}}{T^{1/2}}\frac{\Upsilon_{\rm u,i}(T)}{g_{\rm u}} n_{\rm e}~,
\label{eq:colldeexc}
\end{equation}
where $\Upsilon_{\rm u,i}(T)$ is the effective collision strength.
The rates are normalized to obtain the relative transition probabilities, and one of the deexcitation channels is selected by a random draw. 

If this is a radiative transition (with a wavelength shorter than a pre-selected cut-off value, chosen as 2.1 $\mu$m here) a new photon packet is emitted. The energy of this packet is 
\begin{equation}
E_{\rm out}'=E_{\rm new}'\left(\Delta E_{\rm ui}/E_{\rm u}\right)~,%E_{\rm u}-E_ {\rm i}}{E_{\rm u}},
\label{eq:energypacket}
\end{equation}
where $\Delta E_{\rm ui}$ is the energy difference between levels $u$ and $i$ and $E_{\rm u}$ is the energy of level $u$. A direction is randomly drawn from Eq. (\ref{eq:dircos})% and (\ref{eq:dircostrans})
, and the packet is followed, starting from Eq. (\ref{eq:lifetau}), until it (and its potential offspring packets) are absorbed or escape the ejecta.

If the deexcitation is instead a collisional one, a contribution to the volumetric heating rate 
\begin{equation}
%\Delta H_{\rm cd}^{\rm i} = \frac{1}{V_{\rm i} \Delta t'}E_{\rm out}'~,
\Delta H_{\rm cd}^{\rm i} = \frac{1}{V_{\rm i}}E_{\rm out}'~,
\label{eq:collheat}
\end{equation}
is recorded. We ignore the possibility of collisional excitations in the cascade. This is a good approximation at late times when the low temperatures make upward collisions much rarer than downward ones.

New deexcitation probabilities from level $i$ are now computed from Eqs. (\ref{eq:raddeexc}) and (\ref{eq:colldeexc}), and a channel is again randomly selected. The process repeats until the atom has deexcited to the ground state. The energy in each deexcitation from level $i$ to level $j$ is given by Eq. (\ref{eq:energypacket}) with $\Delta E_{\rm ui}$ replaced by $\Delta E_{\rm ij}$.
%\begin{equation}
%E_{\rm out}'=E_{\rm in}'\frac{E_{\rm i}-E_ {\rm j}}{E_{\rm u}}.
%\label{eq:eout}
%\end{equation}
This treatment means that
\begin{equation}
\sum E_{\rm out}' = E_{\rm new}'~.
\end{equation}
While this becomes only an approximate treatment for the energy distribution of the fluorescence when the absorption occurs from an excited state, this is rare at late times when most atoms are in their ground states. This treatment will likely be modified for calculations at earlier epochs when this may not be the case. Note that the outgoing energy (assuming only radiative deexcitations) is generally not conserved in the rest frame, although it is in the comoving frame. This corresponds to adiabatic losses of the radiation field.

.%. the photospheric phase of the supernova. 

%If a continuum absorption occurs, the photon energy is split in proportion to the continuum opacity of each element (and the dust), and adds to the bookkeeping of ionization rates and radiative heating.

%\\
%We do not deploy either indestructability or indivisibility at this point, as recursion is not a big problem in the nebular phase of the supernova, and temperature and ionization are fixed input parameters. 
%At early times, the optical depth in Ly$\alpha$ is so high that the Sobolev treatment may be inappropriate due to the high optical depth by close by Fe II or H$_2$ lines. At these late times, however, the Ly$\alpha$ optical depth is low enough that this is not an important effect. We therefore treat Ly$\alpha$ with normal Sobolev and continuum escape probabilities.

\subsubsection{Emerging spectrum}
\label{sec:emspec}
When it reaches the edge of the nebula (with expansion velocity $V_{\rm e}$), the observed wavelength and energy of the escaping packet are obtained as
\begin{equation}
%\lambda_{\rm obs} = \lambda'\gamma \left(1-\frac{V_{\rm e}}{c}\mu_{\rm e}\right),
\lambda_{\rm obs} = \lambda'\left(1-\frac{V_{\rm e}}{c}\mu_{\rm e}\right)~,
\end{equation}
\begin{equation}
%E_{\rm obs} = E'\left[\gamma \left(1-\frac{V_{\rm e}}{c}\mu_{\rm e}\right)\right]^{-2},
E_{\rm obs} = E'\left(1-\frac{V_{\rm e}}{c}\mu_{\rm e}\right)^{-1}~,
\end{equation}
%and 
%\begin{equation}
%%E_{\rm obs} = E'\left[\gamma \left(1-\frac{V_{\rm e}}{c}\mu_{\rm e}\right)\right]^{-2},
%\Delta t_{\rm obs} = \Delta t'\left(1-\frac{V_{\rm e}}{c}\mu_{\rm e}\right)~,
%\end{equation}
where $\mu_{\rm e}$ is the direction cosine of the packet upon crossing the outer edge.
The escaping packets are binned according to $\lambda_{\rm obs}$, using a wavelength bin size of $\Delta \lambda/\lambda=0.001$.
%\begin{equation}
%%F_{\rm \lambda} = \frac{1}{4\pi d^2}\sum \frac{E_{\rm obs}}{\Delta t_{\rm obs}}~,
%F_{\rm \lambda} = \frac{1}{4\pi d^2}\sum E_{\rm obs}~,
%\end{equation}
%where $d$ is taken to be 51.4 kpc. 
The final spectrum is obtained by smoothing this distribution with a box-car filter of width $\Delta \lambda/\lambda=0.002$.

An alternative method for computing the emerging spectrum would be to solve for full-size atoms, including photoexcitation rates in the NLTE level population solutions, and compute the formal integral. However, computing large NLTE atoms is time-consuming and also numerically challenging in terms of convergence and accuracy. The line transfer mainly depends on the populations of the ground states and other low-lying states, which are more reliably computed than high-lying states \citep{Lucy1999}. We therefore choose to compute smaller NLTE solutions without photoexcitation/deexcitation rates and treat the scattering/fluorescence process explicitly in the Monte Carlo transfer, which also allows us to investigate its properties from the emerging packet histories.

Because our model assumes steady state, we checked that the typical photon flight times $t_{\rm p}$ are short compared to the dynamic time scale. Crossing the ejecta takes a time of about $\sim V_{\rm e}t/c$ for the photons, which means that $t_{\rm p}/t \sim V_{\rm e}/c \ll 1$. However, UV photons may experience multiple line scatterings before emerging and their flight times could therefore become long. Here, the presence of dust in the ejecta limits the total flight times to a maximum of a few crossing times, and steady state is therefore justifiable from the radiative transfer perspective as well.

\subsection{Dust}
\label{sec:dust1}

There is strong evidence from blue-shifting line peaks, a developing IR excess, and
a simultaneous decrease in the optical flux that dust formed in the
inner parts of the ejecta, starting around day 530 \citep{Lucy1989,Wooden1993}, or possibly as early as day 350 \citep{Meikle1993}. By day 700 most of the dust formation was probably over \citep{Lucy1989}. The apparently weak wavelength dependence of the dust extinction and the absence of
any dust features in the IR emission suggest that the dust was at least partly formed in optically thick clumps \citep{Lucy1991part2}.

The last far-IR observations are from
day 1731 \citep{Bouchet1996}, when the dust temperature was estimated to be $\sim$155 K.
\citet{Fassia2002} find that essentially all the NIR ejecta lines have blue-shifted peaks even at day 2112, which indicates that the dust was still optically thick at that time. We determined the median value of their measured line-shifts to 500 \kms. This agrees well with \citet{Wang1996}, who find blue-shifts of $\sim$400 \kms~for several optical lines at 1862 and 2210 days. 

Between 2000 and 6000 days there were no instruments capable of detecting the FIR emission by the ejecta.
\citet{Bouchet2004} identify a Gemini $N$ filter detection at 6094 days with ejecta dust. 
%Non-detection in the Qa band then limits the dust temperature to $>90$ K. On the other hand, $T<100$ K is required in order not to emit more than the energy deposited by $2\e{-4}~M_\odot$~$^{44}$Ti. Spitzer
Observations at 6526 days were dominated by the heated
dust emission from the ejecta -- ring collision, and no ejecta dust was detected
\citep{Bouchet2006}. But the HST imaging shows a clear 'hole' in
the central parts of the ejecta even at these late times \citep{Larsson2011}, suggesting that the dust is still optically thick. This may, however, also be a result of external X-ray heating from the ring collisions \citep{Larsson2011}.

We examined the observed line profiles of Mg~I] \wl4571 and H$\alpha$
%, [Fe~I] 1.44$\mu$, and [Si~I] 1.607/1.646$\mu$ 
in the eight-year spectrum (see Sect. \ref{sec:TOS} for a description of the observed spectrum). Both show distinct blue-shifts of their peaks of $\sim$700 and $\sim$600 \kms, respectively, 
indicating that the dust is still providing extinction at this epoch. This was also noted from the H$\alpha$ profile by C97. 
The constancy of the line shifts compared to earlier epochs suggests that the dust resides in co-expanding clumps with optical depths much higher than unity also at this late epoch.

Assuming emission and
absorption in a homogeneous sphere of velocity $V_{\rm d}$, the dust radial optical depth $\tau_{\rm d}$
can be found from the blue-shift of the peak $\Delta V$ \citep{Lucy1989}
\begin{equation}
\frac{\Delta V}{V_{\rm d}} = -1 + \frac{\ln\left({1+\tau_{\rm d}}\right)}{\tau_{\rm d}}~.
\label{formula:taud}
\end{equation}
 %Note that this formula can only be directly applied to an emission line; a line resulting from scattering or fluorescence will have contributions from photons having traveled a different average pathway. Thus, in the UV, where several redirections may occur, we can expect to see different values for the line shifts.
%Using this formula with 650 \kms and a core size of 1800 \kms gives an opacity of $\tau_d=1.3$
%Table \ref{table:dustshifts} shows the derived vales of $\tau_d$ from the line profiles in the 8-year spectrum. 
Using this equation %Eq. (\ref{formula:taud}) 
for the Mg~I] \wl4571 and H$\alpha$ line shifts %(which are mainly produced by direct emission as later confirmed by the modeling) 
with $V_{\rm d}=2000$ \kms gives $\tau_{\rm d} = 1.0-1.2$, similar to its value at earlier epochs \citep{Lucy1989,Wooden1993}. Using $\Delta V = 400 ~\kms$, as found from the day 1862 and day 2210 spectra by \citet{Wang1996}, gives a value of $\tau_{\rm d}=0.6$, and using $\Delta V = 500 ~\kms$, as determined in \citet{Fassia2002}, gives $\tau_{\rm d}=0.8$. 

Because it is unknown exactly in which zone(s) the dust resides,  we model the dust as a constant, gray opacity within the core. From the estimates above, $\tau_{\rm d}$ is likely to be in the $0.6-1.2$ range. For our standard model we choose $\tau_{\rm d}=1$.

We also note that dust in the ejecta can have strong effects on the cooling (KF98 a, b), with consequences for the spectrum depending on where the dust formed.

\section{The observed spectrum}
\label{sec:TOS}

For the UV/optical range, we use the post-COSTAR HST Faint Object Spectrograph observations obtained on January 7 1995 (2875 days or 7.87 years after explosion), which covers the wavelength range from 1650 to 9200 \AA. Details on the observations and the data reduction can be found in C97. The flux calibration is stated to be good to $\sim$5\%. The aperture excludes the inner circumstellar ring, but the northern outer loop passes through the same line of sight as the central debris, and therefore contaminates
the spectrum with some narrow lines. This is not a major problem because these lines can be easily identified. The strongest ones are C III] \wl1909,  [O~II] \wl 3727, [O III] \wll 4959, 5007, [N~II] \wll6548, 6584, and H$\alpha$ \wl6563 \citep{Panagia1996}. We did not remove these narrow lines from the observed spectra displayed in the paper.
% with resolution ranging from 1.35 ~\AA~ at the blue
%end to 6.17 ~\AA~ at the red end. 

For the near-infrared, we use the AAT spectrum from \citet{Fassia2002} from day
2952. %Although also a later spectrum exists at day 2952, the resolution and wavelength coverage of that spectrum is poorer. We compared the spectra and found the flux levels to have changed little between these epochs, which is also to be expected from the slow $^{44}$Ti decay. 
The IR spectrum contains contributions from both the ejecta and the inner equatorial ring. Because of the limited spectral resolution of $\lambda/\Delta\lambda=120$, narrow lines from the ring can appear almost as broad as the ejecta lines. In addition,
the removal of both continuum and line emission from star 3 and possibly star 2 adds
a significant uncertainty. The accuracy of the absolute flux levels is stated as $\pm40\%$. The NIR spectrum
is therefore useful for line identifications, but only for approximate quantitative comparisons.
%...), scaled by the total flux as measured from the IR
%photometry in \cite{S2002}.

To compare the model to the observations, we adopt a distance of 51.4 kpc \citep{Panagia2005}, the extinction curve (for $\lambda<8000~\AA~$) derived from the neighboring star 2 in \citet{Scuderi1996}, which has $E_{\rm B-V}=0.19$. For $\lambda>8000~\AA$ we use the extinction curve in \citet{Cardelli1989}.  
A different analysis by \citet{Fitzpatrick1990} find $E_{\rm B-V}=0.16$, and we take $\pm$0.03 to be a reasonable estimate for the extinction uncertainty. 
This dereddening gives a total luminosity  in the 1600-8500 \AA~range of $3.4\e{35}$~\ergs. The value found by
C97 is $3.5\e{35}$~\ergs, using a distance of 50 kpc and $E_{\rm B-V}=0.2$. We also adjust the model spectrum for interstellar UV absorption lines based on \citet{Blades1988}.

The distance uncertainty to SN 1987A is stated as 1.2 kpc in \citet{Panagia2005}, which corresponds to a flux uncertainty of $\pm$4.8\%. 
%Using $E_{\rm B-V}=0.2$ instead of $E_{\rm B-V}=0.16$ raises the 1600-8500 \AA flux by 21\%. 
Dereddening the spectrum with $E_{\rm B-V}=0.16$ or 0.22 instead of 0.19 implies as change of $\sim\pm$ 13\% in the total luminosity in the 2000-8000 \AA~range (which is the range we later use for estimating the \iso{44}Ti mass). Combined, the observed luminosity then has an estimated uncertainty of $\sim (5^2+4.8^2+13^2)^{1/2}=\pm 15\%$.  

\section{Results}
\label{sec:results}

\subsection{Energy deposition and degradation}
At eight years, \iso{44}Ti completely dominates the energy input over \iso{56}Co and \iso{57}Co. Using a $^{44}$Ti mass of \bestti, which we later show to be the best value for reproducing the observed spectrum, the total energy deposition in the ejecta is $1.9\e{36}$ \ergs. Of this, $\sim$85\% is positron energy; the other $\sim$15\% is gamma-ray energy, which represents $\sim$5\% of the $5.2\e{36}$ \ergs gamma-rays emitted. The other $\sim$95\% of the gamma-rays escape the ejecta.

The dominance of positrons, combined with the assumption that these are locally absorbed, means that most of the energy ($\sim$80\%) is deposited in the Fe/He clumps. Most of the gamma-rays are absorbed in the H zones, giving them $\sim$8\% of the total energy deposition. The remaining $\sim$12\% are split between the silicon, oxygen, and helium zones. All energy depositions are given in Table \ref{table:Zoneresults}.

\subsection{Temperature, ionization, and emissivities}
\label{Res:Tandion}
The solutions for the temperature and ionization in the various zones are given in Table \ref{table:Zoneresults}. From these steady-state solutions, we need to check the assumption that the recombination time-scale is shorter than
the time-scale for density change $n/\dot{n}=t/3\sim 2.6$ years. The last column in Table \ref{table:zonemasses} lists the recombination times ($1/\left[n_{\rm e} \alpha\right])$ in years for the dominant element in the various zones. We see that a steady-state treatment of all core zones as well as of the helium envelope zone is justified, whereas the hydrogen envelope is experiencing freeze-out as expected.

The temperature is 70-170 K in all core zones (Table \ref{table:Zoneresults}).  The temperatures could be even lower if molecular cooling or dust cooling are important. This would not have any strong influence on the UV/optical/NIR spectra, however, because non-thermal processes dominate the output at those energies. Table \ref{table:coolers} lists the dominant coolers in the various zones. For the envelope, we use the temperatures listed in Table \ref{table:freeze}.

\begin{table}[htb]
\caption{Dominant cooling transitions for each zone.}
\centering
\begin{tabular}{l l l l}
\hline\hline
Zone & Cooler & Fraction\\
\hline
Fe/He & [Fe~II] 25.99 $\mu$m & 94\% \\
      & [Fe~II] 14.98 $\mu$m* & 5\%\\
      & [Fe I]   24.04 $\mu$m & 1.4\%\\
\hline
Si/S & [Si I] 68.47 $\mu$m &  63\%\\
     & [Si I] 44.81 $\mu$m &  22\%\\
     & [Si I] 129.68 $\mu$m & 15\%\\
\hline
O/Si/S & [O~I] 63.19 $\mu$m & 35\%\\
       & [Si~I] 68.47 $\mu$m & 32\%\\
       & [O I] 44.06 $\mu$m* & 17\%\\
\hline
O/Ne/Mg & [O I] 44.06 $\mu$m & 33\%\\
       & [O I] 63.19 $\mu$m & 29\%\\
       & [Si I]  68.47 $\mu$m & 16\%\\
\hline
O/C    & [O~I] 63.19 $\mu$m & 41\%\\
       & [O~I] 44.06 $\mu$m* & 40\%\\
       & [Si I] 129.68 $\mu$m & 7\%\\
\hline
He (core) & [Fe~II] 25.99 $\mu$m & 46\%\\ 
         & [Si I] 68.47 $\mu$m & 23\%\\
        & [Si II] 34.81 $\mu$m & 16\%\\
\hline
H (core) & [Si~II] 34.81 $\mu$m & 59\%\\
         & [O~I] 63.19 $\mu$m & 19\%\\ 
         & [Fe II] 25.99 $\mu$m & 15\%\\
\hline
\end{tabular}
\tablefoot{Transitions that are strictly forbidden (and therefore give no line emission) are marked with an asterisk.}
\label{table:coolers}
\end{table}

\begin{table*}[htb]
\caption{Energy deposition, temperature, and electron fraction in the various zones in the standard model with M(\iso{44}Ti) = \bestti.}
\centering
\begin{tabular}{l l l l l l l l l l l l }
\hline\hline
Zone & Fe/He & Si/S & O/Si/S & O/Ne/Mg & O/C & He & H \\
\hline
Deposition [\ergs] & $1.6\e{36}$ & $9.1\e{34}$ & $3.6\e{34}$ & $5.4\e{34}$ & $1.7\e{34}$ & $2.6\e{34}$\tablefootmark{a} & $1.5\e{35}$\tablefootmark{a}\\
Fractional deposition & 80\% & 4.7\% & 1.9\% & 2.8\% & 0.88\% & 1.4\%\tablefootmark{a} & $8.0$\%\tablefootmark{a}\\
Deposition external [\ergs] & $1.7\e{34}$ & $7.4\e{33}$ & $1.1\e{33}$ & $1.1\e{34}$ & $2.4\e{33}$ & $7.7\e{33}$\tablefootmark{a} & $1.3\e{34}$\tablefootmark{a}\\ 
\hline
Heating    & 0.59 & 0.31 & 0.23 & 0.19 & 0.22 & 0.20\tablefootmark{b} & 0.14\tablefootmark{b}\\
Excitation & 0.16 & 0.27 & 0.21 & 0.21 & 0.18 & 0.23\tablefootmark{b} & 0.41\tablefootmark{b}\\
Ionization & 0.25 & 0.42 & 0.56 & 0.60 & 0.60 & 0.57\tablefootmark{b}  & 0.45\tablefootmark{b}\\
\hline
Temperature  [K] & 170 & 140 & 140 & 70 & 70 & 140\tablefootmark{b} & 100\tablefootmark{b}\\
Electron fraction & 0.17 & 0.020 & $2.9\e{-3}$ & $9.0\e{-4}$ & $5.6\e{-4}$ & $2.1\e{-4}$\tablefootmark{b} & $1.5\e{-4}$\tablefootmark{b} \\
\hline
%Cooler & [Fe~II] 26 $\mu$m  91\% & [Fe~I] 24 $\mu$m  43\% & [O~I] 63 $\mu$  36\% & [O~I] 63 $\mu$m 63\% & [O~I] 63 $\mu$m  53\% &[Si~I] 68.4 $\mu$m 44\% &  [Si~II] 34.8 $\mu$m  49\% \\
%Cooler & [Fe~II]  91\% & [Fe~I]  43\% & [O~I]  36\% & [O~I]  63\% & [O~I]   53\% &[Si~I]  44\% &  [Si~II]  49\% \\
%       & [Fe~II]  6\%  & [O~I]   33\% & [Si~I] 30\% & [Si~I] 38\% & [O~I]   48\% & [Si~II 35\% & [Fe~II]   33\% \\
%\hline
\end{tabular}
\tablefoot{
'Deposition external' is the amount of absorbed radiation originating in other zones.\\
  \tablefoottext{a}{All zones.}
  \tablefoottext{b}{Core zone.}
}
\label{table:Zoneresults}
\end{table*}

We now discuss the individual zones in more detail to understand the emission lines that are produced. In Sect. \ref{sec:pos} we discuss how the energy deposition in the various zones changes if there is a non-local positron deposition. 

\subsubsection{Fe/He zone}
\label{sec:fehe}
The high electron fraction in the Fe/He zone ($\sim$0.17)  implies that a large part ($\sim$59\%) of the deposited energy goes into heating, re-emerging mainly in the [Fe~II]~26 $\mu$m line (Table 5). The reason that this line dominates over [Fe~I]~24 $\mu$m is that it has a much larger collision strength, $\sim$6 versus $\sim$0.02 \citep{Pelan1997,Zhang1995}. Because the Fe/He zone absorbs $\sim$80\% of the deposited energy (assuming on-the-spot positron absorption), and $\sim$60\% of this goes into heating, $\sim$50\% of the total deposited energy or $\sim 9\e{35}$~\ergs (or a factor two or so less considering the dust) should emerge in the [Fe~II]~26 $\mu$m line at this epoch. We discuss observations of this line in Sect \ref{sec:discussion}. 
%Observation of this line could serve to check the assumption of local positron absorption. ISO observations at Unfortunately, both ISO and Spitzer had too low spatial resolution and sensitivity to detect the line \citep{Borkowski1997}. 

About 25\% of the energy goes into ionizations, with $\sim$14\% going to He I, $\sim$7\% to Fe I and $\sim$4\% to Fe II. Iron reaches an ionization degree of $\sim$40\%, while helium reaches $\sim$6\%. Fe III is largely neutralized by charge transfer reactions, mainly with Fe I. Nickel, which makes up $\sim$4\% of the zone mass, is almost fully ionized owing to charge transfer with Fe II. He I, Fe I, and Ni I therefore emit strong recombination lines. The recombination line spectrum of Fe~I is rather smooth with significant emissivity in the 2000-6000 \AA~range; this constitutes a major part of the emitted flux in this range (Sect. \ref{sec:sprop}). The He I recombination emission is mainly in the form of the two-photon continuum plus He~I \wl626. This flux is mostly continuum-absorbed by other elements and serves to additionally boost the ionization of iron and other metals (see also KF98 b). In the optical, the strongest recombination line is the 5876 \AA~line. Helium also emits IR lines at 1.083 $\mu$m and 2.058 $\mu$m. The 1.083 $\mu$m line is mainly excited by recombinations, while the 2.058 $\mu$m line arises mainly as a result of non-thermal excitations.

The remaining $\sim $16\% of the deposited energy excites resonance lines from the ground states of Fe~I ($\sim$9\%), Fe II ($\sim$3\%) and He I ($\sim$3\%). For the Fe I transitions, we use the approximate Bethe rates, whereas for Fe II and He I we have more accurate cross sections.  
%The $2p^1P$ level in HeI gets most of the energy going into helium, and gives rise to the 2.058$\mu$ line as mentioned above. 
%We note that the fine-structure levels of Fe~I and Fe~II are generally far from LTE, even in the ground multiplets.

Based on modeling of the Fe II lines during the first 800 days \citep{Li1993iron}, the density of the Fe/He zone may be lower than the one we use here, but not higher. Lowering the density of this zone by a factor of two increases the heating fraction by about five percentage units, at the expense of excitations and ionizations. Assuming that the UV/optical/NIR flux scales in proportion to the excitation plus ionization energy, the UV/optical/NIR emission from the Fe/He zone then falls by $\sim$12\% by such a maneuver, and the total UV/optical/NIR emission by $\sim$5-10\%. Using $V_{\rm core}=2500$ \kms~instead of 2000 \kms (which lowers all densities by a factor of $\sim$2) lowers the total ejecta emission in the 2000-8000 \AA~range by $\sim$10\%. %This must be kept in mind when discussing the \iso{44}Ti mass.

\subsubsection{Si/S zone}
\label{sec:sis}
The higher density and the smaller amount of $^{44}$Ti ($\sim$5\% of the total) present in this zone compared to the Fe/He zone implies a lower electron fraction ($\sim$0.02). A smaller part of the deposited energy therefore goes to heating (31\%). Of the rest, 42\% goes into ionizations, mainly of the abundant Si~I and S~I. Still, these elements stay almost fully neutral because of rapid charge transfer with elements such as Fe~I, Mg~I and Ti~I. These recombine by further charge transfer reactions, and it is eventually calcium that is the dominantly ionized species. This means that quite a strong Ca~I recombination spectrum is emitted from this zone. In Sect. \ref{sec:eoct}, we show that this holds true even without charge transfer, because calcium is also strongly photoionized.%This conclusion, however, relies on the chain of transfer reactions  which contains several uncertain rates (see Sect. \ref{sec:eoct}).

About 27\% of the energy goes to non-thermal excitations, mainly for transitions in Si I and S I.% (6\%).%, Fe I (2\%), and Ca II (1\%). %The strongest Si I emission line is at 2970/2987 \AA, at $ 3.8\e{33}$ \ergs. 
%We use the Bethe approximation for the Si I lines. 

\subsubsection{O/Si/S zone}
This is the smallest oxygen zone (0.16 \msun) but contains some $^{44}$Ti ($\sim$2\% of the total). The largest part of the energy (56\%) goes into ionizations, which through charge transfer reactions eventually leave iron and aluminium as the dominantly ionized elements. %Aluminium emits recombination lines at 3961 \AA, 3090 \AA~and 1.31 $\mu$m.
% but these are all too weak to be clearly seen. 
The non-thermal excitations occur in similar amounts for O~I, Si~I and Mg~I.

\subsubsection{O/Ne/Mg zone}
\label{sec:onemg}
The O/Ne/Mg-zone is by far the most massive oxygen zone (1.89~\msun), but contain no $^{44}$Ti. The total energy deposition here is 2.8\% of the total. About half of the energy goes into ionizations of oxygen. A series of charge transfer reactions neutralize the oxygen ions and transfer the ionization to sodium, which becomes the most abundant ion. This zone therefore emits strong Na I recombination lines, including the doublet at 5890, 5896 \AA~as well as at 5683, 5688 \AA, 6154 \AA~and 8183, 8194 \AA. 

In our model, the dominant charge transfer reactant with O II is C I, where a reaction to the excited 2p($^1$D) state in O I is close to resonance with an energy defect of only 0.4 eV. With the recipe for estimating rates given in \citet{Pequignot1986}, we estimate the reaction rate to be $1\e{-9}$ cm$^3$s$^{-1}$, which together with the relatively high carbon abundance in this zone (4.3\% by number) makes it the dominant reaction for neutralizing the oxygen ions. Because 2p($^1$D) is the upper state for the [O I] \wll6300, 6364 lines, this has the interesting effect of maximizing the doublet emission efficiency by producing a photon for each O I ionization, which is equivalent to $\sim$7\% of the deposited energy. Only $\sim$ 10\% of radiative recombinations pass through the 2p($^1$D) state at these densities (giving $\sim$0.7\% of the deposited energy), and since other competing charge transfer reactions likely do not occur to 2p($^1$D), the conversion efficiency is even lower if they would dominate. %As we shall see in Sect. \ref{sec:lineids}, the scenario of charge transfer directly to 2p($^1$D)  the observed doublet luminosity.

The collisional cross sections for the [O I] doublet lines are moderate and only $\sim$1.5\% of the deposited energy goes to direct non-thermal excitation of 2p($^1$D). This state also receives some indirect contribution by non-thermal excitations to higher levels, the most important channel being via the 3d($^3$S) state. Collisions to this state contribute $\sim$0.4\% to 2p($^1$D) by cascading. The total energy emitted in [O~I] \wll 6300, 6364 is then $\sim$8.9\% (7\%+1.5\%+0.4\%) of the deposited energy if the charge transfer reaction with C I dominates, and only $\sim$2.6\% (0.7\%+1.5\%+0.4\%) otherwise. 
%If we switch off the charge transfer and allow recombinations to contribute, the flux is moderately increased. We find that the effective recombination rate to the 2p($^1$D) state is $\sim$10\% of the total rate at the oxygen densities at eight years. The maximum fraction of energy passing through the $^1$D state due to recombinations is then $\sim 0.5\cdot 1.96/13.6\cdot 0.1=0.7\%$. These results are in disagreement with those found in C97. We discuss this later. 
% who calculate that $\sim$8\% of the deposited energy is reemitted in the [O I] \wll6300, 6364 doublet.

Magnesium has an abundance (by number) that is $\sim$25 times lower than that of oxygen. But because the Mg I] \wl4571 transition has a large collisional cross-section, it receives about the same number of non-thermal excitations as the [O I] doublet. 
%The [O I] \wll6300, 6364 to Mg I] \wl 4571 line ratio from this zone should therefore be close to unity if both upper states are collisionally populated. 
About 3.7\% of the deposited energy goes to ionizing Mg I, but Mg II is neutralized by charge transfer with Ni I, which in turn ionizes Na I. 
%Even if it was not, the energy from recombinations would be smaller than the energy from collisions, and the ratio of unity should hold. 
%About 20\% of this should pass through the 4571 \AA~line, giving 0.7\%. Clearly, the Mg I] \wll4571 \AA~line is mainly collisionally pumped.

Sodium is five times less abundant still, but Na I has an even larger cross-section to its first excited state, which also receives $\sim$1.5\% of the energy through non-thermal excitations. This contributes further to the Na I \wll5890, 5896 emission, which is, however, dominated by recombinations.
% While the non-thermal ionizations are unimportant, sodium is ionized by charge transfer reactions as well as by the internal UV field. The result is that Na I emits a very strong recombination spectrum. The 5896 \AA~line is dominated by these recombinations.

\subsubsection{O/C zone}
This is the intermediate-mass oxygen zone (0.58 \msun), which receives 0.9\% of the total energy deposition. The low abundance of sodium in this zone, compared to the O/Ne/Mg zone, makes magnesium the dominantly ionized element. A strong Mg~I] \wl4571 recombination line is therefore emitted from this zone, although more magnesium resides in the O/Ne/Mg zone.

\subsubsection{He zone}
Because of mixing, the helium zone has both a core component and an envelope component (Sect. \ref{sec:modeling}). These components behave similarly.
He II is neutralized by charge transfer, mainly with Si I, and ionizations eventually end up in Na II, Mg II, Si II, and Fe II. 

Non-thermal excitations go mainly to the He~I 2p($^1$P) state ($\sim$15\%), resulting in the 2.058 $\mu$m line, with only small contributions from recombinations for this line. As long as the positrons are locally absorbed in the Fe/He zone, all helium emission lines have stronger contributions from the Fe/He zone than from the He zone. As an example, the fraction of the total energy deposited in the ejecta going into non-thermal excitations of 2p($^1$P) is 0.16\% from the helium zones, and 2.3\% from the Fe/He zone, which means a 93\% contribution to the 2.058 $\mu$m lines from the Fe/He zone. Similar relations hold for all other helium lines as well. In Sect. \ref{sec:discussion} we explore this point in more detail. Because of the strong helium emission from the Fe/He zone, the He zone is not expected to leave any particular imprint on the spectrum at this epoch.
%This conclusion is found also for the 19-year spectrum \citep{Kjaer2010}, indicating that $\alpha$-rich freeze-out has indeed happened in the iron zone, given the assumption of local positron deposition.
%See also Sect. \ref{sec:discussion}. 

\subsubsection{H zone}
As with helium, hydrogen is present both in the envelope and as a mixed-in core component (Sect. \ref{sec:modeling}). The envelope zones are powered mainly by freeze-out recombinations. Here, hydrogen emits a low-temperature case B recombination spectrum. The freeze-out emission from the envelope is generally stronger than the emission from the steady-state core, and it provides the bulk of the hydrogen lines and continua in the model. The core component provides the peaks of the hydrogen lines, but also serves to emit ionizing radiation into the other core zones.

In the core H zone, steady-state still prevails and instantaneous gamma-ray deposition determines the output. Ionization of H I accounts for $\sim$36\% of the deposited energy. Because we are close to case B, most recombinations pass through $n=2$. About 28\% of the deposited energy goes into non-thermal excitations of the H I 2p state. Both in the core and in the (inner) envelope, 2s and 2p are still populated according to their statistical weights, i.e. as 1:3. In the core component, the Ly$\alpha$ optical depth is $\sim 4\e{8}$, which gives a ratio of two-photon to Ly$\alpha$ emissivity of $\sim$5. The two-photon continuum therefore dominates the decay of the $n=2$ state. The two-photon continuum can affect the spectrum in many ways. It can boost the ionization of metals with low ionization treshholds, both in the H zones themselves and in the other zones. It can also provide an important input into optically thick lines and affect the UV/optical/NIR spectrum by fluorescence. In wavelength windows where no or few lines exist, the emission may escape and contribute to the UV continuum. In the envelope, the Ly$\alpha$ optical depth is lower and Ly$\alpha$ dominates over the two-photon continuum.

H$\alpha$ from the core is powered in similar amounts by non-thermal excitations and recombinations ($\sim$4\% of the deposited energy each). About 70\% of recombinations lead to H$\alpha$ emission in a Case B scenario at 100 K \citep{Martin1988}.

\subsection{The spectrum}
\label{sec:sprop}

\begin{figure*}[t]
\centering 
\includegraphics[width=1\linewidth]{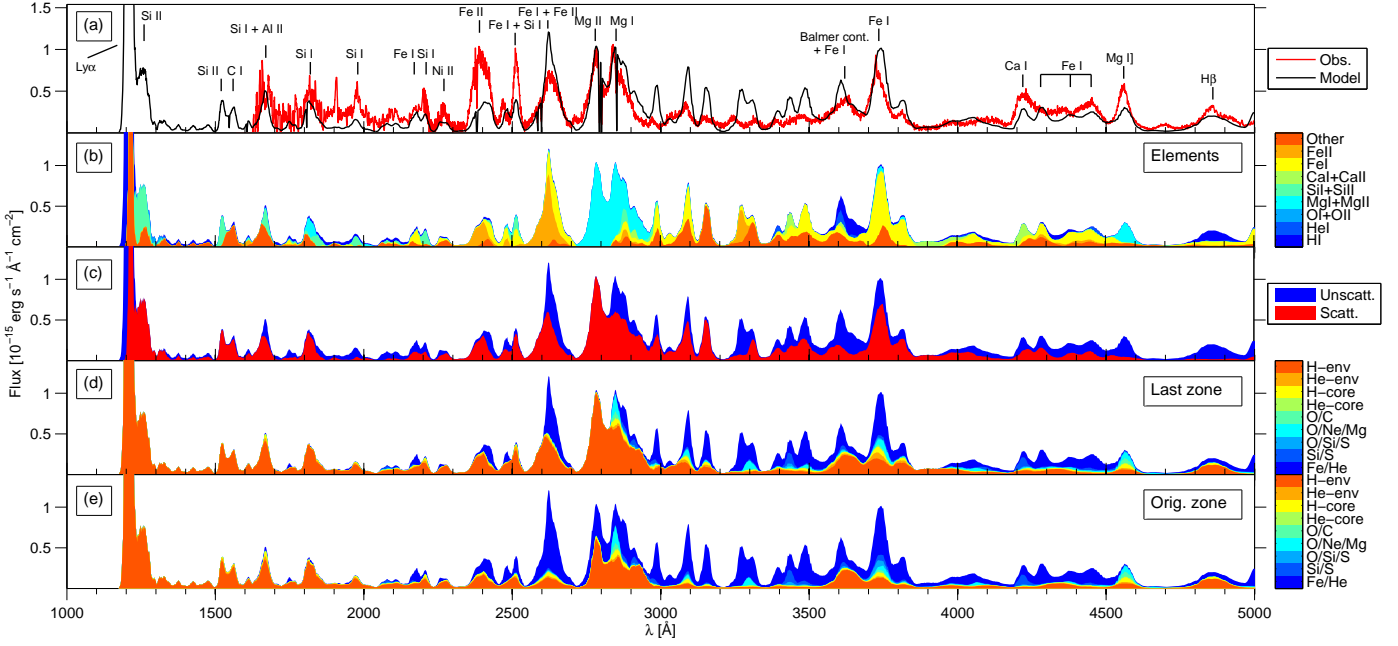}
\caption{Model spectrum at 2875 days for $\mbox{M(\iso{44}Ti)}=\mbox{\bestti}$~and $\tau_{\rm d}=1$, together with the observed spectrum. The 1000-5,000 \AA~range. (a) The observed spectrum (red) versus the model spectrum (black). The observed spectrum was dereddened, corrected for redshift, and smoothed below 2200 \AA~and in the 7500-8500 \AA~range. (b) The contributions by various selected elements to the flux. (c) The part of the emerging flux that comes from scattering/fluorescence (red) and the part that is direct emission (blue). (d) The zones in which the photons had their last interaction; this is the zone of emission if they did not scatter and the zone where they last experienced scattering or fluorescence if they did. (e) The zones from where the photons were emitted before any scattering or fluorescence occurred. The color codings for panels 3-5 are on the right-hand side. Ly$\alpha$ extends to to a peak flux of $1.4\e{-14}$ erg s$^{-1}$ \AA$^{-1}$ cm$^{-2}$. See also Sect. \ref{sec:sprop} in the text.} 
\label{fig:s1} 
\end{figure*} 

\begin{figure*}[htb]
\centering 
\includegraphics[width=1\linewidth]{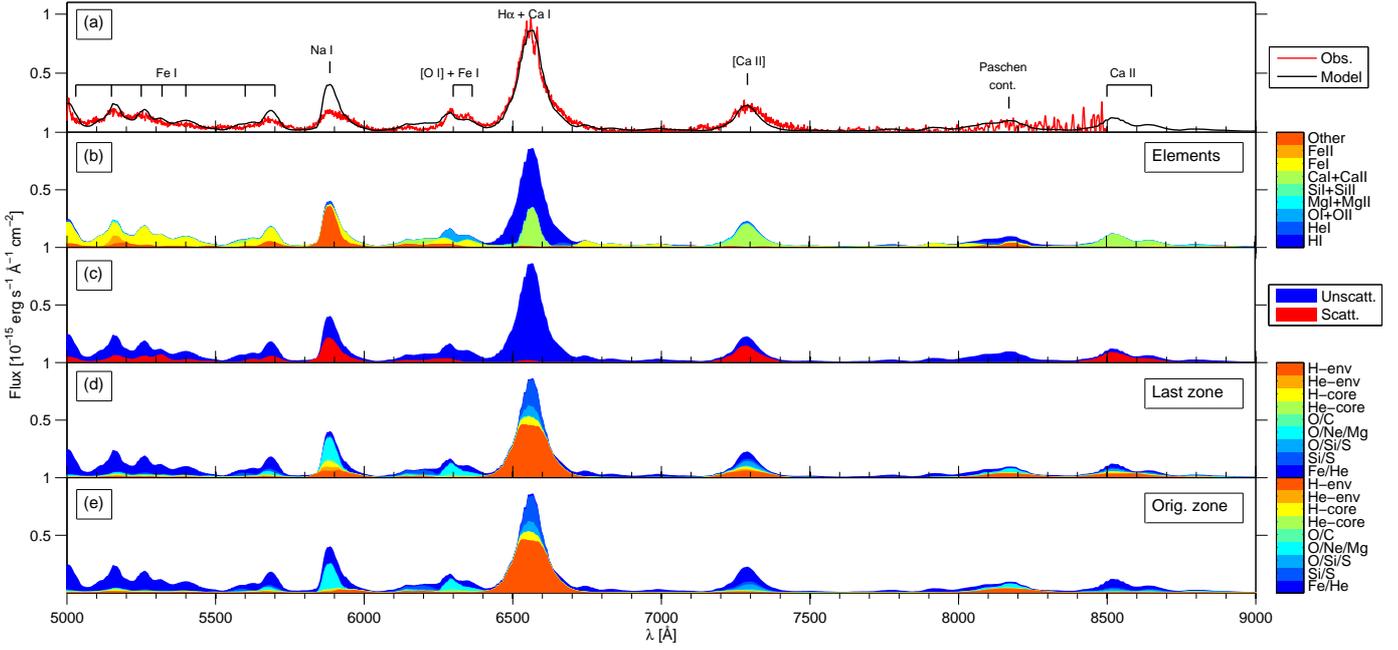} 
\caption{Same as Fig. \ref{fig:s1} but for the 5000-9,000 \AA~range.} 

\label{fig:s2} 
\end{figure*}

%\begin{figure*} 
%\centering 
%%\includegraphics[width=1\linewidth]{spectrum_8years_3.eps} 
%\includegraphics[width=1\linewidth]{spectrum_8years_3b.eps}  % linear
%\caption{Same as Fig. \ref{fig:s1} but for the 6000-10,000 \AA~range.} 
%\label{fig:s3} 
%\end{figure*}

\begin{figure*}[htb]
\centering 
\includegraphics[width=1\linewidth]{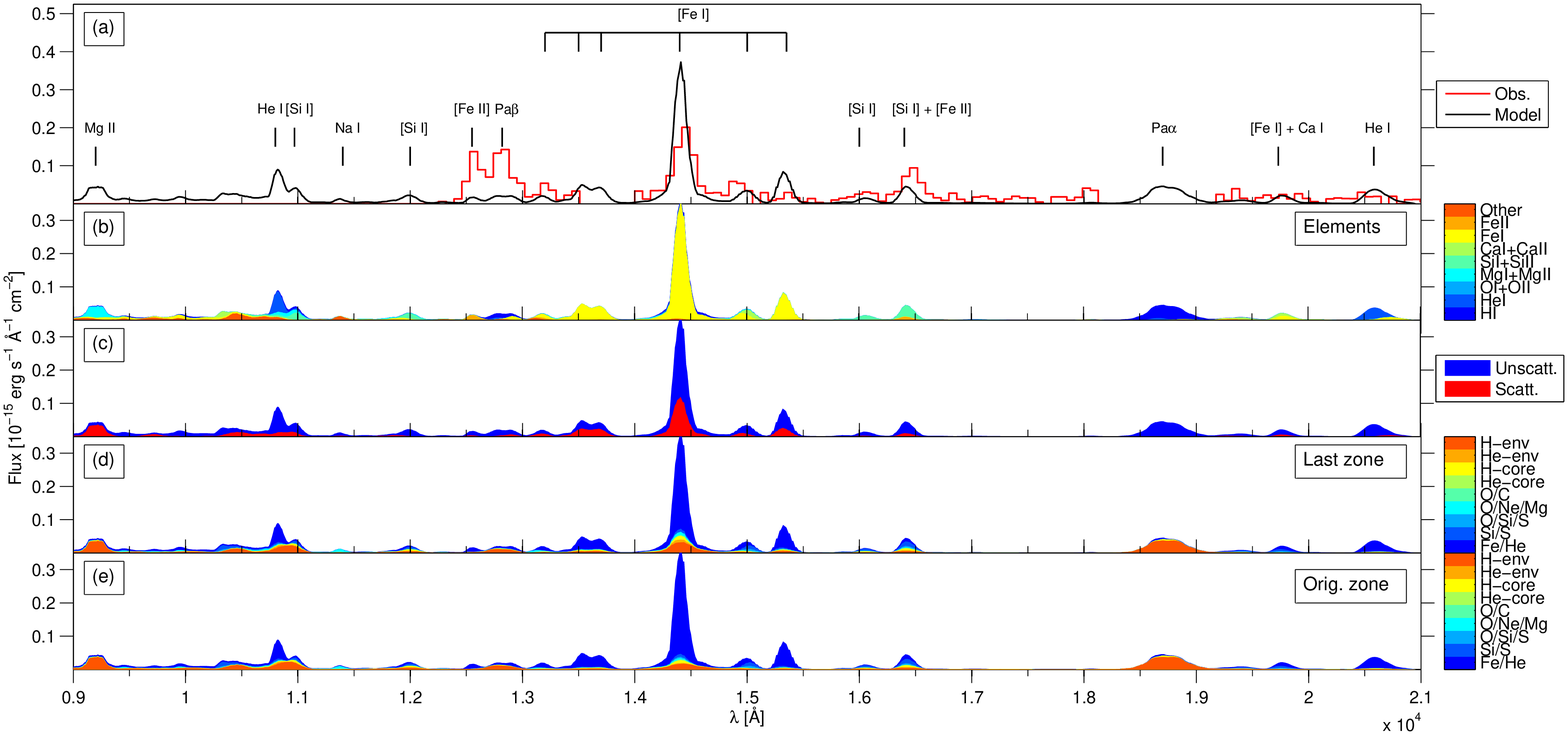}  % linear resub
\caption{Same as Fig. \ref{fig:s1} but for the 0.9-2.1 $\mu$m range.} %The observed spectrum is from day 2112. \textbf{Part of the descrepency between the model and the observations is due to that the observed NIR spectrum has significant contributions from the circumstellar ring, as well as an uncertain flux calibration.}} 
\label{fig:s4} 
\end{figure*}

%\begin{figure}[htb]
%\centering 
%%\includegraphics[width=1\linewidth]{spectrum_8years_4.eps} 
%\includegraphics[width=1\linewidth]{Halpha.eps}  % linear
%\caption{Same as Fig. \ref{fig:s1} but for H$\alpha$ only.} 
%\label{fig:s5} 
%\end{figure}
The model spectrum, for $\mbox{M(\iso{44}Ti)}=\mbox{\bestti}$~and $\tau_{\rm d} = 1$, is shown together with the observed spectrum in Figs. \ref{fig:s1}-\ref{fig:s4}. These figures contain five panels. Panel (a) shows the model spectrum (black) overlaid on the observed spectrum (red). The observed spectrum was dereddened and corrected for redshift (Sect. \ref{sec:TOS}). It was also smoothed below 2200 \AA~and in the 7500-8500 \AA~range. Panel (b) identifies the contribution by individual elements to the spectrum. These are the elements that emitted the photon if it escaped directly, and the element that produced the last scattering/fluorescence if it did not. Panel (c) shows the part of the flux that was processed by scattering and/or fluorescence in red, and the part that comes from direct emission in blue. Panel (d) shows in which zones the photons were last processed. This is the zone of emission for photons that escaped directly, and the zone of the last scattering/fluorescence for those which did not. Finally, panel (e) shows from which zones the photons were originally emitted (which is the same as in panel (d) for the photons that escaped directly). 

We see that the model reproduces most of the features in the observed spectrum, which bolsters our confidence in the fundamental assumptions in the model and to the code. The main exceptions occur in the 3000-3500 \AA~range, which we discuss below. %Considering the complexity and many uncertainties in the modeling, the agreement is satisfactory. 
Apart from this range, all major observed lines are present in the model spectrum, and, equally important, the model does not produce any strong lines that are not observed. 

\subsubsection{Scattering and fluorescence}
\label{sec:saf}

Although the line blocking effect in the photospheric phase has been extensively modeled before \citep[e.g.][]{Karp1977,Lucy1987,Wagoner1991,Mazzali1993}, an analysis of the actual fluorescence in the nebular phase is rarer. 
\citet{Li1996} study the process at 300 and 800 days, using a He I two-photon continuum that scatters in a single-density hydrogen envelope. They conclude that the process transfers a significant fraction of the UV emission into a quasi-continuum
in the optical and NIR up to at least 800 days. The importance of the effect was found to be sensitive to the adopted temperature, which determines the population of the lower levels, and thereby the number of optically thick lines. 

That especially resonance lines can be optically thick even at eight years
can be seen with a simple estimate. The Sobolev optical depth
for a line from the ground state can be written as 
%\eq{ \tau = 0.0265 f_{12} n_i \lambda t \.  }  
%\eq{ \tau = 5.7 f_{12} {X_i \over 10^{-5}} \bar{A}^{-1}
%  \epsilon_i^{-1} {M_i \over \Mo } \left({V_{\rm core} \over 2000
%    \ \kms }\right)^{-3} \left({t \over \ 10 \ \rm yrs }\right)^{-2}
%  \  }
\begin{equation}
\tau \sim 3\e{3} f_{\rm abs} \left(\frac{\lambda}{5000~\AA}\right)\left(\frac{X}{10^{-5}}\right)\left(\frac{n}{10^6~\mbox{cm}^{-3}}\right)~,
\end{equation} 
where $X$ is the number fraction of the element and $n$ is the total number density, which is in the range of $10^4-10^7~\mbox{cm}^{-3}$ in the core at this epoch. Obviously, even elements with low 
abundances can still be optically thick in their ground-state resonance lines. 
%Although not
%important for the cooling, they can therefore have important effects
%by scattering of especially the UV and near-UV photons.

By looking at the panels labeled (b) in Figs. \ref{fig:s1}-\ref{fig:s4}, we see that scattering and fluorescence are indeed important for the spectral formation at eight years. The fraction of the emerging flux that has been processed by line transfer (red) is $\sim$60\% in the UV ($<$4000 \AA), $\sim$30\% in the optical (4000-7500 \AA) and $\sim$30\% in the NIR (7500-21,000 \AA). Running the model with the line transfer switched off resulted in a $\sim$30\% increase in the UV flux, a $\sim$10\% decrease in the optical flux, and a $\sim$30\% decrease in the NIR flux.
%Because at earlier epochs the amount of reprocessing is even higher, we draw the important conclusion that a nebular analysis is \textit{never} particularly suitable for interpreting supernova spectra, if an accuracy better than a factor of a few is desired. 

Few lines are unaffected by the line transfer. Those with the highest fraction of direct emission in the flux are Mg I] \wl4571, H$\beta$, Fe~I] \wl5012, [Fe~I] \wl5697, the [O I]/Fe I complex at $\sim$6300-6400 \AA, H$\alpha$+Ca I \wl6573, He I 1.083 $\mu$m, [Si I] 1.60, 1.64 $\mu$m, He I 2.058 $\mu$m, and a few others in the NIR part of the spectrum.

By inspecting the bottom-most panels in Figs. \ref{fig:s1}-\ref{fig:s4}, we see that a substantial fraction of the emerging flux, especially in the UV, has its origin in the hydrogen envelope. This is mainly freeze-out energy that is emitted as Ly$\alpha$ and two-photon continuum, and is then processed by scattering and fluorescence. Below $\sim$2500 \AA~essentially the whole spectrum originates in the envelope.

An alternative way of illustrating the fluorescence process is to plot the escape wavelengths versus the emitted wavelengths, which we did in Fig. \ref{fig:lambdalambda} for all photons that have been scattered. The transformation of high-energy photons to low-energy ones can clearly be seen here, and it is clear that most of the fluorescent output originates as UV emission ($\lambda_{\rm in} < 4000$ \AA). The low occupation numbers of excited states makes any inverse fluorescence, i.e. conversion to shorter wavelengths, unimportant at this epoch, and the line transfer produces a systematic reddening of the spectrum. The contribution by many input wavelengths to each output wavelength shows that the fluorescence process is insensitive to the input spectrum, as also found by \citet{Li1996}. Some of the more distinct input lines are Ly$\alpha$ and Fe I/Fe II emission lines at $\sim$2500 \AA, $\sim$2900 \AA~and $\sim$3900 \AA. Ly$\alpha$ powers the emission in  Mg II 2795, 2802 (Sect. \ref{sec:lineids}).%This emission is produces by Ly$\alpha$, Fe II ???, and Fe I ??.  %Two-photon emission and iron UV lines escape with wavelengths from UV to the IR.
%Ly$\alpha$ is seen to produce fluorescence in the Mg I and Mg II lines as $\sim$2800 \AA, as well as at other wavelengths. Fe I emission (?) at $\sim$ 3900 \AA~pumps both the [Ca II] \wll 7291,7323 doublet, as well as Fe I at $\sim$ 5300 \AA.
\subsubsection{Line identifications}
\label{sec:lineids}
We now go through the spectrum, identifying and commenting on the most prominent lines. We use the same line peak wavelengths as in C97. 
\begin{figure}[t] 
\centering 
\includegraphics[width=1\linewidth]{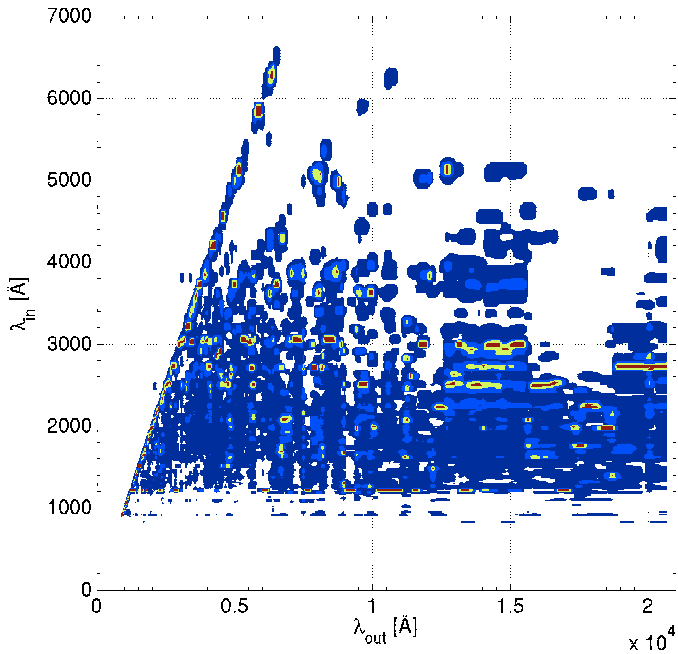} 
\caption{The fluorescence process illustrated by plotting emerging versus emitted wavelengths ($\lambda_{\rm out}$ and $\lambda_{\rm in}$, respectively) for all escaping photons that were absorbed by a line at least once. The intensities are normalized in each $\lambda_{\rm out}$-bin ($\Delta \lambda/\lambda = 0.03$), so that the dominant $\lambda_{\rm in}$ for each $\lambda_{\rm out}$ can be seen as the yellow/red regions. Contours are at 0.1, 1, 10, 50 and 90\%. Very little fluorescence occurs for $\lambda_{\rm  in}> 7000~\AA$, which is therefore excluded from the plot.} 
\label{fig:lambdalambda} 
\end{figure}
\paragraph{UV} As panel (e) in Fig. \ref{fig:s1} illustrates, most of the emerging far-UV flux (below $\sim$ 2400 \AA) has its origin as emission from the envelope. Panel (b) shows that most of this flux has been reprocessed by scattering and fluorescence. 
%We see from panel (c) in Fig. \ref{fig:s1} that several of the emergent lines are from Si I, i.e. 1975 \AA, 2206 \AA~and 2515 \AA.
Longward of $\sim$2400 \AA~the core begins to make significant contributions. The feature at 2400 \AA~is mainly scattering in the Fe II UV 2 multiplet, with similar contributions from the envelope and the iron zone. The feature is blended with an Fe I emission line on the red side. The feature at 2600 \AA~is also a blend of Fe II scattering (UV 1) and Fe I emission. The strong line at 2513 \AA~appears to be scattering in Si I UV 1, complemented by Fe I.

The feature at $\sim$2790 \AA~is scattering by Mg II \wll2795, 2802 in the envelope. We investigated the formation of this line and found it to be dominantly powered by Ly$\alpha$, which is absorbed in the Mg II 1240 \AA~line, which then produces fluorescence at 2795, 2802 \AA,  2798 \AA, 2928, 2936 \AA, 9218 \AA, and 10,914 \AA. %The model has a weak contribution from the core H zone, producing a somewhat flat-topped profile. The reason is that the density here is so high that the two-photon continuum quenches Ly$\alpha$. The core zones could still produce a narrow component by scattering Ly$\alpha$ emitted from the envelope. However, the Fe/He zone has an optically thick Fe I line at 2807 \AA, which, due to the large filling factor of the Fe/He zone, blocks the escape of the scattered photons. %Using a smaller filling factor for the Fe/He zone produced a better fit to the Mg II feature. 
There is some flux missing on the blue side of the Mg II feature, but we find no Fe II contributions that could improve this, as suggested in C97.

Next to the Mg II feature follows emission/scattering in the Mg I \wl2852 line. This feature has a broad component that comes from scattering in the envelope (the line is optically thick out to $\sim$6000 \kms~even though magnesium is $\sim$99\% ionized in the envelope), and a narrow component that is dominated by emission from the O/Ne/Mg zone. On the red side there is some contribution from iron and other elements.

%The model produces a Fe I emission line at $\sim$2986 \AA~, which is not seen in the observed spectrum. In Sect. \ref{sec:eoct} we show that this line is sensitive to the charge transfer. There is a weak observed line at $\sim$2975 \AA, which could be the blue part of this line, if the rest of the line is absorbed by line or dust extinction more strongly than the model predicts.
%, 2821 \AA~(scattering in Mg~II \wll2795,2802 and Mg~I \wl2852, and 2975 \AA~(Fe~I). 

The 3000-3500 \AA~range is not well reproduced by the model, and no definitive identifications can be made. Panels (d) and (e) in Fig. \ref{fig:s1} show that most of the flux here originates in the Fe/He zone. The model produces strong lines at 2986 \AA~(Fe I), 3090 \AA~(Fe I and Al I), 3160 \AA~(Ti II scattering), 3270 \AA~(Fe II), 3310 \AA~(Ni I scattering), 3400 \AA, 3440 \AA~and 3490 \AA~(emission and scattering mainly from iron, cobalt, and nickel). The overproduction of several lines in this region could conceivably be due to the omission of trace elements that could provide additional line absorptions. We surveyed the line lists in the 3000-3500 \AA~range of all omitted elements with $Z<30$, however, and there are no lines from these elements that could be important scatterers at this epoch.
The discrepancy could also be caused by the lack of accurate collision cross-sections for the non-thermal excitations in Fe I (we use the Bethe approximation). Although several of the lines are produced by scattering/fluorescence, the origin is often an Fe I emission line. As we see in Sect. \ref{sec:eoct}, charge transfer can also be responsible by introducing uncertainties in the ionization balance. Finally, it is possible that the dust extinction is wavelength-dependent and should be higher in the UV. This would affect the UV lines coming from the core more than the ones coming from the envelope.%the envelope as much as the lines coming from the core.}

We identify the feature at 3650 \AA~as $\sim$50\% Balmer continuum and $\sim$50\% scattering by mainly Fe I and Cr I in the iron core. The 'contamination' of the Balmer continuum makes the attempts to determine of the hydrogen zone temperature in W96 and C97 unsuitable, since they are based on an assumption of pure Balmer emission. The temperatures derived there are also significantly higher than in the ones calculated in our model ($\sim$100 K). Taking the contaminations of both the Balmer continuum and H$\alpha$ into account (see below), the observed ratio between the Balmer continuum and H$\alpha$ is 0.25, which is close to the theoretical ratio of 0.32 obtained by using the Case B recombination rates at 80 K in \citet{Martin1988}.   

The strong observed line at 3730 \AA~is reproduced by the model and appears to be scattering and emission from Fe~I. We find the [O II] \wl 3727 contribution to be negligible, agreeing with C97 that there is no need for this line, a possibility raised in W96.

\paragraph{Optical} The 4000-4500 \AA~range is a blend of Fe~I, Ca~I, H~I, and others.  The strongest line in this range is at 4223 \AA, which the model identifies as Ca I \wl4226 emission from the Si/S zone. As we discussed in Sect. \ref{sec:sis}, Ca I is likely to emit most of the ionization energy in this zone, a conclusion which is supported by the presence of the 4226 \AA~line in the observed spectrum. This line is sensitive to density; $A\beta$ has a value of $\sim$600 s$^{-1}$ for the density we use here (a number density ten times lower than in the O/Ne/Mg zone), which dominates the rate of $A_{\rm tot}=140$ s$^{-1}$ for other radiative branching paths. But if the Si/S zone would have the same density as the oxygen zones, $A\beta$ would be only $\sim$10 s$^{-1}$ (calcium is also less ionized at higher density and so the scaling is not proportional), and the other deexcitation paths would then dominate. The presence of this line in the spectrum, at this epoch, therefore requires a number density in the Si/S zone at least $\sim$10 times lower than the one of the oxygen zones. 
%This line 'switches on' ($A\beta(4226 \AA)=140$)when the density in the Si/S zone 
%becomes low enough. 
%falls below a density of $\sim 2\e{6}$cm$^{-3}$, and for the line to be as strong as observed at this epoch, the density in the Si/S must be lower than about a third the density in the oxygen zones. 
Because the Si/S zone contains some ($\sim$5\%) of the \iso{56}Ni, according to the explosion model we use, this is consistent with the physical picture that radioactive decays cause expansions in the early phase of the supernova, as was found to have happened for the Fe/He clumps by \citet{Li1993iron}. 
%By inspecting earlier spectra, we find the Ca I \wl4226 line to switch on around early 1992. Assuming that the ionization balance for calcium is about the same at that epoch, this allows us to derive an exact density of the Si/S zone; $2\e{6}$cm$^{-3}$ at $\sim$5 years or $5.1\e{5}$cm$^{-3}$ at the epoch here, which is a factor seven below the number density in the O/Ne/Mg zone.

%A complication is the fact that the 4226 line is on the border of being absorbed by some lines in the 4226-4260 \AA~range; 

%\begin{table}
%\begin{tabular}{|c|c|c|c|c|}
%\hline
%2 & VI 4234 (1-86) &  567 km/s  & 5.4 \\
%  & Cr I 4254 (1-19) & 1916 km/s &1E4\\
%\hline
%1 & Cr I 4254 (1-19) &   & 3.2\\
%\hline
%7 & Cr I 4254 (1-19) &  & 0.6\\
%\hline
%\end{tabular}
%\end{table}

The rest of the 4200-4500 \AA~ 'plateau' is mainly emission and scattering from the Fe/He zone, complemented with H$\gamma$ at 4343 \AA. The Fe II contribution, which is suggested to dominate the features at 4223 \AA, 4339 \AA, and 4453 \AA~in C97, is only at the $\sim$10\% level in our model. %But since our model somewhat under-produces these features we cannot rule out a more dominant role by Fe II either. 

Mg~I] \wl4571 is one of the lines with the smallest contribution from scattering/fluorescence. The model underproduces the flux by a factor of $\sim$2-3, possibly as a result of the uncertain ionization balance in the O/Ne/Mg zone (Sect. \ref{sec:onemg}). Although most of the synthesized magnesium resides in this zone, its contribution to the Mg I] \wl4571 line is moderate because Mg II is largely neutralized by charge transfer with nickel and sodium, leaving only non-thermal excitations to power the line.  It is possible that we use rates that are too high for some of these charge transfer reactions, and that Mg II instead recombines radiatively, in which case the Mg~I] \wl4571 line would become stronger. Our model gives emission of similar magnitudes from the O/Ne/Mg zone, the O/C zone, and the He and H zones. 

The feature at 4850 \AA~is $\sim$2/3 H$\beta$ and $\sim$1/3 Fe~I in the model. Removing the Fe I contamination gives a Balmer decrement of $\sim$5, which agrees well with a Case B recombination scenario at $\sim$100 K.

The 5000-5700 \AA~range is a quasi-continuum dominated by Fe I lines from the iron core, with about 3/4 of the flux being unscattered. The range is well reproduced by the model, apart from an overproduction of the Fe I line at 5007 \AA~(the narrow observed line does not come from the ejecta, but from the northern loop).

The line at 5900 \AA~is emission and scattering in Na~I 5896 \AA. The scattered part is predominantly from emission in Na I 5890 \AA, but also from He I 5876 \AA~emitted in the Fe/He zone. Panel (e) in Fig. \ref{fig:s2} shows that most of the energy originates in the O/Ne/Mg zone, which means that the helium contribution is minor. Na I emits strongly from the O/Ne/Mg zone because of a series of charge transfer reactions and a high photoionization rate that make it the dominant ion in this zone (Sect. \ref{sec:onemg}). Apart from the 5890, 5896 \AA~doublet, the Na I recombination spectrum also includes lines at 5690 \AA, 6154~\AA, 8190 \AA~and 1.14 $\mu$m. There are indeed observed lines at $\sim$5690 and $\sim$6154 \AA, but blending with Fe~I and Ca~I makes identifications uncertain. At 8190 \AA~the spectrum is too noisy to clearly detect a line, and in addition there is also contamination from other lines in the model. There is a clear detection at 1.14 $\mu$m in the day 2112 spectrum in \citet{Fassia2002} (not shown here), but all in all it is difficult to tell whether a significant Na I recombination spectrum is emitted by the ejecta or not. The model overproduces the Na I \wll5890, 5896 doublet, and it is possible that the ionizations instead go to magnesium in the O/Ne/Mg zone, especially because the Mg I] \wl4571 line is underproduced. However, in Sect. \ref{sec:eoct} we show that the Na I doublet is overproduced also if the charge transfer reactions are switched off.
%The 6154 \AA~line is mainly a Ca~I recombination. 

[O I] \wll6300, 6364 is contaminated, which the deviation of the observed ratio from the expected 3:1 value verifies. In the model, most of the contaminating flux is from Fe I. The complex is well reproduced by the model, although the good fit relies on the assumed high rate for the O II + C I $\rightarrow$ O I(2p$^1$D) + C II charge transfer reaction, as discussed in Sect. \ref{sec:onemg}. If this reaction is not the dominant channel for O II recombinations, the [O I] \wll 6300, 6364 emission would decrease by a factor of $\sim$4. The uncertainties regarding this reaction combined with the Fe I contamination make the oxygen doublet unsuitable for drawing any strong conclusions at this epoch, including attempts to determine the oxygen mass.
%As described in Sect. \ref{sec:onemg}, we get a very low contribution from recombinations to the the oxygen emission due to the charge transfer neutralization of O~II. We show in Sect. \ref{sec:eoct} the 6300, 6364 \AA~region with charge transfer switched off. The feature is then much better reproduced. 

Interestingly, H$\alpha$ is significantly contaminated by Ca~I 6572 \AA~in the model. This line originates, like the Ca I 4226 \AA~line, mainly in the Si/S zone, but also in the O/Si/S zone. Being the first transition from the ground state in Ca I, the line has a high effective recombination rate and is about twice as strong as the H$\alpha$ emission from the core H zone. Note that this component is needed to satisfactorily reproduce the 'H$\alpha$' profile, which would otherwise be too flat-topped. For the high-velocity part of H$\alpha$, the good fit of the model is an important verification that our treatment of the freeze-out in the envelope is accurate. %The hydrogen emission is visible out to $\sim$6000 \kms.  

%The 7148 \AA~feature is a mix between [Fe~II] \wl7155 and Ca~I \wl7156. 

The [Ca~II] \wll 7291,7323 doublet is well reproduced and results from pumping in the H and K lines. Fig. \ref{fig:s2}d-e show that the fluorescence is produced by both synthesized and primordial calcium. The H, K lines are optically thick in all core zones, and in the envelope out to $\sim$7000 \kms. Fig. \ref{fig:lambdalambda} shows that most of the photons start as emission close to the H and K lines. The flux in the Ca II \wll 8600 IR triplet is similar to the flux in the doublet, as it must be for a H, K pumping scheme.

\paragraph{Infrared} In the infrared, some of the observed lines are dominated by narrow circumstellar components seen at low resolution. We identify the 1.083 $\mu$m line, observed in the day 2112 spectrum \citep{Fassia2002}, with He I emission from the Fe/He zone. %The narrow CSM component dominates the observed line. 
The broad component has an observed flux (Table 2 in \citet{Fassia2002}) similar to the model flux. Pa$\gamma$ \wl 1.0938 $\mu$m also has a strong narrow component. The model shows the ejecta emission here to be dominated by [Si I] 1.099 $\mu$m. The model flux is 1.5 times the the observed broad component at 2112 days.

Further, we identify 1.14 $\mu$m with Na I, 1.20 $\mu$m with Si I, 1.25 $\mu$m with [Fe II], and 1.28 $\mu$m with Pa$\beta$. %, although the model fluxes in all these lines are on the low side. 
The strong 1.44 $\mu$m line is [Fe I] emission from the core, with a model flux 1.8 times the observed value. Also 1.50 $\mu$m and 1.53 $\mu$m are Fe I emission. 1.60 and 1.64 $\mu$m are dominantly [Si~I] emission from the Si/S zone, and 2.058 $\mu$m is He I from the iron core. The 2.058 $\mu$m line is overproduced in the model. We also refer to \citet{Kjaer2010}, where we discuss the IR spectrum at an age of 19 years, with most line identification being the same as at this epoch.
%As we show in Sect. \ref{sec:pos}, this zone gets about the same deposition independent of the assumption of positron leakage, so the presence of these lines does not constrain that process. 

%\textbf{If the day 2112 IJ-band spectrum in \citet{Fassia2002} is assumed to be representable at day 2875, the model has a quasi-continuum in this range $\sim$10 times lower than the observed level. This is similar to the situation at 19 years \citep{Kjaer2010}. We have found no definite explanation for this, although incorrect removal of contamination from Star 3 and Star 2, or scattered light from the ring are possible alternatives.}%, as discussed in \citet{Kjaer2010}. %We include all recombination continua, the tail of the H and He two-photon emission, and thermal Bremsstrahlung (which is negligible at these low temperatures). We do not include dust emission, but the dust temperature is low enough that the contribution at NIR wavelengths should be negligible.
 
\paragraph{Summary of the spectrum}
In general, we find good agreement with the line identifications in W96 and C97. The most important difference is that we find many lines to be from Fe~I rather than Fe~II. This is what to expect since Fe~I emits not only a recombination line spectrum, but also receives more non-thermal excitations because Fe I atoms outnumber Fe~II ions by a factor of $\sim$1.5 (see Sect. \ref{sec:fehe}). In our model, $\sim$30\% of the flux between 2000 and 8000 \AA~is from Fe~I, and only $\sim$9\% is from Fe~II (including both emission and scattering). We do not, however, include ionizations to excited states in Fe II, which causes some underestimate of the Fe II emission.

%\begin{table*}
%\caption{The total luminosity between 2000 and 8000 \AA~for different combinations of the $^{44}$Ti-mass and the dust extinction  $\tau_{\rm d}$.}
%\centering
%\begin{tabular}{l l | l l l l} 
%\hline\hline
%M($^{44}$Ti) & $\tau_{\rm d}$ & $L_{\rm model}(2000-8000 \AA)$ & $L_{\rm model}(2000-8000 \AA)/L_{\rm obs}(2000-8000 \AA)$\\ 
%($M_\odot$) &  &  [$10^{35}$ \ergs] &\\
%\hline
%0          &  1  & 1.00 &\\
%$0.5\e{-4}$ &  0 & 3.00 & 0.98 \\%& A nice spectrum, but some lines clearly lack blue-shift (3730 ex)\\ 
%             &  0.5 & 2.30 & 0.75 \\
%             &  1 & 1.92 & 0.63\\
%             &  1.5 & 1.69 & 0.55 \\
%\hline 
%$1\e{-4}$ &  0 & 4.50 & 1.47 \\%& UV too weak, rest too strong\\
%     & 0.5 &  3.33 & 1.09 \\%& UV excellent,3000-3500 too stron, 5000-6000 a bit too strong\\
%   %  & 0.75 & 2.3 & 2.8 & 1.01 \\%& UV exc, rest a bit too stront..too stron Fe emission. \\   
%     & 1    & 2.72 & 0.89    \\%&\\
%     & 1.5  & 2.25 & 0.74\\
%\hline
%$1.5\e{-4}$ & 0 & 6.12 & 2.00\\
%           &  0.5 & 4.43 & 1.45 \\
%          & 1 & 3.47 & 1.13\\
%          & 1.5 & 2.90 & 0.95\\
%\hline
%$2\e{-4}$  & 0 & 7.50 & 2.45\\
%      & 0.5  & 5.35 & 1.75\\
%   %  & 0.75 & 3.8 & 4.2 & 1.53 & \\
%      & 1  & 4.16 & 1.36 \\
%     & 1.5  & 3.44 & 1.10 &\\
%  %   & 2 & 5.5 & 2.6 & 0.94 \\%& Quite ok spectrum, definately acceptable. 3000-3500 still bad.\\
%\hline
%\end{tabular}

%\label{table:variousTi}
%\end{table*}

\subsection{The \iso{44}Ti mass}
\label{sec:ti-mass}

The character of the spectrum, i.e. the line ratios and the ratio of lines
to continuum, is not very sensitive to the absolute level of the radioactive powering. The character is
partly determined by the relative amounts of energy that go into
excitations and ionizations, which is insensitive to the total
deposition. It is also influenced by the ionization balance of
the dominant elements in each zone, which is likewise
insensitive to the total deposition; the ionization fraction of
the dominant element approximately scales as the square root of the
deposition. Furthermore, there is a negative feedback present because a higher electron fraction reduces the fraction that goes
into ionizations (and excitations) at the expense of the heating
fraction. 

\begin{figure}[!h]
\centering
\includegraphics[width=1\linewidth]{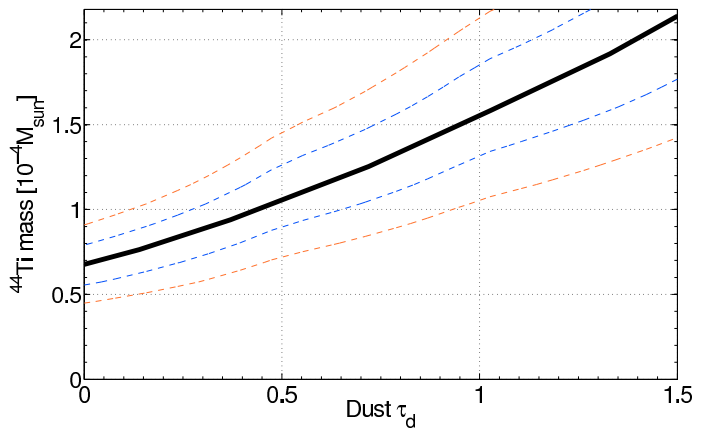}  % switched axes
\caption{Relationship between the $^{44}$Ti-mass and the dust optical depth $\tau_{\rm d}$ required to match the total luminosity in the 2000-8000 \AA~range (solid black line). Also plotted are the 10\% and 20\% contours of $|\left(L_{\rm model}-L_{\rm obs}\right)|/L_{\rm obs}$ (dashed lines).} 
\label{fig:ti44contours}
\end{figure}

The $^{44}$Ti mass therefore mainly determines the absolute
flux level of the spectrum, rather than its form. 
Unfortunately, it is difficult to distinguish the $^{44}$Ti mass from
the uncertain dust properties.  We
therefore calculated the integrated flux in the 2000-8000 \AA~range for
different combinations of M($^{44}$Ti) and the dust optical depth $\tau_{\rm d}$ (Sect. \ref{sec:dust1}). We did not
include the NIR because of the uncertainties in the calibration of
these observations (Sect. \ref{sec:TOS}). Fig. \ref{fig:ti44contours} shows the $^{44}$Ti mass that reproduces the observed flux in the 2000-8000 \AA~range as function of $\tau_{\rm d}$, together with the masses that produce the correct flux plus/minus 10\% and 20\%. 
%that reproduces the 2000-8000 \AA~luminosity.
Note that the emerging flux is less than proportional to the \iso{44}Ti mass. %\textbf{(for a fixed dust extinction, the change in \iso{44}Ti mass corresponding to a flux change of X\% is more than X\%)}. 
The main reason is that a 
significant fraction of the flux comes from freeze-out emission rather than from instantaneous
\iso{44}Ti deposition. Another reason is that a higher deposition increases
the electron fraction, which increases the fraction of the flux emerging in the FIR.
  
A $^{44}$Ti mass of \bestti~is our best match for $\tau_{\rm d}=1$, which is
our best estimate for the dust extinction from the line profiles. This result is only valid
for the assumption of local positron deposition. The dust
optical depth has a $\pm$30\% uncertainty (Sect. \ref{sec:dust1}), which according to Fig. \ref{fig:ti44contours} gives a $\pm$20\% uncertainty for the best fitting \iso{44}Ti mass. %\textbf{(the best mass is 1.1 \msun~(79\% of 1.4) \msun~for $\tau_{\rm d}=0.7$ and 1.7 \msun~(121\% of 1.4) \msun~for $\tau_{\rm d}=1.3$).} 
In Sect. \ref{sec:fehe} we noted
that the uncertainty in the density of the Fe/He zone adds an uncertainty of $\pm$10\% for the model flux, and in Sect. \ref{sec:TOS} we saw that the observed flux level has a $\pm$15\% uncertainty from errors in distance, reddening, and flux calibration. Together, these give a $(15\%^2+10\%^2)^{1/2}=18\%$ uncertainty for the ratio between modeled and  observed flux. According to the contours in Fig. \ref{fig:ti44contours}, this corresponds to a $\pm$30\% uncertainty in the \iso{44}Ti mass. Together with the dust error, the total uncertainty for the \iso{44}Ti-mass 
becomes $\sim (20\%^2+30\%^2)^{1/2}=36\%$. 
%For the reasons mentioned above, the luminosity is unfortunately quite weakly dependent on
%the \iso{44}Ti-mass, and the uncertainties in the dust content combined with the uncertainty
%in the true luminosity 
%Assuming that $\tau_{\rm d}$ is bracketed by the 0.7-1.2 range, 
The $^{44}$Ti mass is then constrained to $(1.0-2.0)\cdot 10^{-4}$~\msun.  
%atthe $\pm$15\% level, which is roughly the uncertainty in the observed luminosity (see Sect. \ref{sec:TOS}). 
%We note, however, that the density in the Fe/He zone could be lower (Sect. \ref{sec:fehe}), which could increase the \iso{44}Ti-mass by another $\sim$10\%. Our final result for the \iso{44}Ti mass is therefore \besttiwlim.

%in Fig. \ref{fig:dustemiss} we plot the total amount of energy absorbed (and re-emitted?) by the dust. For $\tau_d=1$ and M($^{44}$Ti)=$1.3\e{-4}~M_\odot$ the dust emission is $2.2\e{35}$ \ergs. 

%\begin{figure}[!h]
%\includegraphics[width=1\linewidth,height=6cm]{./ti44contours_dustflux.eps}
%\caption{Dust emission in the the M($^{44}$Ti) and $\tau_d$ parameter plane. Contours are in 10\% intervals, the maximum flux (top right corner) is $5.5\e{35}$\ergs.} 
%\label{fig:dustemiss}
%\end{figure}

%In Fig. XX we show spectra for three values of the ${}^{44}$Ti mass,
%$(0.5, 1.0, 2.0) \times 10^{-4} \Mo$. We see that.....

%A higher 44Ti mass leads to a higher degree of ionization. The change seems to be much less than linear though? The Fe/He zones goes from 15 to 20\%.  The heating fraction is increased from 54 to 59\% in the Fe/He zone. Otherwise, the different energy fractions are essentially unchanged. This means that we can expect the spectrum to scale rather linearly with the 44Ti mass in the spectral regions dominated by core emission.

\subsection{Positron trapping}
\label{sec:pos}

An described in Sect. \ref{sec:edep}, an interesting and important question is whether the positrons are indeed locally absorbed, as assumed so far, or if they leak into the other zones in the core before they are absorbed. We recomputed the spectrum for a model where we assume that the positrons are not absorbed on the spot, but in proportion to the electron content (free and bound) of each zone. This treatment is based on the Bethe stopping formula (Eq. (\ref{eq:bethe})), where we neglect the small differences due to different effective ionization potentials. The number of electrons per zone is proportional to $M\bar{Z}/\bar{A}$, where $M$ is the zone mass and $\bar{Z}$ and $\bar{A}$ are the average nuclear charge and atomic weight, respectively. 
%\begin{equation}
%  N_e = N_{\rm tot} \sum x_i Z_i = \frac{M}{\bar{A}} \sum x_i Z_i = M\frac{\bar{Z}}{\bar{A}}
%\end{equation} 

For this alternative model, the \iso{44}Ti mass needed to give the correct flux in the 2000-8000 \AA~range for $\tau_{\rm d}=1$ is only $8\e{-5}~$\msun, which we therefore used. The reason for the lower \iso{44}Ti mass is that a smaller fraction of the positron energy is now emitted by far-IR cooling lines (mainly [Fe II] 26 $\mu$m) by the low-density Fe/He zone, and that the oxygen and hydrogen zones have some strong lines in the 2000-8000 \AA~range. Consequently, less \iso{44}Ti is needed to get the same UV/optical/NIR luminosity. 

In this model, the zones obtain total energy depositions (positrons + gamma-rays) given in Table \ref{table:leakage}, where we also give the relative change in the deposition fraction with respect to the on-the-spot model. From this, we expect positron leakage to lead to a severe weakening of emission lines from the Fe/He zone, and a boost by up to a factor of $\sim$10 for emission lines from the O/Ne/Mg, O/C, core He, and core H zones. The $^{44}$Ti content of the silicon zones (Si/S and O/Si/S) however, is such that the deposition here does not change much between the two scenarios.

\begin{table}[h!]
\caption{Fractional energy deposition per core zone in the positron-leakage model (which roughly corresponds to the fraction of the electrons that reside in each zone), and the change in fractional deposition relative to the on-the-spot model. About 5\% of the total deposition occurs outside the core.}
\centering
\begin{tabular}{l | l l}
\hline\hline
Zone & Fractional deposition & Relative change \\
\hline
Fe/He   & 1.3\% & 0.017 \\ 
Si/S    & 2.1\% & 0.43 \\
O/Si/S  & 2.4\% & 1.2\\
O/Ne/Mg & 28\%  & 10\\
O/C     & 8.7\% & 10\\
He(core)& 4.9\% & 10\\
H(core) & 48\%  & 11\\
\hline
\end{tabular}
%\tablefoot{
%\tablefoottext{a}{Core component.}
%}
\label{table:leakage}
\end{table}

\begin{figure}[h!]
\centering 
\includegraphics[width=1\linewidth]{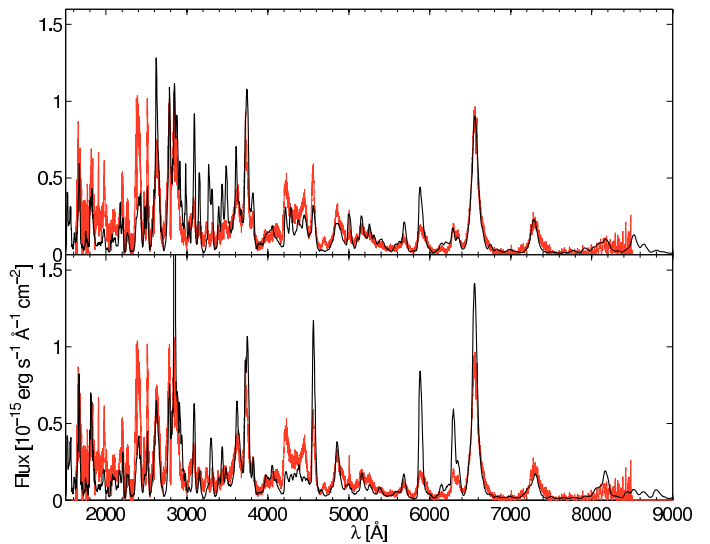} 
\caption{Standard model (assuming on-the-spot positron absorption) (upper, black) compared to a model assuming positron deposition in proportion to the electron content of each core zone (lower, black). The latter uses a 40\% smaller $^{44}$Ti mass (in order to produce the correct total flux in the 2000-8000 \AA~range). The Mg I 2852 \AA~line continues up to $\sim 3.1\e{-15}$ in the bottom panel. The observed spectrum in red, same as in Fig. \ref{fig:s1}.} 
\label{fig:positrons} 
\end{figure}

Fig. \ref{fig:positrons} shows the resulting model spectrum compared to the standard on-the-spot model (which is the same as in Figs. \ref{fig:s1}-\ref{fig:s4}). On the whole, the spectra are quite similar, but contain some important differences as well.

The hydrogen emission from the core is boosted by a factor of several, as expected. H$\alpha$ and the Balmer continuum are now overproduced. However, because the fraction of the hydrogen that we mix into the core (25\%) is not tightly constrained, there is some room to improve the H lines by reducing this in-mixing. That would, however, increase the deposition in the other core zones and strengthen the lines from them.

The [O I] \wll 6300, 6364 doublet becomes much too strong, but this result is sensitive to the uncertain O II + C I $\rightarrow$ O I (2p$^1$D) + C II charge transfer rate used, as discussed in Sect. \ref{sec:onemg}. If this reaction is switched off, the agreement is actually better in this model than in the on-the-spot model. The Mg~I] \wl 4571 line is mainly emitted from the oxygen zones, and becomes about as much overproduced as it was underproduced before. The Mg I \wl 2852 line, however, now becomes much too strong (it is truncated in the figure). This line is powered by non-thermal excitations, and is therefore independent of the uncertain ionization balance in the O/Ne/Mg zone (magnesium is dominantly neutral in all scenarios). This line should therefore be a more reliable indicator than the recombination-driven 4571 \AA~line, and all this suggests that too much energy is now being deposited in the oxygen zones. The Na~I recombination lines at %3302 \AA, 5682 \AA, 
5890, 5896 \AA, and 6160 \AA~from the O/Ne/Mg zone become overproduced as well. The 5890, 5896 \AA~doublet, which was already too strong without leakage, is now a factor $\sim$4 too strong. 

%For the sum of the strongest lines from the oxygen zones ([O I] \wll 6300, 6364, Na I 5896, Mg I] \wl 4571 and Mg I \wl 2852, the ratio of the model flux to the observed flux is 0.84 in the on-the-spot model and 2.0 in the $MZ/A$-model. From the perspective of obtaining the right total flux in these three lines, a moderate positron leakage of order $\sim$ 10\% is then suggested. 

%But if we look only at the oxygen and magnesium lines, the scenario with full leakage is the preferred one. The failure of the model to reproduce the correct relative line strengths makes us cautious in drawing any firm conclusions. Of the chemical abundances in the O/Ne/Mg zone, probably sodium is the most uncertain. 
%are 5.4E33+4.2E33+4.8E32 = 1.0E34 which is a factor 0.67 of the observed. 
%The conclusion from this analysis is that the optimum amount of positron leakage for the oxygen-zone lines is a moderate leakage of order 10\% (of which $\sim$ 3\% goes into the oxygen zones). A 100\% leakage gives too much emission in these lines. This conclusion is in agreement with the analysis made in C97.  
%$1.6\e{34}$, $2.0\e{34}$ and $4.4\e{34}$\ergs versus the observed $5.5\e{33}$, $5.6\e{33}$ and $4.0\e{33}$\ergs. The total is 8E34 in the model and 1.5E34 observed; a factor 5.3. We should be remember that the model with leakage is 30\% offset though and so a fair comparison number is $5.3\cdot 0.7=3.7$.

For the rest of the spectrum, the two models produce surprisingly similar spectra.  Because fluorescence makes important contributions at almost all wavelengths, turning off the iron emission lines from the Fe/He zone typically lowers the line fluxes by no more than a factor of about two. An example of this situation is the 5000-5500 \AA~range. The iron lines with the smallest contamination by fluorescence in the UV/optical are the lines at 3270 \AA, 3490 \AA, 5012 \AA, and 5159 \AA, and 5697 \AA. All of these lines seem somewhat better reproduced in the leakage model.%, but there could be some selection effects, as the most over-produced (for other reasons) emission lines are the ones that stand out the most.

Because the flux calibration in the NIR is uncertain (Sect. \ref{sec:TOS}), we refrain from comparing the models in this range. All in all, we think the on-the-spot model produces a spectrum that better agrees with the observations, although the uncertainties mentioned above prohibit a definitive ruling out of the leakage scenario. The average $\chi^2$-value (per spectral bin) is about twice as high in the leakage model as in the on-the-spot model.

\subsection{Fragmentation level}
\label{sec:fragmentation}
The fragmentation level, $N_{\rm cl}$, is a proxy for the hydrodynamical mixing. This
parameter determines the length of the paths that the photons travel in their parent zone before they have the chance to enter the other core zones. A high fragmentation level, implying quick exposure to other zones, will lead to a higher degree of radiative energy exchange between the zones. If one were able to constrain this parameter it would be very useful for evaluating multi-dimensional explosion models.

To check the sensitivity of the spectrum to this parameter, we compared spectra for models using $N_{\rm cl}=10$ and $N_{\rm cl}=10^5$. The results were very similar, and we therefore conclude that the fragmentation parameter has no significant influence on the spectrum at this epoch. One possible explanation for this is that few line optical depths can be expected to be close enough to unity that an abundance change (by switching zones) between a high value in the zones where the element is synthesized to the (low) primordial values where it is not, will make any difference. Another could be that few of the line absorptions occur before the photon has exited its parent zone, even in the low fragmentation limit. Finally, a significant part of the line transfer occurs in the envelope, where there is no mixing in the model.
%The clump radii for the iron zone are 560 \kms for $N=10$ and 26 \kms for $N=10^5$. For the core H zone, the corresponding numbers are similar. At 3000 \AA, 560 \kms corresponds to 5.6 \AA. This is presumably too small a wavelength range to have any significant impact by changing the line list from the parent zone to the neighboring zones.
\subsection{Effects of charge transfer}
\label{sec:eoct}
\begin{figure}[h!]
\centering
\includegraphics[width=1\linewidth]{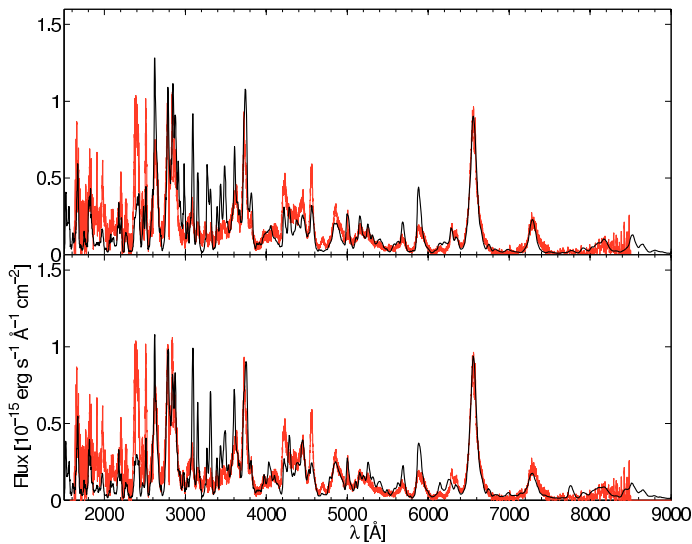}
\caption{The standard model (upper, black) compared to a model with all charge transfer reactions switched off (lower, black). Observed spectrum in red.}
\label{fig:wwoutct}
\end{figure}

Finally, the uncertainty in many of the charge transfer rates prompted us to examine the influence of these reactions on the spectrum. It is also interesting to see how important these reactions are for supernova spectra in general. We recomputed our standard model ($\mbox{M(\iso{44}Ti)}=\mbox{\bestti}$, $\tau_{\rm d}=1$, and on-the-spot positron deposition) with all charge transfer reactions switched off. The resulting spectrum is compared to the standard one in Fig. \ref{fig:wwoutct}. 

The total flux in the 2000-8000 \AA~range changes by only a few percent, so charge transfer should not affect our conclusions regarding the \iso{44}Ti mass. Several lines differ significantly between the two models though. The [O I] \wll 6300, 6364 doublet is weakened for reasons discussed in Sect. \ref{sec:onemg}, and the Fe I lines now dominate the doublet. At the same time, the O I \wl7775~recombination line, which is not observed, emerges in the model. This suggests that the oxygen ions are indeed neutralized by charge transfer reactions, as in our standard model.

The important Ca I lines (4226 \AA~and 6572 \AA) do not decrease by much, which shows that charge transfer is not critical for them. Ca I is significantly ionized by the internal radiation field, and so emits strong recombination lines even without charge transfer. 

%We judged the rates from \citet[][and references therein]{Arnaud1985,Swartz1994,Kingdon1996,Stancil1998,Zhao2004} to be reliable and were kept on.
%Fig. \ref{fig:wandwoutct} shows the impact on the emergent spectrum by comparing the models with and without charge transfer. 
%The differences between the two models were small, and we refrain from showing them. The main difference was in the 6300-6364 range, which increased by $\sim$30\% to $\sim$75\% of the observed value. This was \emph{not} mainly due to the oxygen lines becoming stronger (the 2p($^1$D) state gets only $\sim$10\% of the recombinations), but due to various other lines strengthening. 
The Na I \wl5896 emission is reduced somewhat, but is still too strong. Just as with Ca I, Na I is significantly ionized by the internal radiation field, and consequently strong recombination emission still occurs. An important conclusion is therefore that the emission lines from elements with low ionization thresholds are is not very sensitive to the charge transfer, because low ionization thresholds also correlate with high radiative ionization rates.

For many lines, it is not possible to directly say why they change. The 'picket fence' of absorption lines changes as the ionization balance changes, which opens some wavelength windows and closes others. Consider, for example, the Fe I emission line at $\sim$2990 \AA. This emission line is still there in the model without charge transfer, but now a Ni I absorption line has become optically thick that absorbs it. The UV spectrum is actually somewhat better reproduced in the model without charge transfer, which may indicate that we overestimate some of our adopted rates. Unknown charge transfer rates clearly constitute one of the main obstacles to accurate supernova spectral modeling today, and more calculations and measurements of these would be highly desirable.

\section{Discussion}
\label{sec:discussion}

Although most of the optical and infrared emission escapes freely in the nebular phase of supernovae, the UV emission does not. The freeze-out energy in the envelope, as well as non-thermal excitations and ionizations in the core, produce strong UV emission that is significantly absorbed by atomic lines and emerges by fluorescence at longer wavelengths. By this process, the emerging flux in the optical and NIR is enhanced by $\sim$10\% and $\sim$30\% compared to a purely nebular spectrum at the epoch we have looked at here. While the non-local line transfer can have a large impact on individual lines at late times, the fact that it has a moderate impact on the supernova photometry is important for judging the accuracy of models without this effect, which has previously been difficult.

While the fraction of UV photons that are absorbed decreases with time, the fraction of the optical/NIR spectrum that comes from fluorescence may actually increase \citep{Li1996}. The reason for this is that with time the cooling lines move from the optical/NIR into the FIR, and non-thermal processes do not produce very strong optical/NIR emission. Our finding that fluorescence is important also at eight years illustrates this point well.

We showed that the only FUV flux that escapes comes from far out in the hydrogen envelope. A consequence of this is that the ejecta appear larger in the UV than in the optical \citep{Jakobsen1994, Wang1996}. At longer UV wavelengths both the envelope and the core contribute to the spectrum. 

From a mathematical perspective, it is not a very well-conditioned problem to attempt exact modeling of the UV lines, because the output depends sensitively on the exact shape of the 'picket fence' of absorption lines, which in turn depends on many uncertain parameters such as abundances, densities, and charge transfer rates. Considering this complexity, it is probably not realistic or meaningful to attempt any modeling of the UV lines in much more detail than we have done here.

%During the explosion, chaotic mixing and fragmentation likely disrupt the ordered onion-like structure of the star. 
As discussed in Sect. \ref{sec:edep}, the positrons emitted in the $^{44}$Ti decay will enter the neighboring oxygen, helium, and hydrogen zones at late times unless a magnetic field exists to trap them. This would have significant consequences for the FIR cooling lines. The model flux in the [Fe II] 26 $\mu$m line, discussed in Sect. \ref{sec:fehe}, corresponds to a peak flux of $\sim$1.8 Jy at day 2875. ISO observations at day 3425 failed to detect this line, giving a upper limit for the peak flux of 0.64 Jy \citep{Lundqvist2001}. Spitzer observations at day 6190 \citep{Bouchet2006} show a distinct broad feature (FWHM $\sim$ 3800 \kms) at 26 $\mu$m, but with a peak flux of only $\sim$0.02 Jy. The slow decay of $^{44}$Ti should make the situation at 6190 days similar to the one at 2875 days, and it is therefore surprising that the observed flux in this line is about two orders of magnitude weaker than expected. Possible explanations are that the positrons are indeed leaking out into the other core zones, or that dust is cooling the Fe/He zone. In the positron leakage model, the flux in the 26 $\mu$m line is $\sim$0.16 Jy at 2875 days (for a \iso{44}Ti mass of $8\e{-5}$ \msun~and no dust correction). As we found in Sect. \ref{sec:pos}, a positron leakage scenario produces a UV/optical spectrum with several lines that are clearly too strong. Our favored explanation for the weak [Fe II] 26 $\mu$m line is therefore that dust cooling occurs in the Fe/He zone. The possibilities of having dust cooling in the ejecta was discussed in KF98 a and in \citet{Lundqvist2001}.

In the case of magnetic confinement of the positrons, all helium lines are dominated by helium in the Fe/He zone, which is produced by freeze-out in the explosion. Although we do not identify any distinct observed lines with helium in the eight-year spectrum, the 19-year spectrum in \citep{Kjaer2010} contains a strong and distinct 2.058 $\mu$m line, which, as we speculated there, probably shows that $\alpha$-rich freeze-out has indeed happened in the Fe/He zone; without it the local positron deposition could not produce such a strong line.

Together with the \iso{56}Ni mass of $0.069\ \Mo$ \citep{Bouchet1991a} and the \iso{57}Ni mass of $\sim 3.3\e{-3} \Mo$
\citep{Fransson1993,Varani1990,Kurfess1992}, the determination of the \iso{44}Ti mass should put considerable constraints on
supernova explosion models. Our result of \besttiwlim~ agrees with the result of $(1-2)\e{-4}$~\msun~found in C97,
although that result was based on an analysis of [Fe II] lines, whose identifications are different in our model on
several instances, and on the [O I] \wll 6300, 6364 doublet, whose emission is highly sensitive to uncertain charge transfer reactions, and which is also contaminated by Fe I lines. It is therefore a bit of a coincidence that we get similar results for the \iso{44}Ti mass. \citet{Fransson2002} determine the \iso{44}Ti mass to $(0.5-2)\e{-4}$~\msun~by comparing the observed B and V band HST photometry between 1800 and 3500 days to the output from a time-dependent model without non-local radiative transfer (as in KF98 a, b). We showed here that this transfer, although it is important for individual lines, only boosts the output in the optical/NIR range by 10-30\% at 2875 days (Sect. \ref{sec:saf}), which implies that the results in FK02 should have a corresponding accuracy. An inspection of Fig. 4 in the FK02 paper shows that the best-fitting \iso{44}Ti mass is in the $(1.5-2)\e{-4}$~\msun~range, which agrees well with our results here; ignoring non-local radiative transfer should give a somewhat over-estimated \iso{44}Ti mass, but not by much. Our result also agrees with the range of $(0.8-2.3)\e{-4}$~\msun~found by \citet{Motizuki2004}. Their result is, however, is based on a comparison betwen the deposited energy and the bolometric luminosity at 3600 days, the latter being highly uncertain since most of the flux comes out in the unobservable FIR.

The \iso{44}Ti isotope is a result of the
$\alpha$-rich freeze-out \citep[e.g.][]{Woosley1973}, which is a  consequence of
the slow rate for the triple-alpha reaction, which cannot keep the
\iso{4}He abundance at the NSE value as the temperature and density
decreases. The high \iso{4}He abundance in turn results in a higher
abundance of \iso{44}Ti than is the case in NSE. The
\iso{44}Ti/\iso{56}Ni ratio therefore sensitively depends on the
entropy \citep[e.g.][]{Woosley1991}. High temperature and/or low
density increases this ratio. 
%A consequence of the $\alpha$-rich
%freeze-out is also a high He abundance in the Fe core. From a study of
%the Type II models by \cite{WW95} \cite{Timmes1996} find \iso{44}Ti
%masses of $(2-8) \times 10^{-5}$ \msun \ in their spherically symmetric
%parametrized explosions.  
The explosion model we use (from WH07) has a mass ratio
\iso{44}Ti/\iso{56}Ni = $3.1\e{-4}$, which gives a \iso{44}Ti
mass of only $2.1\e{-5}$ \msun~(for M(\iso{56}Ni) = 0.069 \msun), which is clearly ruled out according to our modeling. For a spherically symmetric explosion of a 20 \msun~star, WW95 find an ejected mass of $(1-2)\e{-5}$~\msun. On the other hand, \citet{Thielemann1996} find a much larger mass of $1.7\e{-4}$~\msun, closer to the value we derive here. The differences are partly caused by the different nuclear reaction rates used, but mainly because WW95 initiate the explosion by injecting momentum (piston-driven), while \citet{Thielemann1996} deposit thermal energy \citep{Timmes1996}. The solar abundance of the decay product \iso{44}Ca requires a \iso{44}Ti-production higher than in WW95 by a factor $\sim$ 3, averaged over all progenitor masses, suggesting that the yields in those models are indeed low \citep{Timmes1996}. 
%The solar \iso{44}Ca/\iso{56}Fe ratio is $1.6
%\times 10^{-3}$ \citep{Asplund2009}, which is fairly close to our \iso{44}Ti/\iso{56}Fe ratio. 

\citet{Nagataki1998}, \citet{Nagataki2000} and \citet{Maeda2003} have investigated
the effects of  asymmetries in the explosion. These (simplified) models have
considerably higher temperatures and entropies in the polar
direction. Consequently, the \iso{44}Ti production occurs mainly in
this direction, and the total \iso{44}Ti mass can be considerably
higher than in a spherically symmetric model. 

Further, the rates of several of the nuclear reactions involved in the
\iso{44}Ti production are uncertain. \cite{Nassar2006} find an
increased rate for the important
\iso{40}Ca$(\alpha,\gamma)$\iso{44}Ti reaction, resulting in a factor of two
higher \iso{44}Ti abundance. The theoretical range of \iso{44}Ti mass
is therefore highly uncertain, as discussed by \cite{The2006} for Cas
A. 

In addition to \iso{56}Ni, \iso{57}Ni and \iso{44}Ti, other radioactive
isotopes may be present in the ejecta. In particular, \iso{60}Co and \iso{22}Na may have sizeable
abundances. These isotopes are produced by neutron and proton capture, respectively, during carbon burning, and have uncertain yields that are sensitive to the convective prescription \citep{Timmes1996}. 

The \iso{60}Co-isotope has a decay time of 7.605 years, and may have a similar mass to \iso{44}Ti (computed as $2.5\e{-5}$~\msun~in WW95  and $6.5\e{-5}$~\msun~ in WH07). The gamma-ray luminosity for these two masses are 56\% and 150\% of the \iso{44}Ti gamma-ray luminosity, respectively, and the electron deposition is 12\% and 30\% of the \iso{44}Ti positron deposition, respectively.%The gamma-ray and electron luminosities at eight years are, compared to the \iso{44}Ti values:
%\begin{equation}
%\frac{L_\gamma(^{60}\mbox{Co})}{L_\gamma(^{44}\mbox{Ti})} = 0.63 \frac{M(^{60}\mbox{Co})/M(^{44}\mbox{Ti})}{0.2}\mbox{erg s}^{-1}
%\end{equation}
%and
%\begin{equation}
%\frac{L_{e-}(^{60}\mbox{Co})}{L_{e+}(^{44}\mbox{Ti})} = 0.13 \frac{M(^{60}\mbox{Co})/M(^{44}\mbox{Ti})}{0.2}\mbox{erg s}^{-1}
%\end{equation}
%\citep{Timmes1996}%, which should be compared to the \iso{44}Ti values
%\begin{equation}
%L_\gamma(^{44}\mbox{Ti}) = 4.1\e{36}e^{-t/78~y}\left(\frac{M}{1\e{-4}M_\odot}\right) \mbox{erg s}^{-1}
%\end{equation}
%and
%\begin{equation}
%L_{e+}(^{44}\mbox{Ti}) = 1.3\e{36}e^{-t/78~y}\left(\frac{M}{1\e{-4}M_\odot}\right) \mbox{erg s}^{-1}
%\end{equation}
~\iso{60}Co may therefore make a non-neglegible contribution to the powering at eight years. % ($\sim$ 12\% for M(\iso{60}Co)=$2\e{-5}$~\msun~,~M(\iso{44}Ti)=$1\e{-4}$~\msun~ and 28\% for M(\iso{60}Co)=$6\e{-5}$~\msun~,~M(\iso{44}Ti)=$1\e{-4}$~\msun~) . 
%The gamma-ray emission is mainly in lines at 1.173 MeV
%and 1.332 MeV, similar to that of \iso{44}Ti, and the opacity should be
%similar. %\citep{Timmes1996}. 
The $\beta$-decays occur as electrons with energies
up to 317 keV, with an average energy of 96.5 keV. This is significantly
lower than the energy of \iso{44}Ti-positrons, and the stopping range is therefore
much shorter. A major difference is that most of the \iso{60}Co electron
energy should be deposited in the oxygen-rich zones. These account for
approximately half of the total column density of the core, and from
Fig. \ref{fig:stopping_power} it can be seen that they can easily 
trap most of these electrons. The increased deposition in the oxygen-rich regions means that an appreciable \iso{60}Co mass may mimic the
effects of the positron leakage in the \iso{44}Ti dominated models
(Sect. \ref{sec:pos}). It will in particular increase the O I, Mg I,  and Na I line fluxes noticeably, further complicating the question of where the positrons are deposited.

\iso{22}Na has a decay time of 3.75 years, emitting a gamma-ray
at 1.274 MeV and in 90\% of the cases a positron, while 10 \% of the
cases occur by electron capture. The positrons have an average energy
of 215.5 keV and a maximum energy of 545.4 keV. %The contribution to the powering at eight years is
%\begin{equation}
%\frac{L_\gamma(^{22}\mbox{Na})}{L_\gamma(^{44}\mbox{Ti})} = 0.054 \frac{M(^{22}\mbox{Na})/M(^{44}\mbox{Ti})}{0.01}\mbox{erg s}^{-1}
%\end{equation}
%and
%\begin{equation}
%\frac{L_{e+}(^{22}\mbox{Na})}{L_{e+}(^{44}\mbox{Ti})} = 0.018 \frac{M(^{22}\mbox{Na})/M(^{44}\mbox{Ti})}{0.01}\mbox{erg s}^{-1}
%\end{equation}
The \iso{22}Na mass in the explosion model is $4.5\e{-6}$ \msun, in \citet{Woosley1989} it is $2.0\e{-6}$, in WW95 it is $3.0\e{-7}$~\msun~and in \citet{Thielemann1996} it is $1.3\e{-7}$~\msun. This gives gamma-ray luminosities of 0.5\%-17\% of the \iso{44}Ti gamma-ray luminosity, and positron depositions of 0.2\%-5.8\% of the \iso{44}Ti positron deposition. Based on these models it is therefore unlikely that \iso{22}Na is important for the energy budget at the epoch studied here.

It is, unfortunately, difficult to constrain the contribution from \iso{60}Co and \iso{22}Na to the eight-year-spectrum  because of the uncertainties in the modeling of the oxygen zones, and the possibility of non-local positron/electron absorption occurring. We assume both of these isotopes to make neglegible contributions to the powering at eight years. %If their masses are instead on the high side, the \iso{44}Ti mass has to be revised down by as much as a factor $\sim$1.5. We allow for this in our low-side error bar.
The shorter decay time scales of \iso{60}Co and \iso{22}Na, compared to \iso{44}Ti, means that their influence varies with time. The modeling of the spectrum and light
curve at other epochs should therefore have the potential to distinguish between
the \iso{44}Ti and the \iso{60}Co/\iso{22}Na input. Unfortunately, only a few
spectra exist at epochs later than the one studied here, and they are
increasingly contaminated by the flux from the ring collisions. Modeling of the NIR 
spectrum at 19 years has shown that the combination of $\mbox{M(\iso{44}Ti)}=1\e{-4}$~\msun~
and $\tau_{\rm d}=1$ matches the observed NIR spectrum at this epoch reasonably well \citep{Kjaer2010}.
However, this spectrum may be affected both by scattered light and X-ray input from the circumstellar ring \citep{Larsson2011}. 

Small or moderate masses of \iso{60}Co and \iso{22}Na are supported by the modeling of the SN 1987A photometry (FK02), which shows
that the light curves do not deviate significantly from a pure \iso{44}Ti input between 1800 and 3500 days.

%This shows that potential contributions by \iso{60}Co and \iso{22}Na cannot
%affect the powering at eight years by more than a maximum of $\sim$30\%; the \iso{44}Ti mass would then
%have to be lower than $\sim 1\e{-4}$~\msun~(for $\tau_{\rm d}=1$) which is refuted by the 19-year
%modeling. 

%These models did not include
%scattering and fluorescence, whose impact on the optical spectrum has been shown here to be $\sim$30\%. We include the 30\% uncertainty in our low-side error bar.

\section{Conclusions}
\label{sec:conclusions}

Our main conclusions are:

\begin{itemize}
%\item A model based on non-thermal, radioactive input by $^{44}$Ti and an ejecta model based on a realistic explosion model provides a good fit of the spectrum of SN 1987A from UV to near-IR. 

%\item A spectral model based on a state-of-the-art explosion model of a 19 \msun~star reproduces most of the observed lines in the 8-year UV/optical/NIR spectrum of SN 1987A to good accuracy. 

\item A spectral model based on the determination of the positron and gamma-ray degradation, the ionization, excitation and temperature structure of the ejecta, and the multi-line radiative transfer, reproduces most of the observed lines in the eight-year UV/optical/NIR spectrum of SN 1987A to good accuracy. We showed that many regions in the spectrum are produced by Fe I lines. The contribution by Fe II lines is small.

\item Even many years after explosion, scattering and fluorescence are important processes for the spectral formation in Type-II supernovae. At an age of eight years, $\sim$30\% of the emerging optical/NIR flux and $\sim$60\% of the UV flux in SN 1987A is produced by fluorescence. Non-local line transfer decreases the UV flux by $\sim$30\%, increases the optical flux by $\sim$10\%, and increases the NIR flux by $\sim$30\%, compared to a nebular treatment.

\item Most of the flux in the hydrogen lines and continua, as well as significant parts of the UV spectrum, are at eight years produced by freeze-out emission from the envelope rather than by instantaneous  \iso{44}Ti deposition. In the far-UV ($\lesssim 2500~\AA$), essentially the whole spectrum is produced in the envelope.

\item {A scenario of positron leakage from the Fe/He clumps produces too strong emission in several lines from the hydrogen and oxygen zones. We therefore support the conclusion in \citet{Chugai1997} that the positrons remain trapped by a magnetic field.}

\item{ The [Fe II] 26 $\mu$m line in Spitzer observations at day 6190 is much weaker than expected. Possible explanations are dust cooling or that positron leakage is occuring at that epoch.}

%\item With local positron deposition into the largely neutral iron gas,.

\item {Modeling the dust as a gray absorber with radial optical depth $\tau_{\rm d}=1$, which is our best estimate from the line peak blue-shifts, the amount of $^{44}$Ti synthesized in the explosion that best reproduces the total flux between 2000 and 8000 \AA~is \bestti. We estimate the uncertainty in this result to $\pm 0.5\e{-4}$~\msun.} %A positron leakage scenario gives a smaller best fitting \iso{44}Ti mass of $8\e{-5}$ \msun.

\item One of our line identifications is the Ca I \wl4226 line. If this identification is correct, the density in the Si/S zone must be at least $\sim$10 times lower than in the oxygen zones for the line to be as strong as observed at this epoch. Because the Si/S zone contains some of the $^{56}$Ni, this adds further evidence to the picture that the early radioactive decays cause expansion in the ejecta. %Neutral calcium also has a strong emission line that blends with H$\alpha$.

%\item The strength of especially the [O I] \wll 6300, 6364 and Mg I] \wl4571 lines are indicative of positron deposition outside of the Fe/He clumps at eight years. However, for the spectrum as a whole the two assumptions of local and non-local positrons depositions produce fits of similar quality and no definite coclusions about positron trapping can be made.  

\end{itemize} 

\begin{acknowledgements}
We thank Alexander Heger for providing us with the explosion model, Peter Meikle for providing the NIR data, Cathy Ramsbottom for providing Fe II collisional cross sections, Sultana Nahar for consultations on radiative oxygen recombination, and Claes-Ingvar Bjornsson for useful discussions. We especially thank Mattias Ergon for comments on the manuscript and many hours of useful discussions. This work is supported by the Swedish Research Council and the Swedish National Space Board.
\end{acknowledgements}

\appendix

\section{Atomic data}
\label{app:atomicdata}
Table \ref{tab:atomdata} lists a summary of the model atoms used and the sources for the atomic data. The number of levels listed are the number of levels (including fine-structure)  whose populations are solved for, followed by the total number of levels included for making the line list. For the population solutions, a size of 50 levels for the neutrals and 10 levels for the ions were typically used, with the larger number for the neutrals because these emit recombination spectra, whereas no ion does in any significance. No lines with optical depths higher than $10^{-3}$ have their lower levels above these limits at this epoch. Elements that have non-thermal excitations or specific recombination rates to higher levels, have more levels included in the population solution. Two examples are Fe I and Fe II, where the populations of all $\sim$500 levels are calculated. 

All atoms are solved for with fine-structure, which is especially appropriate at late times. For allowed transitions that lack published collision strengths, we use the van Regemorter formula \citep{Regemorter1962}. For forbidden transitions we used the van Regemorter formula with an absorption oscillator strength of $1\e{-2}g_u$. This treatment is based on the fact that forbidden transitions usually have collision strengths about an order of magnitude smaller than allowed transitions. This treatment also allows a reproduction of the typically very large collision strengths for transitions of the type $nl\rightarrow nl'$, since the van Regemorter formula contains a $\Delta E^{-1}$ dependence. We checked the sensitivity to this treatment by comparing the spectrum with one using a constant $\Upsilon_{\rm u,l}=0.1g_lg_u$ for the forbidden lines. Only the Mg I] \wl 4571 line was sensitive to the treatment. The fine-structure collision strengths in this atom is on the order of $10^5$ \citep[e.g.][]{Sundqvist2008}, and the first treatment is better.

If specific recombination rates are lacking, we allocate 30\% of the total recombination rate to the ground multiplet and the remaining 70\% in proportion to the statistical weights. If some specific recombination rates are available but not all, we allocate the difference between the total rate and the sum of the known rates to the states with unknown rates in proportion to their statistical weights. If all specific recombination rates are known, but do not match the total rate, we scale them so that they do. We do not include dielectronic recombinations, which are unimportant at the low temperatures here.

All ground-state photo-ionization cross sections are from a routine based on \citet{Verner1996}. In addition, cross sections for some selected excited states are included (n=2 in H I, 2$^1$S and 2$^1$P$^0$ in He I, $^1$D, $^1$S and $^5$S in C I, $^1$D and $^1$S in O I, $^3$P in Mg I, $^2$P in Mg II, $^1$D in Si I, $^1$D in S I, $^2$D in Ca II, a$^5$ F in Fe I and a$^4$F in Fe II), taken from TOPBASE\footnote{http://cdsweb.u-strasbg.fr/topbase/xsections.html}.

For information on the cross sections used in the Spencer-Fano routine, see KF92 and K98 a, b. An important update is the addition of detailed cross sections for Fe II \citep{Ramsbottom2005,Ramsbottom2007}, and for Ca I \citep{Samson2001}.

%Penning ionizations (He I(2$^3$S) + H I $\rightarrow$ He I + H II) occur with a rate of $7.5\e{-10}$ cm$^3$ s$^{-1}$ \citep{Bell1970}.

\begin{table*}
\caption{Summary of the model atoms used and sources for the atomic data. Col. 1: Element, Col. 2: number of fine-structure levels solved for and the number included in the radiative transfer, Col. 3: highest state included, Col. 4: source for level energies and transition A-values, Col. 5: source for the total radiative recombination rate, Col. 6: source for the specific radiative recombination rates, and the number of levels for which these rates are included in parenthesis, Col. 7: source for the (thermal) collision strengths, and the number of levels with collision strength included in parenthesis.}
\centering
\begin{tabular}{l l l l l l l}
\hline\hline
Element & Levels & Highest state  & Energies and A-values & Tot. rr. & Spec. rr. & Coll. strengths \\
%   &        &          & E. \& A.  & Tot. rr. & Spec. rr & Coll. strengths \\
\hline
H I & 210/465 & n=20/n=30 & 23 & 18 & 4 (all) & 1, 5 (all)\\

He I & 29/29 & 5p($^1$P) & 23 & 6 & 3 (all) & 25 (all)\\

C I & 90/90 & 5d($^1$D) & 23 & 11 & 11 (first 17) & 16 (first 6)\\
C II & 7/77 & 8g($^2$G) & 23 & 10 & -- & 16 (first 5)\\

N I & 68/68 & 5s($^2$P) & 23 & 19 & -- & 16 (first 5)\\
\hline
N II & 10/148 & 2p$^3$($^5$S)/3p($^3$D) & 23 & 19 & -- & 16 (all)\\

O I & 135/135 & 8d($^3$D*) & 23 & 10 & 10 (first 49) & 16 (first 7)\\
O II & 16/16 & 2p$^4$($^2$S) & 23 & 10 & -- & 16 (first 5)\\

Ne I & 3/3 & 3s($^2$[3/2]) & 23 & 19 & -- & --\\
Ne II & 3/3 & 2p$^6$($^2$S) & 23 & 19 & -- & 16 (first 1)\\

\hline
Na I & 66/66 & 9f($^2$F) & 23 & 14 & 24 (first 60) & 20 (first 2)\\

Mg I & 200/371 & 37d($^3$D)/80d($^3$D) & 7 & 14, 19 & 24 (first 18) & 8 (first 21) \\
Mg II & 35/35 & 7p($^2$P) & 23 & 21 & -- & 16 (first 2)  \\ 

Al I & 50/86 & 10d($^2$D)/21s($^2$S) & 7  & 14 & -- & --\tablefootmark{c}\\
Al II & 10/154 & 3p$^2$($^3$P)/3pd($^3$D) & 7  & --\tablefootmark{b} & -- & --\tablefootmark{c}\\

\hline

Si I & 200/494 & 10s($^3$P)/21s($^1$P) & 7 & 11 & 11 (first 49) & --\tablefootmark{c}\\
Si II & 77/77 & 9p($^2$P) & 23 & 10 & -- & 16 (all)\\

S I & 125/125 & 8f($^3$F) & 7 & 14, 19  & -- & --\tablefootmark{c} \\ 
S II & 5/5 & 3p$^3$($^2$P) & 23 & 10 & -- & 16 (all with gs)\\

Ar I & 3/3 & 3p$^5$($^4$S) & 23 & 19 & -- & --\tablefootmark{c}\\

\hline
Ar II & 2/2 & 3p$^5$($^2$P) & 23 & 19 & -- & 16\\ 

Ca I & 198/198 & 6d($^3$D) & 7 & 14 & 24 (first 11) & --\tablefootmark{c}\\
Ca II & 69/69 & 16d($^2$D) & 23 & 14 & -- & 9 (all) \\

Sc I & 50/254 & 4p($^4$F)/7p($^2$P) & 7 & --\tablefootmark{a} & -- & --\tablefootmark{c}\\
Sc II & 10/165 & 3d$^2$($^3$P)/6f($^1$H) & 7 & --\tablefootmark{b} & -- & --\tablefootmark{c}\\

\hline
Ti I & 50/394 & 4s(c$^3$P)/4p$^2$($^1$P) & 7 & 14 & -- & --\tablefootmark{c}\\
Ti II & 10/212 & 4s(a$^2$F)/5d($^4$H) & 7 & --\tablefootmark{b} & -- & --\tablefootmark{c}\\ 

V I & 50/502 & sp(z$^6$D)/4p$^2$(r$^2$H) & 7 & --\tablefootmark{a} & -- & --\tablefootmark{c}\\
V II & 10/323 & 4s(a$^5$F)/4f(2(11/2)) & 7 & --\tablefootmark{b} & -- & --\tablefootmark{c}\\

Cr I & 50/392 & sp(z$^7$D)/21p(e$^5$F) & 7 & --\tablefootmark{a} & -- & --\tablefootmark{c}\\
\hline
Cr II & 10/500 & 4s(a$^4$D)/4f($^4$H) & 7 & --\tablefootmark{b} & -- & --\tablefootmark{c}\\

Mn I & 50/293 & 4s(b$^4$G)/4p(z$^2$I) & 7 & 14 & -- & --\tablefootmark{c}\\
Mn II & 10/509 & 4s(a$^5$G)/9h($^7$H) & 7 & --\tablefootmark{b} & -- & --\tablefootmark{c}\\ 

Fe I & 496/496 & 4p($^1$P$^0$)($^3$F) & 7 & 10 & 10 (first 356) & 2, 15 (first 120)\\
Fe II & 578/578 & ($^7$S)4d($^8$D) & 12, 23  & 10 & 10 (first 381) & 22 (first 372) \\
\hline
Fe III & 168/168 & 4p($^3$I) & 7 & --\tablefootmark{b} & -- & --\tablefootmark{c} \\
Co I & 50/317 & d9(c$^2$D)/5s(37$^0$) & 7 & --\tablefootmark{a} & -- & --\tablefootmark{c}\\
Co II & 10/108 & 4s(b$^3$F)/7d($^5$H) & 7, 17 & --\tablefootmark{b} & -- & --\tablefootmark{c}\\ 

Ni I & 50/136 & 4p(F*)/4d($^2$[1/2]) & 23 & 19 & -- & --\tablefootmark{c}\\
Ni II & 10/500 & 4s($^2$D)/sp($^4$D) & 7 & 19 & -- & --\tablefootmark{c}\\  
  
\hline
\end{tabular}
\tablebib{
(1) \citet{Anderson2000}; (2) \citet{Axelrod1980}; (3) \citet{Benjamin1999}; (4) \citet{Brocklehurst1971}; (5) \citet{Johnson1972}; (6) \citet{Hummer1998}; (7) R. Kurucz (http://www.cfa.harvard.edu/amp/ampdata/kurucz23/sekur.html); (8) \citet{Mauas1988}; (9) \citet{Melendez2007}; (10) S. Nahar (www.astronomy.ohio-state.edu/~nahar/nahar\_radiativeatomicdata/index.html); (11) \citet{Nahar1995}; (12) \citet{Fuhr2006}; (13) \citet{Nahar1998_FeII}; (14) \citet{Pequignot1986}; (15) \citet{Pelan1997}; (16) A. Pradhan (www.astronomy.ohio-state.edu/~pradhan/table2.ps); (17) \citet{Quinet1998}; (18) \citet{Seaton1959}; (19) \citet{Shull1982}; (20) \citet{Trail1994}; (21) \citet{Verner1996}; (22) \citet{Zhang1995}; (23) NIST (www.nist.gov); (24) TOPBASE (http://cdsweb.u-strasbg.fr/topbase/xsections.html); (25) \citet{Almog1989}; 
}
\tablefoot{
%  \tablefoottext{a}{Includes also forbidden transitions}
  \tablefoottext{a}{Fe I rate used}
  \tablefoottext{b}{Fe II rate used}
%  \tablefoottext{d}{Power law to $10^4$ K}
%  \tablefoottext{c}{Default rate for neutrals : $5\e{-12}T_{100}^{-1/2}$}
%  \tablefoottext{d}{Default rate for singly ionized : $2\e{-11}T_{100}^{-1/2}$}
%  \tablefoottext{g}{Rates exist but not implemented}
%  \tablefoottext{c}{Only allowed transitions.}
  \tablefoottext{c}{Regemorter's approximation.}
%  \tablefoottext{i}{Temperature-dependence exists?}
%  \tablefoottext{j}{SOME forbidden transitions included?}
%  \tablefoottext{k}{SS82 for $t>10^3$ K, PA86 otherwise}
}
\label{tab:atomdata}
\end{table*}

\bibliographystyle{aa}
\bibliography{./references}

%\bibliographystyle{/home/andersj/latex/aa}
%\bibliography{references.bib}%,/home/andersj/latex/references.bib}

\end{document}